\documentclass[aps,prx,reprint,groupedaddress,superscriptaddress,floatfix]{revtex4-2}

\usepackage[T1]{fontenc}
\usepackage{newtxtext}
\usepackage{newtxmath}
\usepackage{bm}

\usepackage{xcolor}

\usepackage{titlesec}

\usepackage{amsmath}
\usepackage{dsfont}
\usepackage{mathtools}
\usepackage{amsfonts}
\usepackage{bbold}

\usepackage{graphicx}
\usepackage[english]{babel}
\usepackage{ragged2e}
\usepackage{epsfig}
\usepackage{subcaption}

\graphicspath{{./img/}}

\usepackage{enumitem}

\usepackage{tabularx}
\usepackage{makecell}
\usepackage{multirow}
\usepackage{booktabs}
\usepackage{array}
\usepackage{dcolumn}

\usepackage{lineno}
\usepackage{ifpdf}
\usepackage{bbm}
\usepackage{amscd}
\usepackage[mathscr]{eucal}
\usepackage{float}
\usepackage{etoolbox}

\usepackage[
    colorlinks,
    citecolor=blue,
    linkcolor=blue,
    urlcolor=blue
]{hyperref}

\usepackage[all]{hypcap}
\usepackage{cleveref}

\makeatletter
\let\addcontentslineOriginal\addcontentsline
\let\addcontentsline\@gobblethree
\makeatother

\begin{document}


\title{Higher-order spacing distributions in non-Hermitian random matrix ensembles and the Sachdev-Ye-Kitaev model}

\author{Sashmita Rout}
\email{sashmitaa111@gmail.com}
\affiliation{Department of Physics, Visvesvaraya National Institute of Technology, Nagpur 440010, India}
 \author{Udaysinh T. Bhosale}
\email{udaysinhbhosale@phy.vnit.ac.in}
\affiliation{Department of Physics, Visvesvaraya National Institute of Technology, Nagpur 440010, India}
\date{\today}

\begin{abstract}
The distribution of higher-order ($k \geq 1$) level spacings (HOS), i.e., the distribution of $k$-th nearest-neighbor spacings of complex eigenvalues for non-Hermitian random matrix ensembles is studied. 
We use the generalized Wigner-Dyson distribution as the functional form for spacing distributions to find the best fit. It is found that, for $k>3$, this approximation works very well. The spacing distributions for three classes of Ginibre ensembles such as GinOE ($AI$), GinUE ($A$), and GinSE ($AII$) agree very well with each other beyond nearest-neighbor in the bulk of the spectrum. Further, the three non-Hermitian random matrix universality classes, GinUE ($A$), $AI^{\dagger}$, and $AII^{\dagger}$ exhibit distinct HOS statistics in the spectral bulk. We verify the universality of our results on the non-Hermitian $q$-body Sachdev-Ye-Kitaev (nHSYK) model. In addition to this, the universality of HOS statistics of GinUE in the bulk is verified by the product of $M$ independent matrices from the GinUE, 
and it is found to be independent of $M$.

\end{abstract}
 
\maketitle                                       
\section{Introduction}
\label{sec:Introduction} 
Spectral statistics has been used as a valuable tool for probing the dynamics of a system, providing correspondence between random matrix theory (RMT) and physical systems \cite{mehta2004random}. For Hermitian systems, this correspondence is described by the Bohigas-Giannoni-Schmit conjecture (BGS) \cite{PhysRevLett.52.1} and Berry-Tabor conjecture \cite{berry1977level} to characterize chaos and integrability, respectively. According to them, a quantum system whose classical limit is chaotic, follows RMT statistics in their level spacing distribution, whereas an integrable system follows Poisson statistics. Among the various spectral fluctuation measures used in RMT, nearest neighbor spacing distribution (NNSD) and nearest neighbor spacing ratio distribution (NNSRD) of eigenvalues are most widely used to study short-range spectral correlations \cite{mehta2004random, oganesyan2007localization, atas2013distribution}. However, the level correlations over large spectral intervals are useful in many cases, such as probing short-time dynamics in chaotic quantum systems having a well-defined classical limit \cite{stockmann2000quantum}, and investigating phenomena accompanying the many-body localization (MBL) transition \cite{sierant2019level,agarwal2017rare}. 
Higher-order spacing distribution (HOSD) and higher-order spacing ratio distribution (HOSRD) of eigenvalues \cite{rao2020higher,tekur2018higher} serve as measures of long-range spectral correlations.
They are comparatively easier for computing numerically and simpler to analyze than other measures of long-range spectral correlations, such as spectral form factor (SFF), number variance, and spectral rigidity. 
In addition to that, the number variance and spectral rigidity are very sensitive to the well-defined unfolding method and can sometimes lead to misleading signatures \cite{gomez2002misleading}. However, HOSD and HOSRD can also help to characterize the system correctly by studying the composite spectrum (collection of all eigenvalues from different symmetry sectors), without resolving the symmetry and to find the symmetry structure of the Hamiltonian, as demonstrated in Refs.~ \cite{tekur2020symmetry,bhosale2021superposition,rout2026higherorderspacingssuperposedspectra}. This requires that the constituent spectra in the composite spectrum should belong to the same random matrix class and should be of equal or nearly equal size.   

The BGS and Berry-Tabor conjectures are valid for isolated quantum systems, which are Hermitian and have $1$D spectra. Therefore, an interesting direction is to investigate the extension of these conjectures to non-Hermitian systems.
Similar to Hermitian RMT, there exists a conjecture that relates the spectral fluctuations of non-Hermitian random matrices with that of open chaotic quantum systems, known as Grobe-Haake-Sommers (GHS) conjecture \cite{grobe1988quantum, grobe1989universality}. According to this conjecture, open quantum systems whose classical limits are chaotic obey non-Hermitian random matrix statistics and integrable systems with complex spectra or systems whose corresponding classical dynamics are integrable obey $2$D Poisson statistics. This conjecture has been verified in various systems such as the dissipative quantum kicked top \cite{jisha2024universality}, dissipative many-body quantum spin chain \cite{akemann2025two, akemann2019universal}, etc. Apart from these, many applications of non-Hermitian operators with $2$D spectra are found. For example, random neural networks, complex networks, ecology, etc. (refer to Ref.~\cite{akemann2022spacing, rotter2015review} and references therein). 

Symmetry and universality play an important role in RMT \cite{mehta2004random, stockmann2000quantum, akemann2011oxford}. 
In the case of Hermitian random matrices, there exists Dyson’s three-fold classification based on conventional time reversal symmetry (TRS), and they are GOE, GUE, and GSE. Altland and Zirnbauer extended this three-fold classification to $10$ classes by incorporating additional symmetries \cite{altland1997nonstandard}. Whereas the generic local bulk spectral statistics of these ten classes can be classified into the three classes of Dyson. 
In the case of non-Hermitian random matrices, a $38$-fold symmetry classification has been proposed, which are based on the algebraic structures of matrix ensembles \cite{Kawabata2019Symmetry}.
Among these $38$ classes, GinOE ($AI$), GinUE ($A$), and GinSE ($AII$) ensembles are analogous to the three ensembles of Dyson's classification of Hermitian random matrices, in terms of their conventional time reversal symmetry, which involves complex conjugation. They consist of real, complex, and quaternion entries, respectively \cite{ginibre1965statistical, forrester2024Progress}. However, these three different symmetry classes show the same universal nearest-neighbor level spacing statistics in the generic bulk of the spectrum, unlike the Hermitian counterpart. Moreover, only three distinct universality classes of two-dimensional local bulk statistics were observed among the $38$ non-Hermitian symmetry classes \cite{hamazaki2020universality}. They are GinUE ($A$), $AI^{\dagger}$, and $AII^{\dagger}$, which are characterized by transposition symmetry (TRS$^{\dagger}$) \cite{hamazaki2020universality}. We adopt the nomenclature used in Ref.~\cite{hamazaki2020universality} (Cartan classification) throughout this paper for non-Hermitian symmetry classes. The $AI^{\dagger}$ and $AII^{\dagger}$ consist of complex symmetric and complex quaternion self-dual matrices. This universality classification is based on the local bulk statistics, specifically, NNSD \cite{hamazaki2020universality}. Later, this finding is strengthened by the two-dimensional Coulomb gas ($2$DCG) analysis \cite{akemann2022spacing}, as well as by the analytical results for both bulk and edge statistics \cite{akemann2025complex}.
The expression for NNSD of complex eigenvalues is given as follows:
\begin{small}
\begin{equation}
s_l^{(1)} = |z_{l}^{(1)}- z_{l}|,\hspace{.5cm}  l = 1,2,3,\ldots 
\label{Eq: HOS} 
\end{equation}
\end{small}
Here, $z_{l}$ are the complex eigenvalues such that $z_{l}=x_{l}+iy_{l}$. And $z_{l}^{(1)}$ is the nearest neighbor of $z_{l}$, with respect to the distance in $\mathbb{C}$.

Among these three universality classes in $2$D, for any $N$, the closed-form expressions for the NNSD exist only for the GinUE ensembles. This is given as follows \cite{grobe1988quantum}:
\begin{equation}
P_{\mathrm{GinUE}}(s)=\prod_{j=1}^{N-1}\frac{\Gamma\!\left(1+j,s^{2}\right)}{\Gamma\!\left(1+j\right)}\sum_{h=1}^{N-1}
\frac{2s^{2h+1}e^{-s^{2}}}{\Gamma\!\left(1+h,s^{2}\right)}.
\label{Eq: GinUE_NNS}
\end{equation}
For $AI^{\dagger}$ and $AII^{\dagger}$, such expressions exist only for $N=2$ \cite{hamazaki2020universality}. Some studies investigated HOSD for a $D$-dimensional Poisson point process \cite{sakhr2006wigner, haake2010quantum, akemann2024interactions}. The expression of NNSD for $2$D Poisson is given by:
\begin{equation}
P_{\mathrm{Poisson}}^{2\mathrm{D}}(s)=\frac{\pi}{2}\, s\,\exp\left(-\frac{\pi}{4}s^{2}\right)
\end{equation}
Further, in Ref.~\cite{akemann2022spacing}, a random matrix representation is proposed for the two-dimensional Coulomb gas ($2$DCG), which is known as the non-Hermitian $\beta$ ensemble using complex normal non-Hermitian random matrices. There, the authors have tried to approximate the NNSD of the three universality classes in terms of the $2$DCG.
For $N=2$, an expression for NNSD is derived as follows \cite{akemann2022spacing}:
\begin{equation}
\tilde{P}(s,\beta)
=\frac{2\alpha^{1+\beta/2}}{\Gamma\!\left(1+\beta/2\right)}\, s^{1+\beta}\exp\!\left(-\alpha s^2\right),
\end{equation}
where $\beta$ is known as the inverse temperature of the $2$DCG. For GinUE, $\beta=2$, and the $2$D Poisson distribution, $\beta=0$.
Here, 
\begin{equation}
\alpha=\left[\frac{\Gamma\!\left(\dfrac{3+\beta}{2}\right)}
{\Gamma\!\left(1+\dfrac{\beta}{2}\right)}\right]^2.
\end{equation}
However, this distribution does not act as a good approximation for large $N$. 
For large $N$, this distribution $P_{N=2}(s,\beta)$ was modified as follows with $\beta_{\mathrm{eff}}$ and consequently $\alpha_{\mathrm{eff}}$ \cite{akemann2022spacing}:
\begin{equation}
P(s,\beta_{\mathrm{eff}})
=\frac{2\alpha_{\mathrm{eff}}^{1+\beta_{\mathrm{eff}/2}}}{\Gamma\!\left(1+\beta_{\mathrm{eff}}/2\right)}\, s^{1+\beta_{\mathrm{eff}}}\exp\!\left(-\alpha_{\mathrm{eff}} s^2\right),
\label{eqn:2D_surmise}
\end{equation}
where, 
\begin{equation}
\beta_{\mathrm{eff}}(\beta)=2.108\,\beta-0.190\,\beta^{2}+0.030\,\beta^{3}
\end{equation}
and
\begin{equation}
\alpha_{\mathrm{eff}}=\left[\frac{\Gamma\!\left(\dfrac{3+\beta_{\mathrm{eff}}}{2}\right)}
{\Gamma\!\left(1+\dfrac{\beta_{\mathrm{eff}}}{2}\right)}\right]^2.
\end{equation}
They also found that NNSD of $AI^{\dagger}$ and $AII^{\dagger}$ are very well fitted by $\beta=1.4$ and $\beta=2.6$ from a $2$DCG, respectively.

To the best of our knowledge, nearest neighbor ($k=1$) spacing and ratio statistics, and next-to-nearest neighbor ($k=2$) spacing statistics of complex eigenvalues are studied explicitly in Refs.~\cite{sa2020complex, akemann2026three, akemann2022spacing,akemann2024interactions, akemann2025two}. While HOSRD (any $k \geq 1$) has been studied in the context of open systems, that deal with singular values rather than complex eigenvalues themselves \cite{tekur2024higher}.
To date, a systematic study on HOSD ($k>2$) of complex eigenvalues of non-Hermitian random matrix ensembles is still lacking, be it numerical or analytical. Therefore, in this paper, our main objective is to numerically study the HOSD of the complex spectra of the non-Hermitian ensembles.
We are interested specifically in GinOE ($AI$), GinUE ($A$), and GinSE ($AII$) ensembles, and the $3$ universality classes $A$, $AI^\dagger$, and $AII^\dagger$.
We now introduce some definitions and concepts that are used in our work.
In the case of Hermitian random matrices, nearest neighbor spacing (NNS) is defined as $s_{l}=E_{l+1}-E_{l}$, where $E_{l},\,\,\, l=1,2,3,\ldots$ are the eigenvalues of the corresponding matrix and $E_{l+1}>E_{l}$ for all $l$. Its distribution for random matrices of large size $N$ can be approximated by \cite{mehta2004random}:
\begin{equation}
P(s,\beta)=A_{\beta}s^{\beta}\exp(-C_{\beta}s^2),\,\,\,\,\beta=1, 2, 4,
\label{Eq:PSBeta}
\end{equation}
where $A_{\beta}$ and $C_{\beta}$ are normalization constants that depend on $\beta$, which is known as the Dyson index.
Furthermore, the expression for HOS for complex eigenvalues can be defined as:
\begin{equation}
s_l^{(k)} = |z_{l}^{(k)}- z_{l}|,\hspace{.5cm}  l,k = 1,2,3,\ldots
\label{Eq:HOS_Non_hermitian}
\end{equation}
Here, $z_{l}^{(k)}$ is the $k$-th nearest neighbor of $z_{l}$, with respect to the distance in $\mathbb{C}$.


We denote the distribution of $s_l^{(k)}$ as $P^{k}(s)$. We now discuss the motivation underlying this study below. First, our study of HOS of complex-valued eigenvalues investigates additional information of spectral correlations beyond the NNSD. Second, Eq.~(\ref{eqn:2D_surmise}) has the same mathematical form as Eq.~(\ref{Eq:PSBeta}). This motivates us to approximate the HOSD using Eq.~(\ref{Eq:PSBeta}) with modified index $\beta'$, where $\beta'$ is a positive integer. This type of concept has already been used in earlier works \cite{tekur2018higher,tekur2020symmetry, bhosale2021superposition,rout2026higherorderspacingssuperposedspectra}. Our approach is important because it acts as a one-parameter good fit, and we are giving a simpler form of the analytical distribution function that can act as an approximation to HOSD. Third, 
we are interested in exploring the behavior of the three Ginibre ensembles and three non-Hermitian universality classes at large spectral gap. Fourth, whether HOSD of other known non-Hermitian ensembles, such as the product of independent complex Ginibre matrices, agree with that of GinUE? Additionally, we want to exemplify the universality of our results by applying them to physical systems.

The rest of the paper is organized as follows: In Sec.~\ref{sec: HOSD_non_Hermitian}, we describe the numerical method used in this paper to find the best fit. Along with that, we study HOSD for various non-Hermitian random matrices like GinOE ($AI$), GinUE ($A$), GinSE ($AII$), $AI^\dagger$, $AII^\dagger$, and the product of independent complex Ginibre matrices. In Sec.~\ref{sec: physical system}, we verify the universality of our results by studying the nHSYK model. Finally, in Sec.~\ref{sec: Conclusion}, we discuss and conclude our work and give some future directions as well.
\section{HOSD in non-Hermitian random matrix ensembles}
\label{sec: HOSD_non_Hermitian}
Among the various non-Hermitian random matrix symmetry classes, we consider the Ginibre, $AI^{\dagger}$, and $AII^{\dagger}$. In addition to that, the product of independent matrices from GinUE are also taken into consideration. In all of these cases, we consider the statistics only in the bulk of the spectrum, that means sufficiently away from the edge and the center or the critical line. We first discuss the numerical method used in this paper to find the best fit $P(s, \beta')$ for the $P^{k}(s)$ quantitatively. Here, $P^{k}(s)$ and $P(s, \beta')$ represent the observed histogram corresponding to the HOS and the postulated functional form (Eq.~(\ref{Eq:PSBeta}), in which $\beta$ is substituted by $\beta'$), respectively. We have chosen the measure $D(\beta')$ to find the best. Because it uses the full distribution and
avoids binning bias compared to raw histogram-based methods.
For HOSD of the spectra, $D(\beta')$ is defined as \cite{tekur2018higher}:
\begin{equation}
D(\beta')=\sum_{l=1}^n\left|F^{k}(s_l)-F(s_l, \beta')\right|,
\label{Eq:DBeta_HOS_Superposed}
\end{equation}
where $F^{k}(s)$ and $F(s, \beta')$ denote cumulative distribution functions corresponding to the $P^{k}(s)$ and $P(s, \beta')$, respectively. It has been used in earlier works \cite{tekur2018higher,bhosale2018scaling,tekur2020symmetry,bhosale2021superposition, rout2026higherorderspacingssuperposedspectra}. Here, $l$ corresponds to the histogram bins. The number of bins is taken as $200$ throughout this paper. Although the value of $D(\beta'$) can take any positive value, depending on the range of $l$, it attains its minimum only for that value of $\beta'$ for which $P(s,\beta')$ is the best fit to the corresponding histogram $P^{k}(s)$.
\subsection{Ginibre, $AI^{\dagger}$, $AII^{\dagger}$ classes}
\label{subsec: Ginibre, AI_dag, AII_dag}
\begin{figure}[H]
\begin{center}
\includegraphics*[scale=0.325]{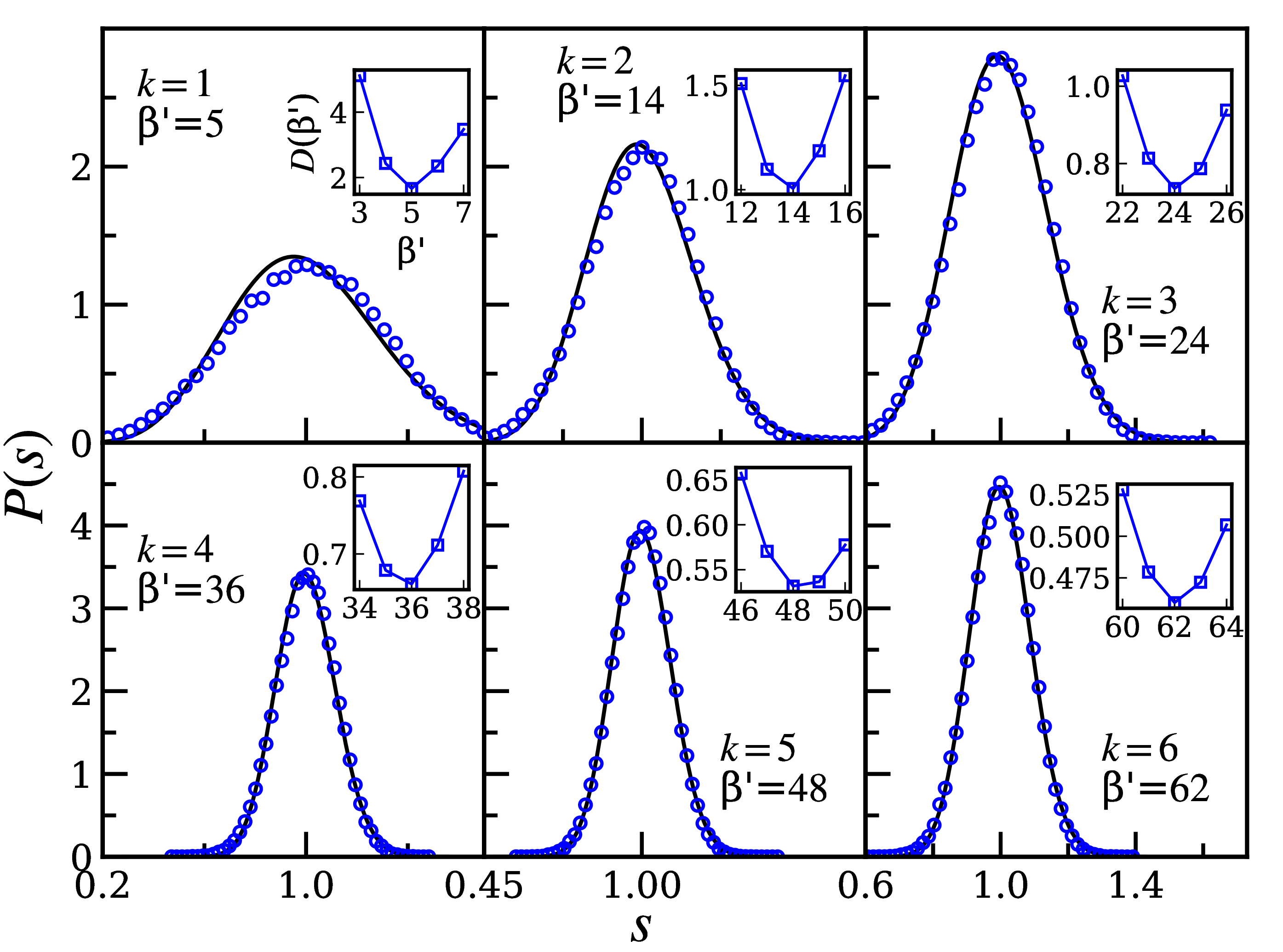}
\caption{\justifying  Distribution of the $k$-th order spacings (circles) for the GinOE spectra. Here, $N=10000$ and $n=500$. The solid curve corresponds to $P(s,\beta')$ as given in Eq.~(\ref{Eq:PSBeta}), in which $\beta$ is substituted by $\beta'$ and $\beta'$ is given in Table~\ref{Table:HOS_Ginibre_AI_AII_dagger}. The insets show $D(\beta')$ as a function of $\beta'$.}
\label{fig:GinOE1_SPACINGS}
\end{center}
\end{figure}

In this subsection, we study the HOSD of the complex spectra of the Ginibre ensemble, $AI^{\dagger}$, and $AII^{\dagger}$ for various values of $k$. We use $D(\beta')$ for determining the best fit. We tabulate those values of $\beta'$ in Table~\ref{Table:HOS_Ginibre_AI_AII_dagger}, for which the corresponding $P(s,\beta')$ agrees well with $P^{k}(s)$ for a given order $k$. Here, the $k$-th order spacing distribution ($P^{k}(s)$) is modeled using Eq.~(\ref{Eq:PSBeta}) with modified index $\beta'$. We use this method throughout the paper, wherever $\beta'$ is calculated.
From Table~\ref{Table:HOS_Ginibre_AI_AII_dagger}, we observe that for a particular ensemble, as $k$ increases, the values of $\beta'$ also increases. For a given $k$, the value of $\beta'$ is the smallest for $AI^{\dagger}$ and the largest for $AII^{\dagger}$.
The results are shown in Figs.~\ref{fig:GinOE1_SPACINGS}-\ref{fig:AII_dagger1_SPACINGS}. For additional illustrations, refer to Sec. SIIIA- Sec. SIIIE of the supplementary material \cite{supplementry2026}. In the figures, the insets show the variation of $D(\beta')$ with $\beta'$, where the minima correspond to the values of best fit $\beta'$.

From Table~\ref{Table:HOS_Ginibre_AI_AII_dagger}, it is found that the observed values of $\beta'$ for the three symmetry classes GinOE, GinUE, and GinSE are the same, except in a few cases, where they differ by $\pm 1$. The difference of $\pm 1$ can be attributed to statistical fluctuations or the finite size of the matrices. For GinUE, $AI^{\dagger}$, and $AII^{\dagger}$, they are found to be different from each other. These results further corroborate the similarity of local bulk statistics of the three classes of Ginibre ensembles and their distinction for GinUE ($A$), $AI^{\dagger}$, and $AII^{\dagger}$ using spacing statistics beyond nearest neighbors. These observations are well supported by the figures in the Sec. SIIIF of the supplementary material \cite{supplementry2026}. From the figures, it is observed that for $k\leq3$, there is a deviation between numerical data and the analytical function, and this deviation decreases with an increase in $k$ for all the non-Hermitian systems we consider. Beyond $k=3$, the agreement is very good. 
\begin{table}[t]
\centering
\renewcommand{\arraystretch}{1.6} 
\setlength{\tabcolsep}{11pt}  
\begin{tabular}{|c|ccc|c|c|}
\hline 
\rule{0pt}{12pt}  
Order&\multicolumn{3}{c|}{Ginibre}&$AI^{\dagger}$&$AII^{\dagger}$\\
\cline{2-4}  
&\multicolumn{1}{c|}{$AI$}&\multicolumn{1}{c|}{$A$}&\multicolumn{1}{c|}{$AII$}& &\\
\rule{0pt}{10pt}  
$k$&\multicolumn{1}{c|}{$\beta'$}&\multicolumn{1}{c|}{$\beta'$}&\multicolumn{1}{c|}{$\beta'$}&$\beta'$&$\beta'$\\ 
\hline
1&\multicolumn{1}{c|}{5}&\multicolumn{1}{c|}{5}&\multicolumn{1}{c|}{5}&4&6\\
2&\multicolumn{1}{c|}{14}&\multicolumn{1}{c|}{14}&\multicolumn{1}{c|}{14}&10&17\\
3&\multicolumn{1}{c|}{24}&\multicolumn{1}{c|}{24}&\multicolumn{1}{c|}{24}&19&30\\
4&\multicolumn{1}{c|}{36}&\multicolumn{1}{c|}{36}&\multicolumn{1}{c|}{36}&28&44\\
5&\multicolumn{1}{c|}{48}&\multicolumn{1}{c|}{48}&\multicolumn{1}{c|}{48}&38&59\\
6&\multicolumn{1}{c|}{62}&\multicolumn{1}{c|}{62}&\multicolumn{1}{c|}{62}&49&75\\
7&\multicolumn{1}{c|}{77}&\multicolumn{1}{c|}{77}&\multicolumn{1}{c|}{77}&60&92\\
8&\multicolumn{1}{c|}{93}&\multicolumn{1}{c|}{92}&\multicolumn{1}{c|}{93}&73&111\\
9&\multicolumn{1}{c|}{108}&\multicolumn{1}{c|}{109}&\multicolumn{1}{c|}{108}&86&130\\
10&\multicolumn{1}{c|}{127}&\multicolumn{1}{c|}{126}&\multicolumn{1}{c|}{126}&100&150\\
11&\multicolumn{1}{c|}{145}&\multicolumn{1}{c|}{144}&\multicolumn{1}{c|}{145}&114&172\\
12&\multicolumn{1}{c|}{163}&\multicolumn{1}{c|}{163}&\multicolumn{1}{c|}{164}&129&194\\
13&\multicolumn{1}{c|}{183}&\multicolumn{1}{c|}{183}&\multicolumn{1}{c|}{183}&144&218\\
14&\multicolumn{1}{c|}{204}&\multicolumn{1}{c|}{203}&\multicolumn{1}{c|}{203}&160&242\\
15&\multicolumn{1}{c|}{225}&\multicolumn{1}{c|}{225}&\multicolumn{1}{c|}{225}&177&267\\                            
\hline
\end{tabular}
\caption{\justifying Tabulation of higher-order indices $\beta'$ for spacing distributions of order $k$ for the three universality classes such as Ginibre (GinOE ($AI$), GinUE ($A$), and GinSE ($AII$)), $AI^{\dagger}$, and $AII^{\dagger}$. Here, the dimension of each matrix of classes $AI$ and $A$ is $N=10000$, of $AII$ is $2N=10000$, of $AI^{\dagger}$ is $N=10000$, and of $AII^{\dagger}$ is $2N=20000$. The number of realizations is $n=500$ for all these classes.}
\label{Table:HOS_Ginibre_AI_AII_dagger}
\end{table}
We also study HOS statistics of eigenvalues of Bernoulli complex symmetric random matrices, and it is found to satisfy the statistics of $AI^{\dagger}$. This again confirms the universality of HOS statistics of $AI^{\dagger}$. For a few $k$-values, there is a difference of $\pm 1$ in the values of $\beta'$, obtained for these two ensembles, which can be attributed to statistical fluctuations. For detailed information on this, one can refer to  Sec. SIV of the supplementary material \cite{supplementry2026}. 
\begin{figure}[H]
\begin{center}
\includegraphics*[scale=0.32]{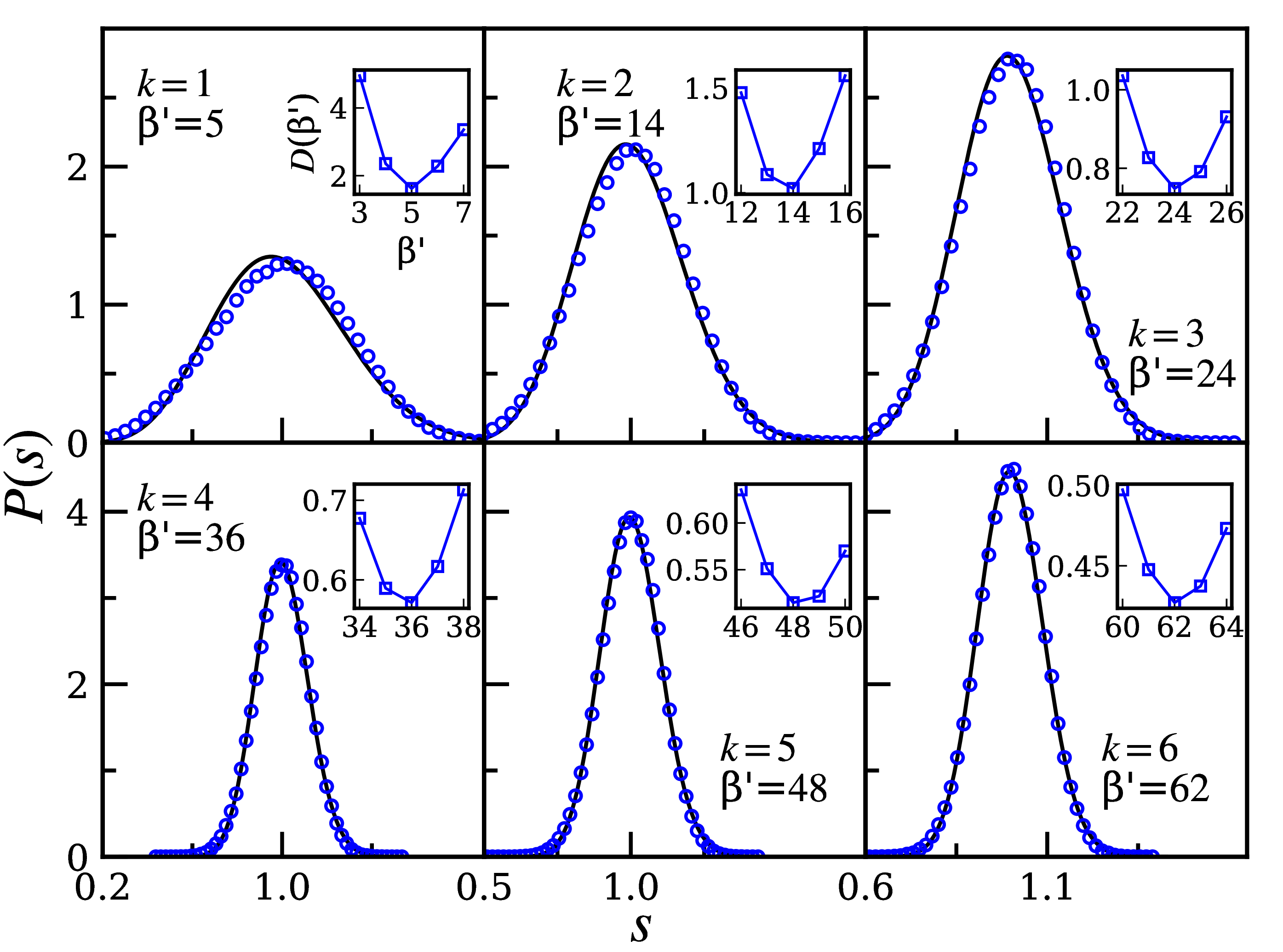}
\caption{\justifying Same as Fig.~\ref{fig:GinOE1_SPACINGS}, but for GinUE.}
\label{fig:GinUE1_SPACINGS}
\end{center}
\end{figure}
\subsection{Product of independent complex Ginibre matrices}
\label{subsec: Prod_GinUE}
In this subsection, we consider the product of independent random matrices from GinUE of equal size. Specifically, the product of two, three, and four such matrices, denoted by $M=2, 3,$ and $4$, respectively. The product of random matrices has received considerable interest in RMT due to their applications in various fields. For example, in the study of Lyapunov exponents in diffusion problems \cite{Forrester2013Lyapunov}, Anderson localization and disordered systems \cite{Comtet2013Lyapunov}, neural networks and deep learning\cite{Pennington2017Resurrecting}, and  fluid dynamics and mixing processes \cite{Vanneste2010Estimating}.
\begin{figure}[H]
\begin{center}
\includegraphics*[scale=0.32]{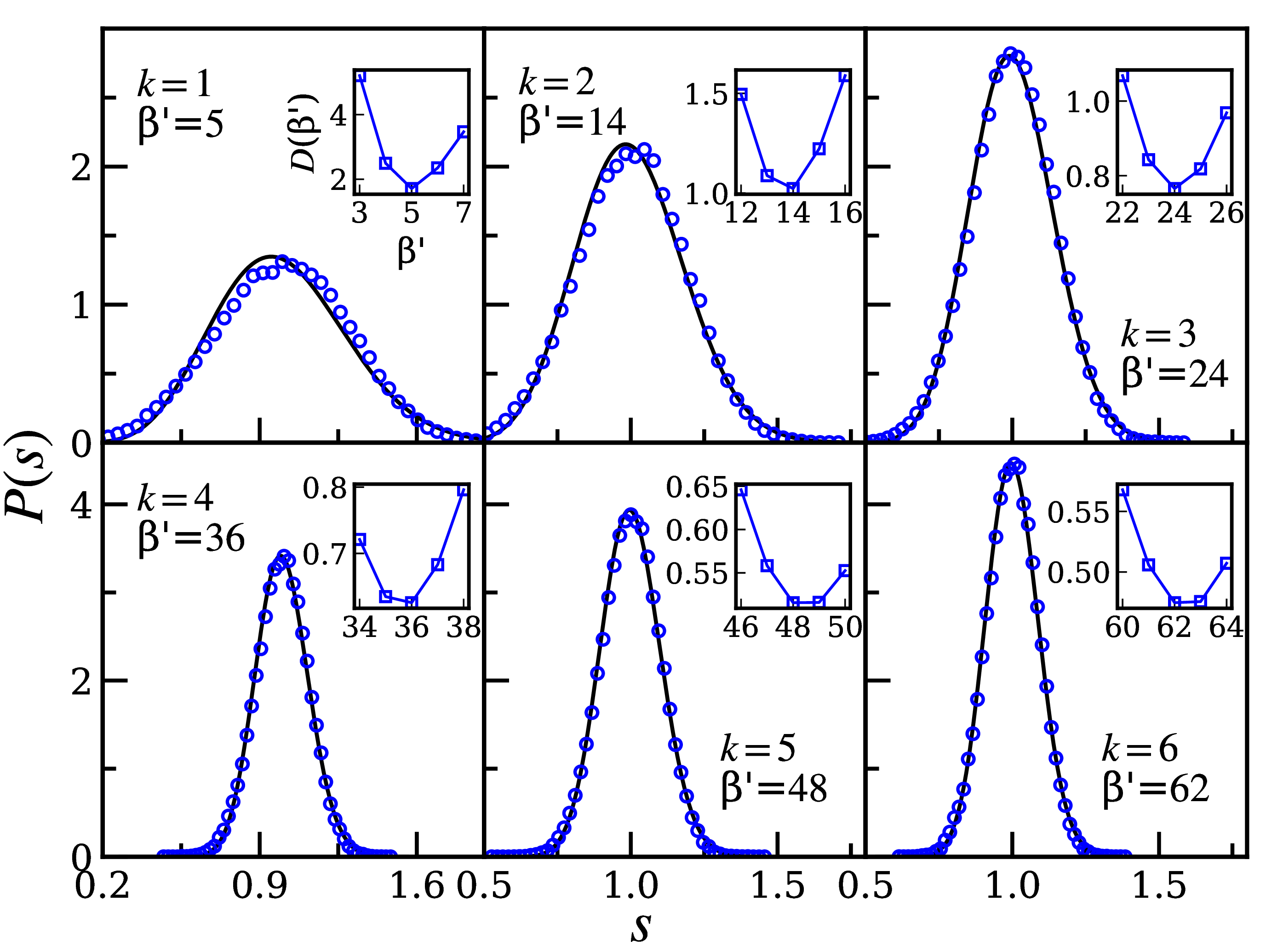}
\caption{\justifying Same as Fig.~\ref{fig:GinOE1_SPACINGS}, but for GinSE. Here, $2N=10000$.}
\label{fig:GinSE1_SPACINGS}
\end{center}
\end{figure}
\begin{figure}[H]
\begin{center}
\includegraphics*[scale=0.32]{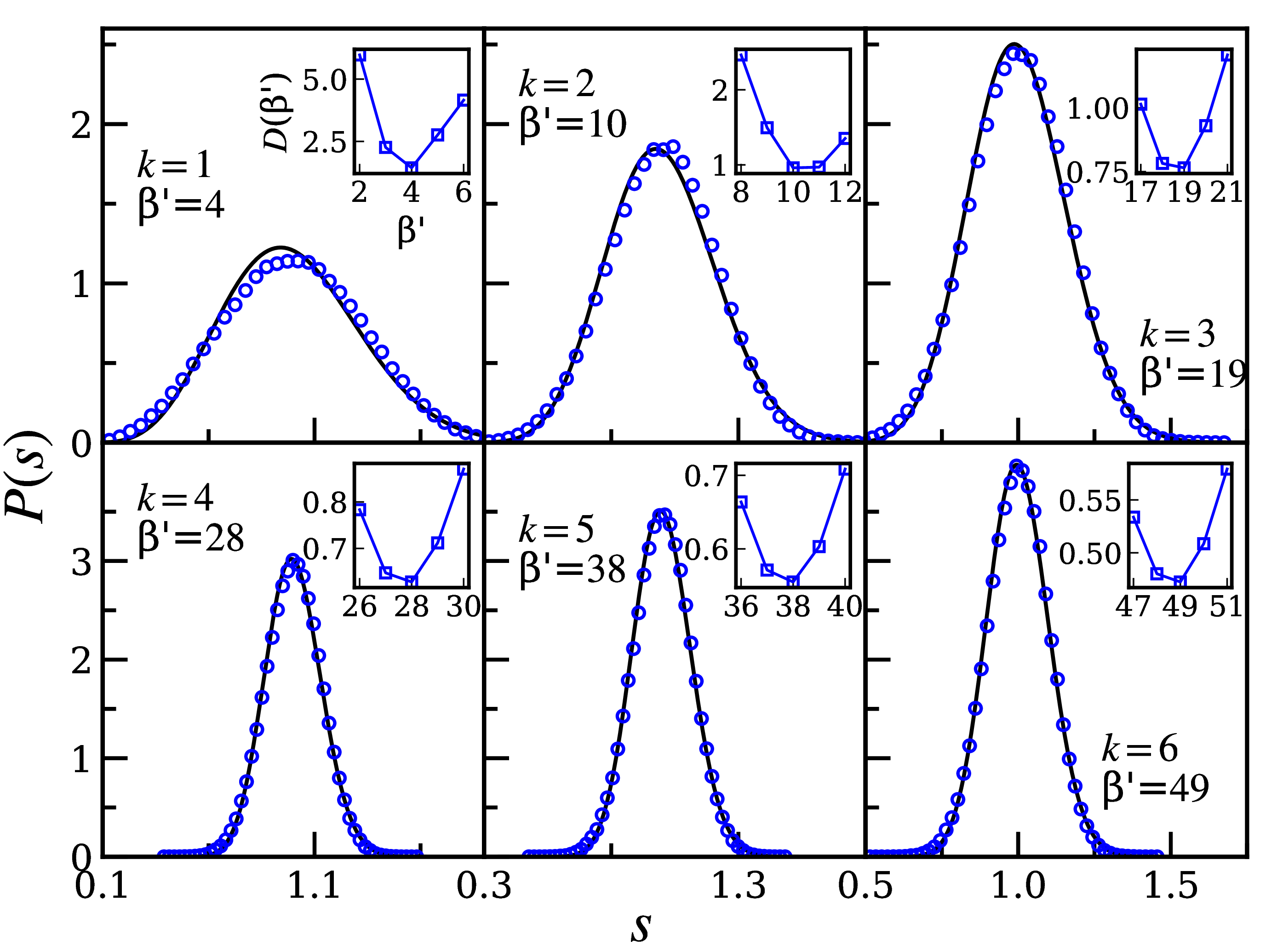}
\caption{\justifying Same as Fig.~\ref{fig:GinOE1_SPACINGS}, but for $AI^{\dagger}$.}
\label{fig:AI_dagger1_SPACINGS}
\end{center}
\end{figure}
\begin{figure}[H]
\begin{center}
\includegraphics*[scale=0.32]{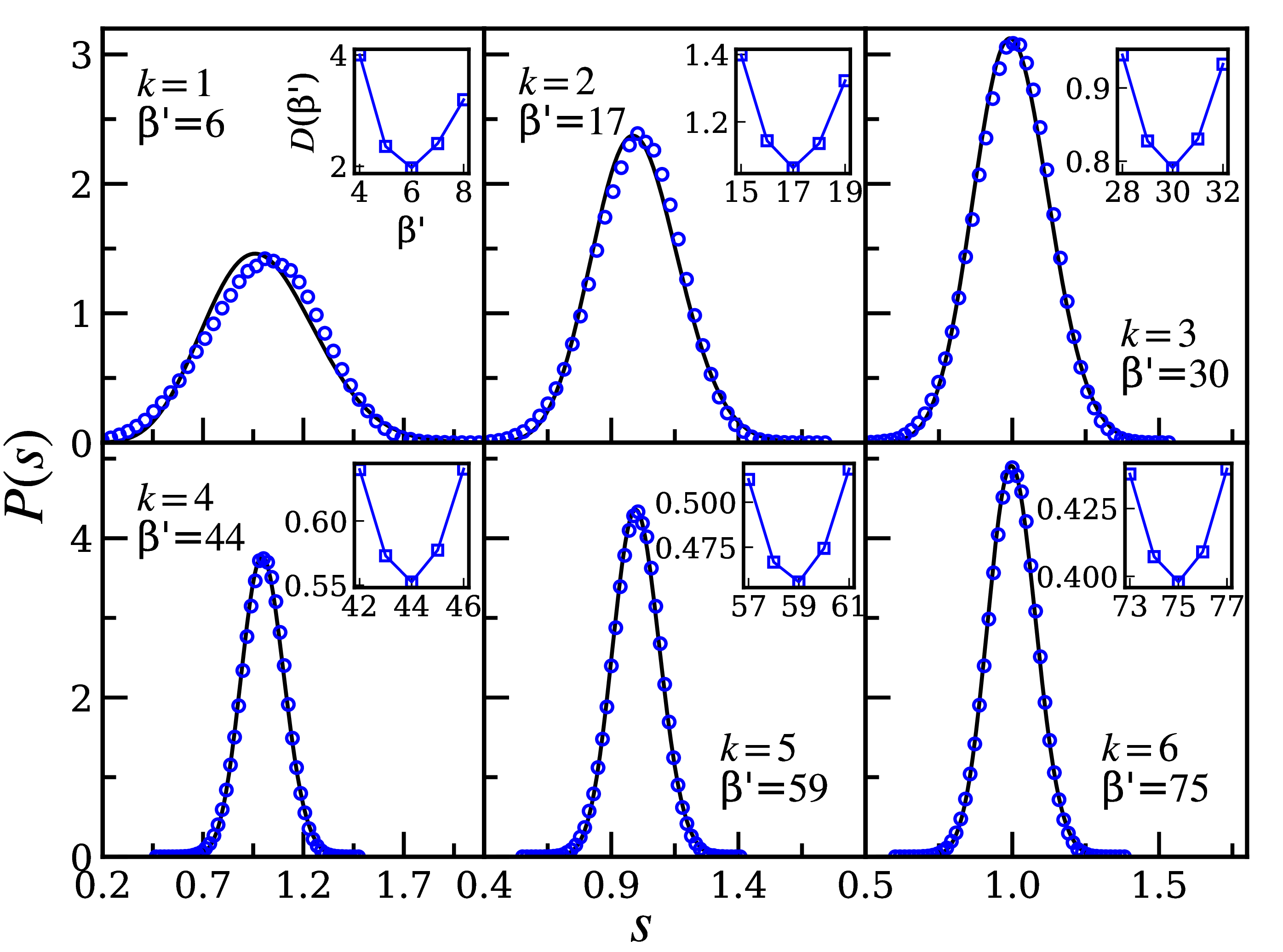}
\caption{\justifying Same as Fig.~\ref{fig:GinOE1_SPACINGS}, but for $AII^{\dagger}$. Here, $2N=20000$.}
\label{fig:AII_dagger1_SPACINGS}
\end{center}
\end{figure}
The average eigenvalue density for this product of $M$ independent matrices from GinUE is given by \cite{Burda2010, akemann2019universal}:
\begin{equation}
\rho_{\mathrm{av}}(x,y)
= \frac{1}{M\pi}|z|^{\frac{2(1-M)}{M}}
\Theta(1-|z|).
\label{Eq: Product_density}
\end{equation}
This density is non-uniform, as it depends on $\left| z \right|$ and is radially symmetric or rotationally invariant due to the independence of the angular coordinate. It also has a singularity at the origin for $M>1$. Here, the dependency of spectral density on $M$ indicates that as $M$ increases, the eigenvalue density becomes more concentrated near the origin with increasing $M$. The scatter plots of the complex eigenvalues for various $M$ are shown in the supplementary material \cite{supplementry2026}. Moreover, it is found that the local bulk statistics of the spectrum agree with those of the GinUE \cite{akemann2012Universal, akemann2019universal}.
This leads to a natural question of how the spectral statistics beyond nearest-neighbor depends on $M$. Here, we study the HOSD in the $2$D spectra of these product matrices in the bulk. Since the eigenvalue density is not uniform, we unfold the spectra using the method proposed in Ref.~\cite{akemann2019universal}. We find that the proposed method also works for $k \geq 1$.
We then find HOS of the unfolded spectra and fit its distribution by using Eq.~(\ref{Eq:PSBeta}). The $\beta'$ values corresponding to the best fits are noted in Table~\ref{Table: HOS_Product_of_M=2_3_4_GinUE}. These results are also shown in Figs.~\ref{fig: M_2_GinUE_Spacings1}-\ref{fig: M_4_GinUE_Spacings1}. Additional plots can be found in the supplementary material \cite{supplementry2026}. Here, we observe the same behavioral pattern of spacing statistics for $k \leq 3$, as described in Sec.~\ref{subsec: Ginibre, AI_dag, AII_dag}.
The observed values of $\beta'$ increase with an increase in $k$. The HOSD in the bulk is found to be independent of $M$, and it obeys the statistics of a GinUE ensemble ($M=1$). However, there is a difference of $\pm 1$ or $\pm 2$ for a few higher $k$ values, among the results corresponding to various $M$. We believe that it is due to statistical fluctuations.
\begin{table}[t]
\centering
\renewcommand{\arraystretch}{1.6}
\setlength{\tabcolsep}{9pt}
\begin{tabular}{|c|c|c|c|}
\hline
\rule{0pt}{12pt}
Order & $M=2$ & $M=3$ & $M=4$\\
$k$ & $\beta'$ & $\beta'$ & $\beta'$\\
\hline
1&5&5&5\\
2&14&14&14\\
3&24&24&24\\
4&36&36&36\\
5&48&49&49\\
6&62&62&62\\
7&77&77&77\\
8&93&93&93\\
9&109&109&109\\
10&126&127&127\\
11&144&145&146\\
12&163&164&165\\
13&183&184&185\\
14&204&205&205\\
15&225&227&227\\
\hline
\end{tabular}
\caption{\justifying Tabulation of higher-order indices $\beta'$ for spacing distributions of order $k$ for the spectra of the product of two, three, and four complex Ginibre square matrices denoted as $M=2$, $M=3$, and $M=4$. Here, the dimension of each matrix before the product is $5000$, and the number of realizations is $n=1000$ for $M=2, 3,$ and $4$.}
\label{Table: HOS_Product_of_M=2_3_4_GinUE}
\end{table}
\begin{figure}[H]
\begin{center}
\includegraphics*[scale=0.34]{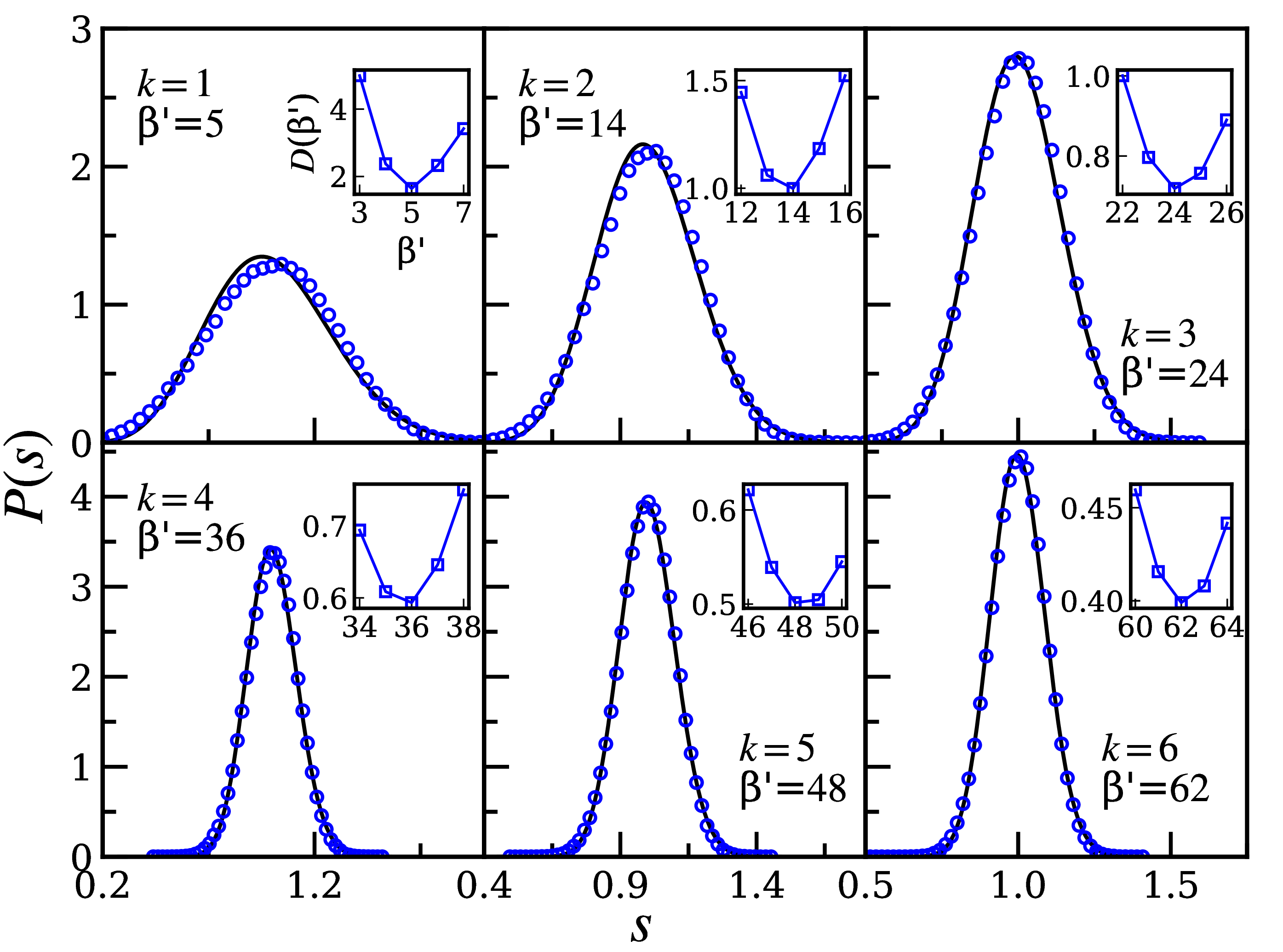}
\caption{\justifying Distribution of the $k$-th order spacings (circles) for the $M=2$ spectra. Here, $N=5000$ and $n=1000$. The solid curve corresponds to $P(s,\beta')$ as given in Eq.~(\ref{Eq:PSBeta}), in which $\beta$ is substituted by $\beta'$ and $\beta'$ is given in Table~\ref{Table: HOS_Product_of_M=2_3_4_GinUE}. The insets show $D(\beta')$ as a function of $\beta'$.}
\label{fig: M_2_GinUE_Spacings1}
\end{center}
\end{figure}
\begin{figure}[H]
\begin{center}
\includegraphics*[scale=0.33]{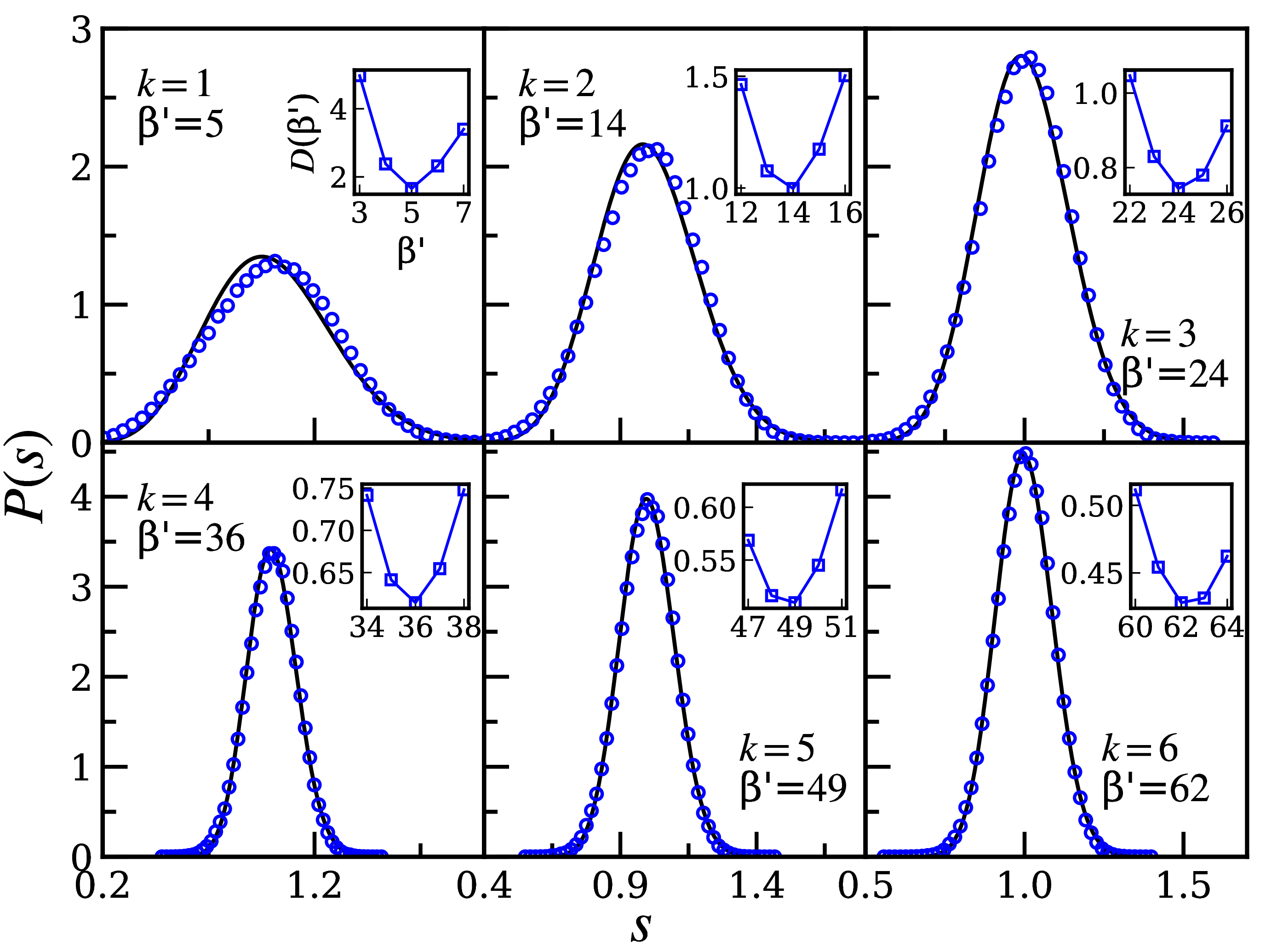}
\caption{\justifying Same as Fig.~\ref{fig: M_2_GinUE_Spacings1}, but for $M=3$.}
\label{fig: M_3_GinUE_Spacings1}
\end{center}
\end{figure}
\section{Application to physical system}
\label{sec: physical system}
In this section, we test our RMT results on the nHSYK model. The nHSYK model 
is a non-Hermitian extension of SYK model, which has played an 
important role in holographic studies \cite{Garcia2022Replica}.
It has been found that both Hermitian and non-Hermitian variants of the SYK model show quantum
chaotic behavior, and their spectral correlations are very well described by RMT \cite{garcia2022symmetry,Garcia2016Spectral, Garcia2017Analytical}. 
\begin{figure}[H]
\begin{center}
\includegraphics*[scale=0.33]{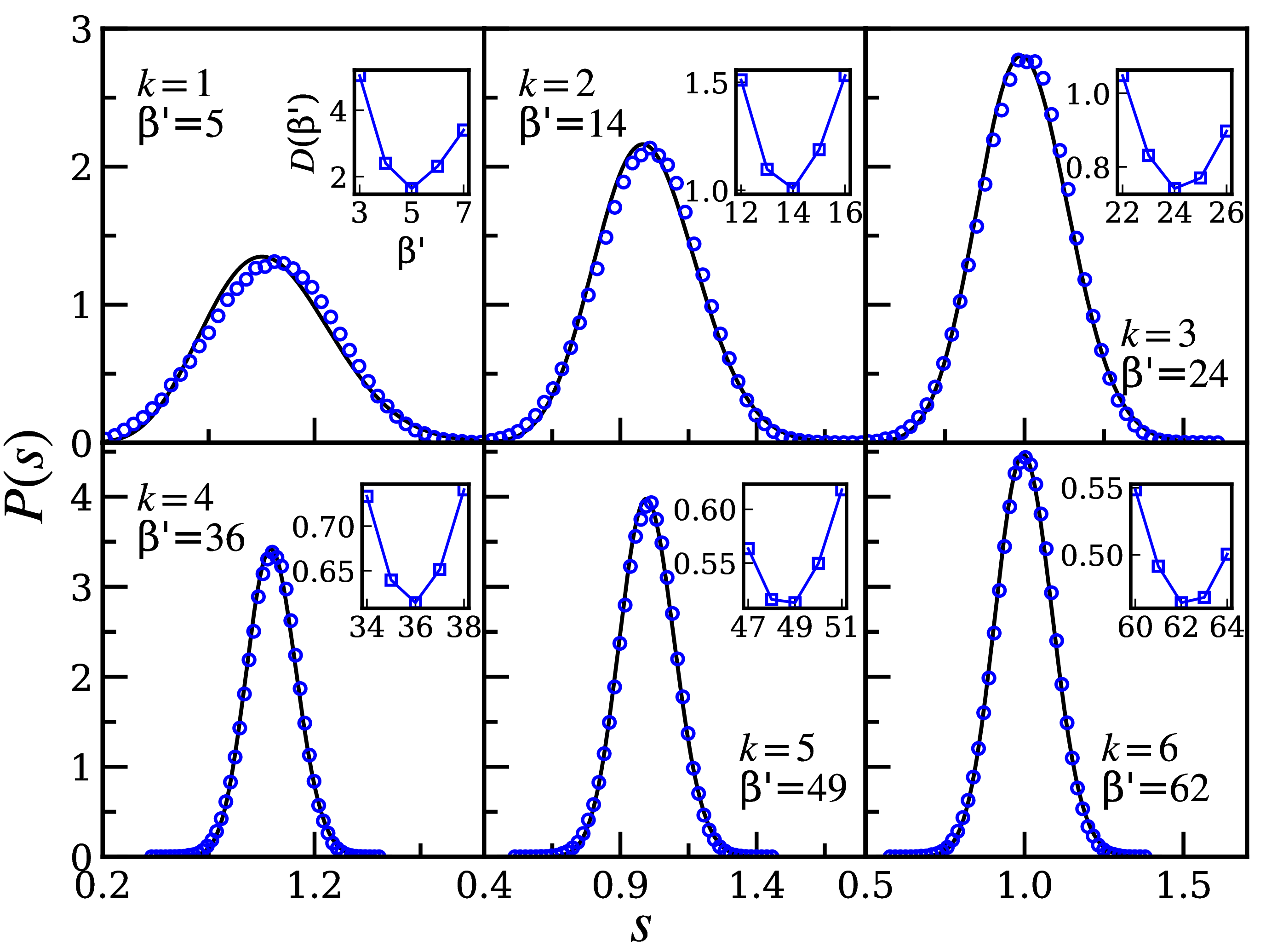}
\caption{\justifying Same as Fig.~\ref{fig: M_2_GinUE_Spacings1}, but for $M=4$.}
\label{fig: M_4_GinUE_Spacings1}
\end{center}
\end{figure}
\subsection{nHSYK model: SYK model with complex couplings}
\label{subsec: nhsyk}
Here, we consider a single nHSYK model of $N$ interacting Majorana fermions with $q$-body interactions in the Fock space and with complex couplings. The nHSYK Hamiltonian is defined as \cite{Kitaev2015, Maldacena2016Remarks, garcia2022symmetry, garcia2023universality}:
\begin{equation}
H = \sum_{\substack{i_1 < i_2 < \cdots < i_q}}^{N}
\left(J_{i_1 i_2 \cdots i_q}
+ iM_{i_1 i_2 \cdots i_q}\right)
\psi_{i_1}\psi_{i_2}\cdots\psi_{i_q},
\label{Eq: nHSYK}
\end{equation}
where, $N$ and $q$ are integers, $J_{i_1 i_2 \cdots i_q}$ and $M_{i_1 i_2 \cdots i_q}$ are the Gaussian random variables with zero mean and variance
\begin{equation}
\sigma^2 = \frac{1}{6(2N)^{q-1}},
\end{equation}
and $\psi_i$ are Majorana fermions satisfying 
\begin{equation}
\{\psi_i,\psi_j\}=2\delta_{ij}.
\end{equation}  
The non-Hermiticity of this model is due to the introduction of the $M$ term. Here, we have considered only even values of $N$. From the symmetry classification of the nHSYK model \cite{garcia2022symmetry}, it is found that depending on both $q$ and $N$, $19$ out of the $38$ classes have been identified for non-Hermitian systems. The bulk universality classes for various $q$ and $N$ are given in Refs.~\cite{garcia2022symmetry, garcia2023universality}. It is observed that the cases of $N=24$ and $q=6$, $N=24$ and $q=8$, and $N=28$ and $q=8$ belong to the classes GinUE, $AI^{\dagger}$, and $AII^{\dagger}$, respectively. Hence, it will be a natural choice for our problem. It is known that for $q\geq6$, the density of eigenvalues of the nHSYK model is almost constant; therefore, unfolding can be performed by just rescaling the eigenvalues \cite{garcia2023universality}. We calculate the eigenvalues corresponding to the mentioned combinations of $q$ and $N$ values. Scatter plots of these eigenvalues are shown in the supplementary material \cite{supplementry2026}. 
\begin{table}[t]
\centering
\renewcommand{\arraystretch}{1.6}
\setlength{\tabcolsep}{4pt}
\begin{tabular}{|c|c|c|c|}
\hline
\rule{0pt}{18pt}  
Order & \makecell{$N=24$ and $q=6$ \\ (GinUE)} & \makecell{$N=24$ and $q=8$ \\($AI^{\dagger}$)} & \makecell{$N=28$ and $q=8$ \\($AII^{\dagger}$)}\\
$k$ & $\beta'$& $\beta'$& $\beta'$\\
\hline
1&5&4&6\\
2&14&10&17\\
3&24&18&30\\
4&36&28&44\\
5&48&38&59\\
6&62&48&75\\
7&77&60&92\\
8&93&73&111\\
9&110&86&130\\
10&127&100&151\\
11&145&114&173\\
12&164&129&195\\
13&184&144&218\\
14&203&161&243\\
15&225&177&268\\
\hline
\end{tabular}
\caption{\justifying Tabulation of higher-order indices $\beta'$ for spacing distributions of order $k$ for the spectra of the nHSYK model. Here, $n=700, 2000,$ and $935$, corresponding to $N=24$ and $q=6$, $N=24$ and $q=8$, and $N=28$ and $q=8$, respectively.}
\label{Table: HOS_nHSYK model}
\end{table}
\begin{figure}[H]
\begin{center}
\includegraphics*[scale=0.32]{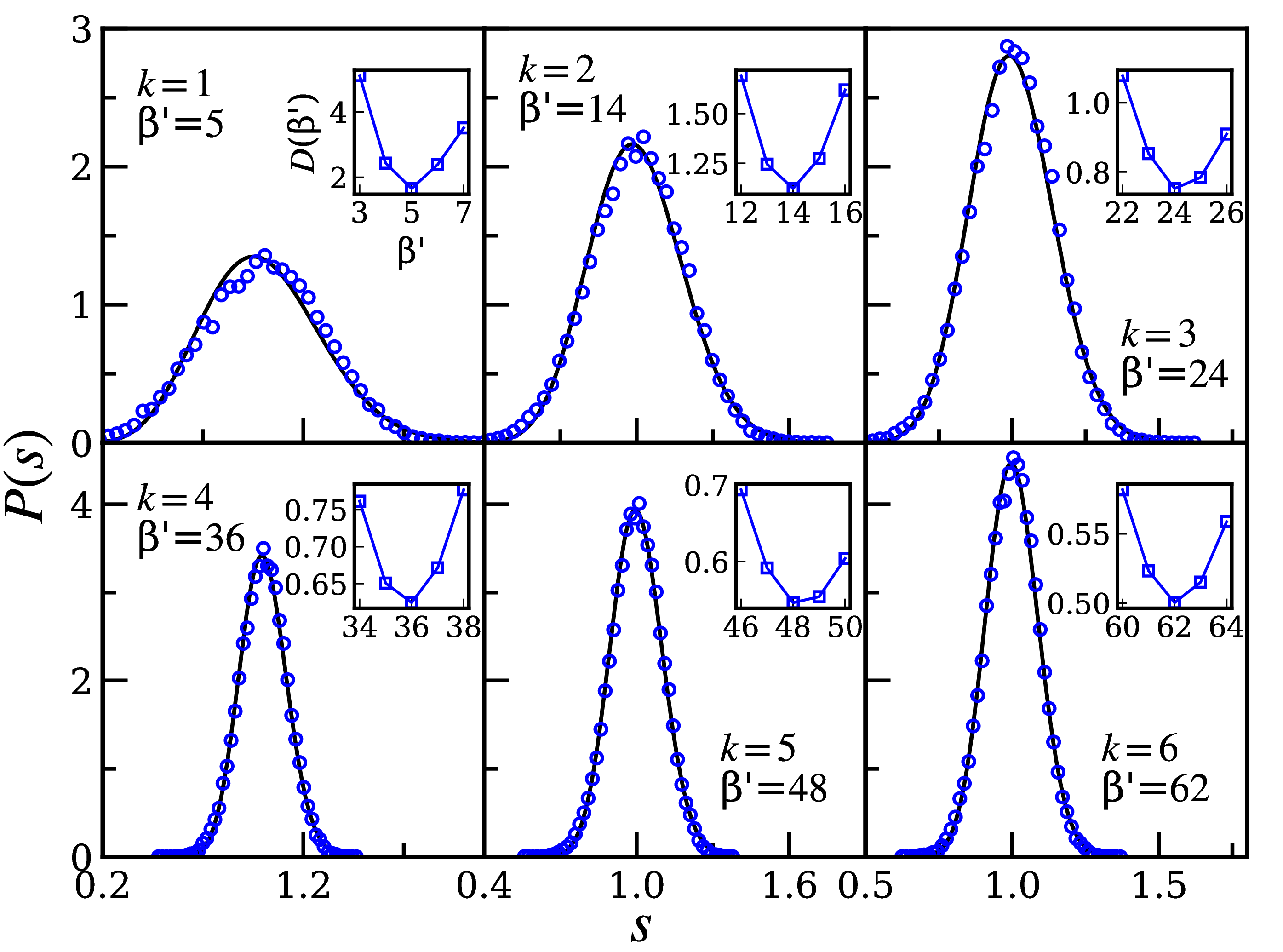}
\caption{\justifying Distribution of the $k$-th order spacings (circles) for the spectra of the nHSYK model for $N=24$ and $q=6$. Here, $n=700$. The solid curve corresponds to $P(s,\beta')$ as given in Eq.~(\ref{Eq:PSBeta}), in which $\beta$ is substituted by $\beta'$ and $\beta'$ is given in Table~\ref{Table: HOS_nHSYK model}. The insets show $D(\beta')$ as a function of $\beta'$.}
\label{fig: k_1_to_6_Nhsyk_GinUE}
\end{center}
\end{figure}
Then, Eq.~(\ref{Eq:PSBeta}) is used to find the best fit for spacing distributions corresponding to various values of $k$. We determine the values of $\beta'$ corresponding to the best fit, and these values are tabulated in Table~\ref{Table: HOS_nHSYK model}. These results are shown in Figs.~\ref{fig: k_1_to_6_Nhsyk_GinUE}-\ref{fig: k_1_to_6_Nhsyk_AII_dagger}. For more illustrations, refer to Sec. SVA of the supplementary material \cite{supplementry2026}. Here, from the figures, we observe the same behavioral pattern for $k\leq3$, as mentioned in Sec.~\ref{subsec: Ginibre, AI_dag, AII_dag}. Further, it is found that the HOS statistics of nHSYK follows the statistics of the corresponding non-Hermitian random matrix ensemble, which is not obvious. However, there is a deviation of $\pm 1$ between the results of the model and random matrices, which can be attributed to statistical fluctuations.
\begin{figure}[H]
\begin{center}
\includegraphics*[scale=0.32]{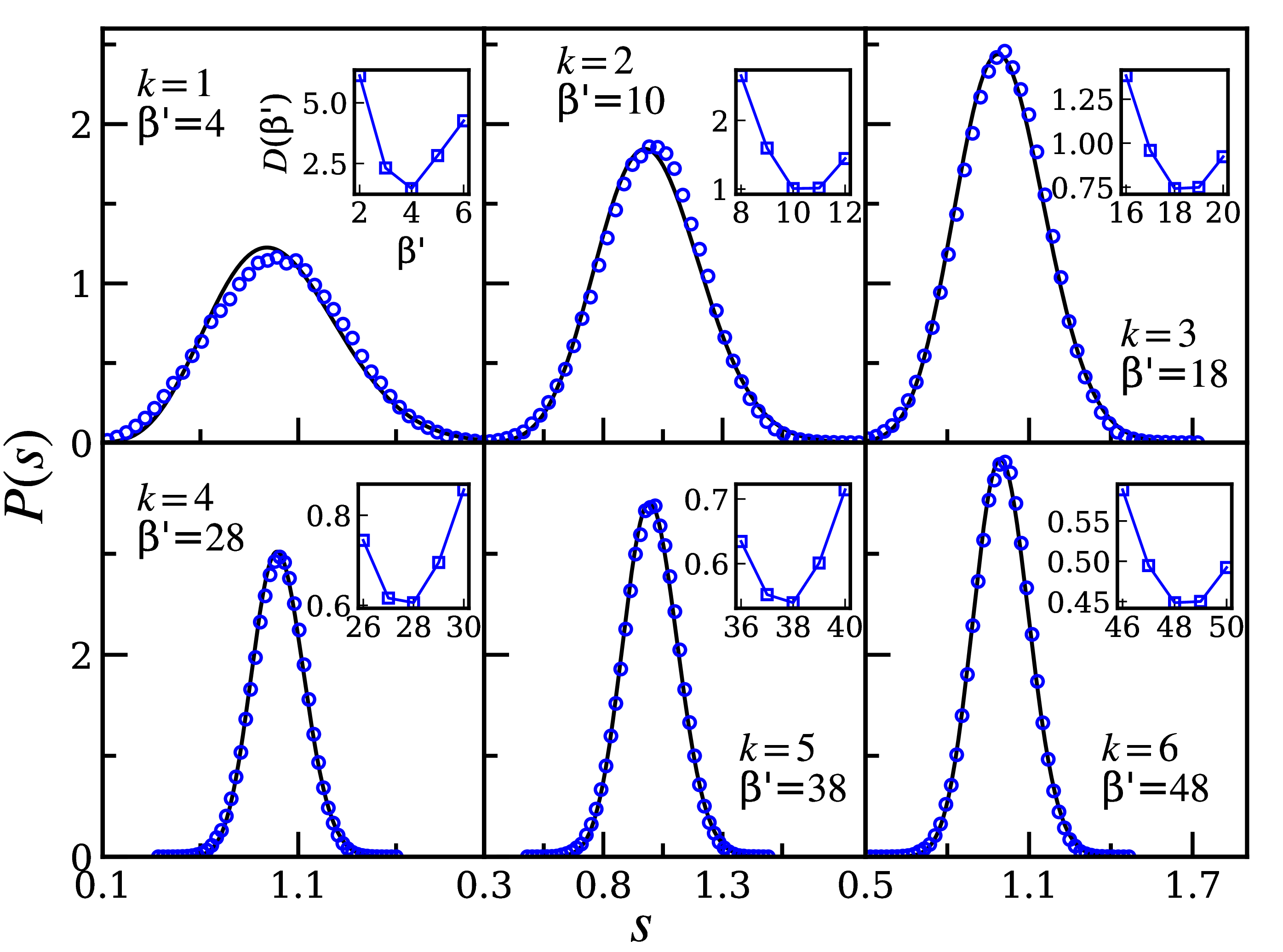}
\caption{\justifying Same as Fig.~\ref{fig: k_1_to_6_Nhsyk_GinUE}, but for $N=24$ and $q=8$. Here, $n=2000$.}
\label{fig: k_1_to_6_Nhsyk_AI_dagger}
\end{center}
\end{figure}
\begin{figure}[H]
\begin{center}
\includegraphics*[scale=0.32]{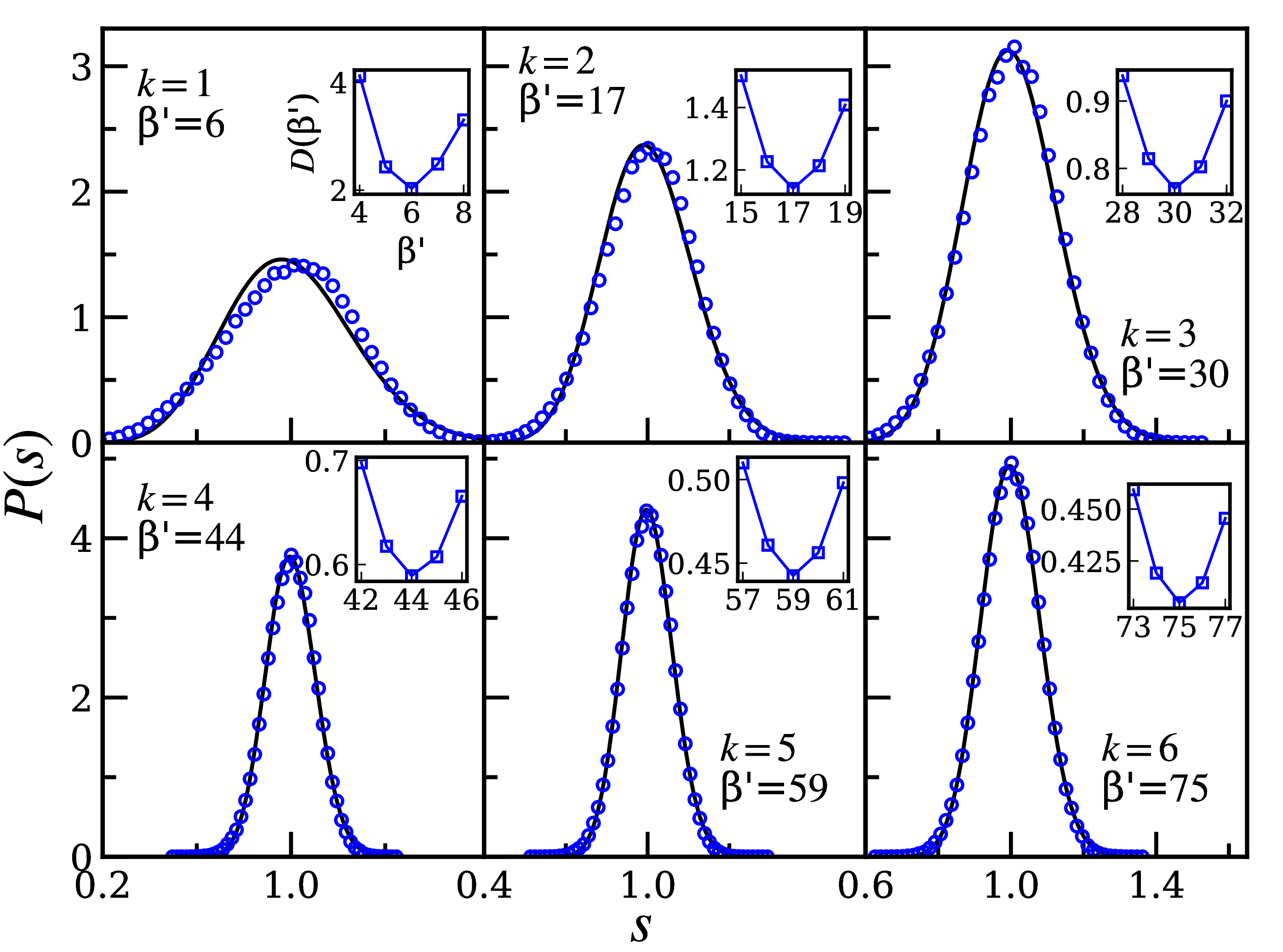}
\caption{\justifying Same as Fig.~\ref{fig: k_1_to_6_Nhsyk_GinUE}, but for $N=28$ and $q=8$. Here, $n=935$.}
\label{fig: k_1_to_6_Nhsyk_AII_dagger}
\end{center}
\end{figure}
\section{Discussion and Conclusion}
\label{sec: Conclusion}
In this work, we study the HOSD of complex eigenvalues in various non-Hermitian random matrix ensembles. 
Since there are studies on NNS and next-to-nearest neighbor spacing (NNNS) of complex eigenvalues, however, studies are lacking on spacing distributions beyond next-to-nearest neighbor. 
Moreover, the analytical derivation of spacing distribution beyond NNNS in $2$D is cumbersome, and the expression becomes complicated with an increase in $k$. Hence, in this study we have approximated the HOSD by using the generalized Wigner-Dyson distribution, as proposed for that in $1$D. This will act as a one- parameter approximation, and we are giving a simpler form of the distribution function that acts as a very good fit to HOSD. 

The measure, denoted as $D(\beta')$, is defined as the sum of differences between the cumulative distributions corresponding to the numerical data and the theoretical distribution function, is used to find the best fit. We find that there is a deviation between numerical data and analytical function for $k\leq3$ and this deviation decreases with increase in $k$ for all the non-Hermitian systems we have considered. Beyond $k=3$, the agreement is very good.
Although deviations between numerical HOS data and predicted analytical functions are observed for $k\leq3$, the spacing distributions and corresponding $\beta'$ values show excellent agreement among GinOE, GinUE, and GinSE for various values of $k$. In contrast, the three universality classes, GinUE ($A$), $AI^\dagger$ and $AII^\dagger$, exhibit distinct behavior. 
It is discussed in Ref.~\cite{hamazaki2020universality} that transposition symmetry influences the repulsion between two neighboring eigenvalues and consequently modifies their spacing distribution. Our results further demonstrate that the influence extends beyond nearest neighbors, which in turn affect HOSD.

We verify the universality of these obtained results by applying them to the nHSYK model. 
Hence, our results strengthen and complement the proposed result on classification of non-Hermitian random matrices, based on local bulk statistics, by using the statistics beyond nearest-neighbor. These findings give additional information on spectral correlations beyond NNSD. 
Furthermore, to the best of our knowledge, the joint probability density functions (jpdf) of complex eigenvalues of $AI^\dagger$ and $AII^\dagger$ remain unknown and only NNSD of $N=2$ matrix are known. Therefore, our results are significant as they provide approximate analytical expressions that describe the HOSDs ($k>3$) accurately in these two classes for large matrix dimension $N$. 

In addition to this, we consider the HOSD of the product of equal-dimensional independent complex Ginibre matrices. Similar behavior is observed corresponding to various values of $k$. The HOSD in the bulk is found to be independent of the number of matrices constituting a product matrix, and they obey the corresponding statistics of GinUE. This result shows the universality of GinUE statistics beyond nearest neighbor. We also study the HOSD of eigenvalues of the Bernoulli complex symmetric random matrices. We find that the HOSD of this agrees with that of $AI^\dagger$. It shows that the statistics does not depend on the distribution of matrix elements but rather on the overall symmetry of the matrix and hence confirms the universality of the HOSD result of $AI^\dagger$.

Our work also opens up new research directions, such as HOSD of other symmetry classes of non-Hermitian random matrices among $38$ classes can also be studied; hence, one can verify whether this extends the three non-Hermitian universality classes of local bulk statistics. It will be interesting to explore whether HOSD can be used for finding hidden symmetries and the number of independent spectra in the composite spectrum of non-Hermitian systems. 
\section{Acknowledgments}
\label{acknowledgement}
We acknowledge National Super computing Mission (NSM) for providing computing resources of ‘PARAM SMRITI’ at NABI, Mohali, which is implemented by C-DAC and supported by the Ministry of Electronics and Information Technology (MeitY) and Department of Science and Technology (DST), Government of India. We are grateful to the Department of Science and Technology (DST) for their generous financial support through sanctioned Project No. SR/FST/PSI/2017/5(C) to the Department of Physics of VNIT, Nagpur. We would like to thank Y. Lanjewar for carefully reading the manuscript and providing invaluable comments.


\bibliography{REFERENCES}

\begin{thebibliography}{49}%
\makeatletter
\providecommand \@ifxundefined [1]{%
 \@ifx{#1\undefined}
}%
\providecommand \@ifnum [1]{%
 \ifnum #1\expandafter \@firstoftwo
 \else \expandafter \@secondoftwo
 \fi
}%
\providecommand \@ifx [1]{%
 \ifx #1\expandafter \@firstoftwo
 \else \expandafter \@secondoftwo
 \fi
}%
\providecommand \natexlab [1]{#1}%
\providecommand \enquote  [1]{``#1''}%
\providecommand \bibnamefont  [1]{#1}%
\providecommand \bibfnamefont [1]{#1}%
\providecommand \citenamefont [1]{#1}%
\providecommand \href@noop [0]{\@secondoftwo}%
\providecommand \href [0]{\begingroup \@sanitize@url \@href}%
\providecommand \@href[1]{\@@startlink{#1}\@@href}%
\providecommand \@@href[1]{\endgroup#1\@@endlink}%
\providecommand \@sanitize@url [0]{\catcode `\\12\catcode `\$12\catcode
  `\&12\catcode `\#12\catcode `\^12\catcode `\_12\catcode `\%12\relax}%
\providecommand \@@startlink[1]{}%
\providecommand \@@endlink[0]{}%
\providecommand \url  [0]{\begingroup\@sanitize@url \@url }%
\providecommand \@url [1]{\endgroup\@href {#1}{\urlprefix }}%
\providecommand \urlprefix  [0]{URL }%
\providecommand \Eprint [0]{\href }%
\providecommand \doibase [0]{https://doi.org/}%
\providecommand \selectlanguage [0]{\@gobble}%
\providecommand \bibinfo  [0]{\@secondoftwo}%
\providecommand \bibfield  [0]{\@secondoftwo}%
\providecommand \translation [1]{[#1]}%
\providecommand \BibitemOpen [0]{}%
\providecommand \bibitemStop [0]{}%
\providecommand \bibitemNoStop [0]{.\EOS\space}%
\providecommand \EOS [0]{\spacefactor3000\relax}%
\providecommand \BibitemShut  [1]{\csname bibitem#1\endcsname}%
\let\auto@bib@innerbib\@empty
\bibitem [{\citenamefont {Mehta}(2004)}]{mehta2004random}%
  \BibitemOpen
  \bibfield  {author} {\bibinfo {author} {\bibfnamefont {M.~L.}\ \bibnamefont
  {Mehta}},\ }\href@noop {} {\emph {\bibinfo {title} {Random matrices}}}\
  (\bibinfo  {publisher} {Elsevier},\ \bibinfo {year} {2004})\BibitemShut
  {NoStop}%
\bibitem [{\citenamefont {Bohigas}\ \emph {et~al.}(1984)\citenamefont
  {Bohigas}, \citenamefont {Giannoni},\ and\ \citenamefont
  {Schmit}}]{PhysRevLett.52.1}%
  \BibitemOpen
  \bibfield  {author} {\bibinfo {author} {\bibfnamefont {O.}~\bibnamefont
  {Bohigas}}, \bibinfo {author} {\bibfnamefont {M.~J.}\ \bibnamefont
  {Giannoni}},\ and\ \bibinfo {author} {\bibfnamefont {C.}~\bibnamefont
  {Schmit}},\ }\href@noop {} {\bibfield  {journal} {\bibinfo  {journal} {Phys.
  Rev. Lett.}\ }\textbf {\bibinfo {volume} {52}},\ \bibinfo {pages} {1}
  (\bibinfo {year} {1984})}\BibitemShut {NoStop}%
\bibitem [{\citenamefont {Berry}\ and\ \citenamefont
  {Tabor}(1977)}]{berry1977level}%
  \BibitemOpen
  \bibfield  {author} {\bibinfo {author} {\bibfnamefont {M.~V.}\ \bibnamefont
  {Berry}}\ and\ \bibinfo {author} {\bibfnamefont {M.}~\bibnamefont {Tabor}},\
  }\href@noop {} {\bibfield  {journal} {\bibinfo  {journal} {Proc. R. Soc.
  Lond. A.}\ }\textbf {\bibinfo {volume} {356}},\ \bibinfo {pages} {375}
  (\bibinfo {year} {1977})}\BibitemShut {NoStop}%
\bibitem [{\citenamefont {Oganesyan}\ and\ \citenamefont
  {Huse}(2007)}]{oganesyan2007localization}%
  \BibitemOpen
  \bibfield  {author} {\bibinfo {author} {\bibfnamefont {V.}~\bibnamefont
  {Oganesyan}}\ and\ \bibinfo {author} {\bibfnamefont {D.~A.}\ \bibnamefont
  {Huse}},\ }\href@noop {} {\bibfield  {journal} {\bibinfo  {journal} {Phys.
  Rev. B}\ }\textbf {\bibinfo {volume} {75}},\ \bibinfo {pages} {155111}
  (\bibinfo {year} {2007})}\BibitemShut {NoStop}%
\bibitem [{\citenamefont {Atas}\ \emph {et~al.}(2013)\citenamefont {Atas},
  \citenamefont {Bogomolny}, \citenamefont {Giraud},\ and\ \citenamefont
  {Roux}}]{atas2013distribution}%
  \BibitemOpen
  \bibfield  {author} {\bibinfo {author} {\bibfnamefont {Y.~Y.}\ \bibnamefont
  {Atas}}, \bibinfo {author} {\bibfnamefont {E.}~\bibnamefont {Bogomolny}},
  \bibinfo {author} {\bibfnamefont {O.}~\bibnamefont {Giraud}},\ and\ \bibinfo
  {author} {\bibfnamefont {G.}~\bibnamefont {Roux}},\ }\href@noop {} {\bibfield
   {journal} {\bibinfo  {journal} {Phys. Rev. Lett.}\ }\textbf {\bibinfo
  {volume} {110}},\ \bibinfo {pages} {084101} (\bibinfo {year}
  {2013})}\BibitemShut {NoStop}%
\bibitem [{\citenamefont {St{\"o}ckmann}(1999)}]{stockmann2000quantum}%
  \BibitemOpen
  \bibfield  {author} {\bibinfo {author} {\bibfnamefont {H.-J.}\ \bibnamefont
  {St{\"o}ckmann}},\ }\href@noop {} {\emph {\bibinfo {title} {Quantum chaos: An
  Introduction}}}\ (\bibinfo  {publisher} {University Press, Cambridge},\
  \bibinfo {year} {1999})\BibitemShut {NoStop}%
\bibitem [{\citenamefont {Sierant}\ and\ \citenamefont
  {Zakrzewski}(2019)}]{sierant2019level}%
  \BibitemOpen
  \bibfield  {author} {\bibinfo {author} {\bibfnamefont {P.}~\bibnamefont
  {Sierant}}\ and\ \bibinfo {author} {\bibfnamefont {J.}~\bibnamefont
  {Zakrzewski}},\ }\href@noop {} {\bibfield  {journal} {\bibinfo  {journal}
  {Phys. Rev. B}\ }\textbf {\bibinfo {volume} {99}},\ \bibinfo {pages} {104205}
  (\bibinfo {year} {2019})}\BibitemShut {NoStop}%
\bibitem [{\citenamefont {Agarwal}\ \emph {et~al.}(2017)\citenamefont
  {Agarwal}, \citenamefont {Altman}, \citenamefont {Demler}, \citenamefont
  {Gopalakrishnan}, \citenamefont {Huse},\ and\ \citenamefont
  {Knap}}]{agarwal2017rare}%
  \BibitemOpen
  \bibfield  {author} {\bibinfo {author} {\bibfnamefont {K.}~\bibnamefont
  {Agarwal}}, \bibinfo {author} {\bibfnamefont {E.}~\bibnamefont {Altman}},
  \bibinfo {author} {\bibfnamefont {E.}~\bibnamefont {Demler}}, \bibinfo
  {author} {\bibfnamefont {S.}~\bibnamefont {Gopalakrishnan}}, \bibinfo
  {author} {\bibfnamefont {D.~A.}\ \bibnamefont {Huse}},\ and\ \bibinfo
  {author} {\bibfnamefont {M.}~\bibnamefont {Knap}},\ }\href@noop {} {\bibfield
   {journal} {\bibinfo  {journal} {Ann. Phys. (Berlin)}\ }\textbf {\bibinfo
  {volume} {529}},\ \bibinfo {pages} {1600326} (\bibinfo {year}
  {2017})}\BibitemShut {NoStop}%
\bibitem [{\citenamefont {Rao}(2020)}]{rao2020higher}%
  \BibitemOpen
  \bibfield  {author} {\bibinfo {author} {\bibfnamefont {W.-J.}\ \bibnamefont
  {Rao}},\ }\href@noop {} {\bibfield  {journal} {\bibinfo  {journal} {Phys.
  Rev. B}\ }\textbf {\bibinfo {volume} {102}},\ \bibinfo {pages} {054202}
  (\bibinfo {year} {2020})}\BibitemShut {NoStop}%
\bibitem [{\citenamefont {Tekur}\ \emph {et~al.}(2018)\citenamefont {Tekur},
  \citenamefont {Bhosale},\ and\ \citenamefont {Santhanam}}]{tekur2018higher}%
  \BibitemOpen
  \bibfield  {author} {\bibinfo {author} {\bibfnamefont {S.~H.}\ \bibnamefont
  {Tekur}}, \bibinfo {author} {\bibfnamefont {U.~T.}\ \bibnamefont {Bhosale}},\
  and\ \bibinfo {author} {\bibfnamefont {M.}~\bibnamefont {Santhanam}},\
  }\href@noop {} {\bibfield  {journal} {\bibinfo  {journal} {Phys. Rev. B}\
  }\textbf {\bibinfo {volume} {98}},\ \bibinfo {pages} {104305} (\bibinfo
  {year} {2018})}\BibitemShut {NoStop}%
\bibitem [{\citenamefont {G{\'o}mez}\ \emph {et~al.}(2002)\citenamefont
  {G{\'o}mez}, \citenamefont {Molina}, \citenamefont {Rela{\~n}o},\ and\
  \citenamefont {Retamosa}}]{gomez2002misleading}%
  \BibitemOpen
  \bibfield  {author} {\bibinfo {author} {\bibfnamefont {J.}~\bibnamefont
  {G{\'o}mez}}, \bibinfo {author} {\bibfnamefont {R.~A.}\ \bibnamefont
  {Molina}}, \bibinfo {author} {\bibfnamefont {A.}~\bibnamefont {Rela{\~n}o}},\
  and\ \bibinfo {author} {\bibfnamefont {J.}~\bibnamefont {Retamosa}},\
  }\href@noop {} {\bibfield  {journal} {\bibinfo  {journal} {Phys. Rev. E}\
  }\textbf {\bibinfo {volume} {66}},\ \bibinfo {pages} {036209} (\bibinfo
  {year} {2002})}\BibitemShut {NoStop}%
\bibitem [{\citenamefont {Tekur}\ and\ \citenamefont
  {Santhanam}(2020)}]{tekur2020symmetry}%
  \BibitemOpen
  \bibfield  {author} {\bibinfo {author} {\bibfnamefont {S.~H.}\ \bibnamefont
  {Tekur}}\ and\ \bibinfo {author} {\bibfnamefont {M.}~\bibnamefont
  {Santhanam}},\ }\href@noop {} {\bibfield  {journal} {\bibinfo  {journal}
  {Phys. Rev. Res.}\ }\textbf {\bibinfo {volume} {2}},\ \bibinfo {pages}
  {032063} (\bibinfo {year} {2020})}\BibitemShut {NoStop}%
\bibitem [{\citenamefont {Bhosale}(2021)}]{bhosale2021superposition}%
  \BibitemOpen
  \bibfield  {author} {\bibinfo {author} {\bibfnamefont {U.~T.}\ \bibnamefont
  {Bhosale}},\ }\href@noop {} {\bibfield  {journal} {\bibinfo  {journal} {Phys.
  Rev. B}\ }\textbf {\bibinfo {volume} {104}},\ \bibinfo {pages} {054204}
  (\bibinfo {year} {2021})}\BibitemShut {NoStop}%
\bibitem [{\citenamefont {Rout}\ and\ \citenamefont
  {Bhosale}(2026)}]{rout2026higherorderspacingssuperposedspectra}%
  \BibitemOpen
  \bibfield  {author} {\bibinfo {author} {\bibfnamefont {S.}~\bibnamefont
  {Rout}}\ and\ \bibinfo {author} {\bibfnamefont {U.~T.}\ \bibnamefont
  {Bhosale}},\ }\href@noop {} {\bibfield  {journal} {\bibinfo  {journal} {arXiv
  preprint arXiv:2510.00503}\ } (\bibinfo {year} {2026})}\BibitemShut {NoStop}%
\bibitem [{\citenamefont {Grobe}\ \emph {et~al.}(1988)\citenamefont {Grobe},
  \citenamefont {Haake},\ and\ \citenamefont {Sommers}}]{grobe1988quantum}%
  \BibitemOpen
  \bibfield  {author} {\bibinfo {author} {\bibfnamefont {R.}~\bibnamefont
  {Grobe}}, \bibinfo {author} {\bibfnamefont {F.}~\bibnamefont {Haake}},\ and\
  \bibinfo {author} {\bibfnamefont {H.-J.}\ \bibnamefont {Sommers}},\
  }\href@noop {} {\bibfield  {journal} {\bibinfo  {journal} {Phys. Rev. Lett.}\
  }\textbf {\bibinfo {volume} {61}},\ \bibinfo {pages} {1899} (\bibinfo {year}
  {1988})}\BibitemShut {NoStop}%
\bibitem [{\citenamefont {Grobe}\ and\ \citenamefont
  {Haake}(1989)}]{grobe1989universality}%
  \BibitemOpen
  \bibfield  {author} {\bibinfo {author} {\bibfnamefont {R.}~\bibnamefont
  {Grobe}}\ and\ \bibinfo {author} {\bibfnamefont {F.}~\bibnamefont {Haake}},\
  }\href@noop {} {\bibfield  {journal} {\bibinfo  {journal} {Phys. Rev. Lett.}\
  }\textbf {\bibinfo {volume} {62}},\ \bibinfo {pages} {2893} (\bibinfo {year}
  {1989})}\BibitemShut {NoStop}%
\bibitem [{\citenamefont {Jisha}\ and\ \citenamefont
  {Prakash}(2024)}]{jisha2024universality}%
  \BibitemOpen
  \bibfield  {author} {\bibinfo {author} {\bibfnamefont {C.}~\bibnamefont
  {Jisha}}\ and\ \bibinfo {author} {\bibfnamefont {R.}~\bibnamefont
  {Prakash}},\ }\href@noop {} {\bibfield  {journal} {\bibinfo  {journal}
  {Europhys. Lett.}\ }\textbf {\bibinfo {volume} {146}},\ \bibinfo {pages}
  {11001} (\bibinfo {year} {2024})}\BibitemShut {NoStop}%
\bibitem [{\citenamefont {Akemann}\ \emph
  {et~al.}(2025{\natexlab{a}})\citenamefont {Akemann}, \citenamefont
  {Balducci}, \citenamefont {Chenu}, \citenamefont {P{\"a}{\ss}ler},
  \citenamefont {Roccati},\ and\ \citenamefont {Shir}}]{akemann2025two}%
  \BibitemOpen
  \bibfield  {author} {\bibinfo {author} {\bibfnamefont {G.}~\bibnamefont
  {Akemann}}, \bibinfo {author} {\bibfnamefont {F.}~\bibnamefont {Balducci}},
  \bibinfo {author} {\bibfnamefont {A.}~\bibnamefont {Chenu}}, \bibinfo
  {author} {\bibfnamefont {P.}~\bibnamefont {P{\"a}{\ss}ler}}, \bibinfo
  {author} {\bibfnamefont {F.}~\bibnamefont {Roccati}},\ and\ \bibinfo {author}
  {\bibfnamefont {R.}~\bibnamefont {Shir}},\ }\href@noop {} {\bibfield
  {journal} {\bibinfo  {journal} {Phys. Rev. Res.}\ }\textbf {\bibinfo {volume}
  {7}},\ \bibinfo {pages} {013098} (\bibinfo {year}
  {2025}{\natexlab{a}})}\BibitemShut {NoStop}%
\bibitem [{\citenamefont {Akemann}\ \emph {et~al.}(2019)\citenamefont
  {Akemann}, \citenamefont {Kieburg}, \citenamefont {Mielke},\ and\
  \citenamefont {Prosen}}]{akemann2019universal}%
  \BibitemOpen
  \bibfield  {author} {\bibinfo {author} {\bibfnamefont {G.}~\bibnamefont
  {Akemann}}, \bibinfo {author} {\bibfnamefont {M.}~\bibnamefont {Kieburg}},
  \bibinfo {author} {\bibfnamefont {A.}~\bibnamefont {Mielke}},\ and\ \bibinfo
  {author} {\bibfnamefont {T.}~\bibnamefont {Prosen}},\ }\href@noop {}
  {\bibfield  {journal} {\bibinfo  {journal} {Phys. Rev. Lett.}\ }\textbf
  {\bibinfo {volume} {123}},\ \bibinfo {pages} {254101} (\bibinfo {year}
  {2019})}\BibitemShut {NoStop}%
\bibitem [{\citenamefont {Akemann}\ \emph {et~al.}(2022)\citenamefont
  {Akemann}, \citenamefont {Mielke},\ and\ \citenamefont
  {P{\"a}{\ss}ler}}]{akemann2022spacing}%
  \BibitemOpen
  \bibfield  {author} {\bibinfo {author} {\bibfnamefont {G.}~\bibnamefont
  {Akemann}}, \bibinfo {author} {\bibfnamefont {A.}~\bibnamefont {Mielke}},\
  and\ \bibinfo {author} {\bibfnamefont {P.}~\bibnamefont {P{\"a}{\ss}ler}},\
  }\href@noop {} {\bibfield  {journal} {\bibinfo  {journal} {Phys. Rev. E}\
  }\textbf {\bibinfo {volume} {106}},\ \bibinfo {pages} {014146} (\bibinfo
  {year} {2022})}\BibitemShut {NoStop}%
\bibitem [{\citenamefont {Rotter}\ and\ \citenamefont
  {Bird}(2015)}]{rotter2015review}%
  \BibitemOpen
  \bibfield  {author} {\bibinfo {author} {\bibfnamefont {I.}~\bibnamefont
  {Rotter}}\ and\ \bibinfo {author} {\bibfnamefont {J.}~\bibnamefont {Bird}},\
  }\href@noop {} {\bibfield  {journal} {\bibinfo  {journal} {Rep. Prog. Phys.}\
  }\textbf {\bibinfo {volume} {78}},\ \bibinfo {pages} {114001} (\bibinfo
  {year} {2015})}\BibitemShut {NoStop}%
\bibitem [{\citenamefont {Akemann}\ \emph {et~al.}(2011)\citenamefont
  {Akemann}, \citenamefont {Baik},\ and\ \citenamefont
  {Di~Francesco}}]{akemann2011oxford}%
  \BibitemOpen
  \bibfield  {author} {\bibinfo {author} {\bibfnamefont {G.}~\bibnamefont
  {Akemann}}, \bibinfo {author} {\bibfnamefont {J.}~\bibnamefont {Baik}},\ and\
  \bibinfo {author} {\bibfnamefont {P.}~\bibnamefont {Di~Francesco}},\
  }\href@noop {} {\emph {\bibinfo {title} {The Oxford handbook of random matrix
  theory}}}\ (\bibinfo  {publisher} {Oxford University Press},\ \bibinfo {year}
  {2011})\BibitemShut {NoStop}%
\bibitem [{\citenamefont {Altland}\ and\ \citenamefont
  {Zirnbauer}(1997)}]{altland1997nonstandard}%
  \BibitemOpen
  \bibfield  {author} {\bibinfo {author} {\bibfnamefont {A.}~\bibnamefont
  {Altland}}\ and\ \bibinfo {author} {\bibfnamefont {M.~R.}\ \bibnamefont
  {Zirnbauer}},\ }\href@noop {} {\bibfield  {journal} {\bibinfo  {journal}
  {Phys. Rev. B}\ }\textbf {\bibinfo {volume} {55}},\ \bibinfo {pages} {1142}
  (\bibinfo {year} {1997})}\BibitemShut {NoStop}%
\bibitem [{\citenamefont {Kawabata}\ \emph {et~al.}(2019)\citenamefont
  {Kawabata}, \citenamefont {Shiozaki}, \citenamefont {Ueda},\ and\
  \citenamefont {Sato}}]{Kawabata2019Symmetry}%
  \BibitemOpen
  \bibfield  {author} {\bibinfo {author} {\bibfnamefont {K.}~\bibnamefont
  {Kawabata}}, \bibinfo {author} {\bibfnamefont {K.}~\bibnamefont {Shiozaki}},
  \bibinfo {author} {\bibfnamefont {M.}~\bibnamefont {Ueda}},\ and\ \bibinfo
  {author} {\bibfnamefont {M.}~\bibnamefont {Sato}},\ }\href@noop {} {\bibfield
   {journal} {\bibinfo  {journal} {Phys. Rev. X}\ }\textbf {\bibinfo {volume}
  {9}},\ \bibinfo {pages} {041015} (\bibinfo {year} {2019})}\BibitemShut
  {NoStop}%
\bibitem [{\citenamefont {Ginibre}(1965)}]{ginibre1965statistical}%
  \BibitemOpen
  \bibfield  {author} {\bibinfo {author} {\bibfnamefont {J.}~\bibnamefont
  {Ginibre}},\ }\href@noop {} {\bibfield  {journal} {\bibinfo  {journal} {J.
  Math. Phys.}\ }\textbf {\bibinfo {volume} {6}},\ \bibinfo {pages} {440}
  (\bibinfo {year} {1965})}\BibitemShut {NoStop}%
\bibitem [{\citenamefont {Forrester}\ and\ \citenamefont
  {Sung-Soo}(2024)}]{forrester2024Progress}%
  \BibitemOpen
  \bibfield  {author} {\bibinfo {author} {\bibfnamefont {P.~J.}\ \bibnamefont
  {Forrester}}\ and\ \bibinfo {author} {\bibfnamefont {B.}~\bibnamefont
  {Sung-Soo}},\ }\href@noop {} {\emph {\bibinfo {title} {Progress on the study
  of the Ginibre ensembles}}}\ (\bibinfo  {publisher} {Springer},\ \bibinfo
  {year} {2024})\BibitemShut {NoStop}%
\bibitem [{\citenamefont {Hamazaki}\ \emph {et~al.}(2020)\citenamefont
  {Hamazaki}, \citenamefont {Kawabata}, \citenamefont {Kura},\ and\
  \citenamefont {Ueda}}]{hamazaki2020universality}%
  \BibitemOpen
  \bibfield  {author} {\bibinfo {author} {\bibfnamefont {R.}~\bibnamefont
  {Hamazaki}}, \bibinfo {author} {\bibfnamefont {K.}~\bibnamefont {Kawabata}},
  \bibinfo {author} {\bibfnamefont {N.}~\bibnamefont {Kura}},\ and\ \bibinfo
  {author} {\bibfnamefont {M.}~\bibnamefont {Ueda}},\ }\href@noop {} {\bibfield
   {journal} {\bibinfo  {journal} {Phys. Rev. Res.}\ }\textbf {\bibinfo
  {volume} {2}},\ \bibinfo {pages} {023286} (\bibinfo {year}
  {2020})}\BibitemShut {NoStop}%
\bibitem [{\citenamefont {Akemann}\ \emph
  {et~al.}(2025{\natexlab{b}})\citenamefont {Akemann}, \citenamefont
  {Ayg{\"u}n}, \citenamefont {Kieburg},\ and\ \citenamefont
  {P{\"a}{\ss}ler}}]{akemann2025complex}%
  \BibitemOpen
  \bibfield  {author} {\bibinfo {author} {\bibfnamefont {G.}~\bibnamefont
  {Akemann}}, \bibinfo {author} {\bibfnamefont {N.}~\bibnamefont {Ayg{\"u}n}},
  \bibinfo {author} {\bibfnamefont {M.}~\bibnamefont {Kieburg}},\ and\ \bibinfo
  {author} {\bibfnamefont {P.}~\bibnamefont {P{\"a}{\ss}ler}},\ }\href@noop {}
  {\bibfield  {journal} {\bibinfo  {journal} {J. Phys. A: Math. Theor.}\
  }\textbf {\bibinfo {volume} {58}},\ \bibinfo {pages} {125204} (\bibinfo
  {year} {2025}{\natexlab{b}})}\BibitemShut {NoStop}%
\bibitem [{\citenamefont {Sakhr}\ and\ \citenamefont
  {Nieminen}(2006)}]{sakhr2006wigner}%
  \BibitemOpen
  \bibfield  {author} {\bibinfo {author} {\bibfnamefont {J.}~\bibnamefont
  {Sakhr}}\ and\ \bibinfo {author} {\bibfnamefont {J.~M.}\ \bibnamefont
  {Nieminen}},\ }\href@noop {} {\bibfield  {journal} {\bibinfo  {journal}
  {Physical Review E—Statistical, Nonlinear, and Soft Matter Physics}\
  }\textbf {\bibinfo {volume} {73}},\ \bibinfo {pages} {047202} (\bibinfo
  {year} {2006})}\BibitemShut {NoStop}%
\bibitem [{\citenamefont {Haake}(2010)}]{haake2010quantum}%
  \BibitemOpen
  \bibfield  {author} {\bibinfo {author} {\bibfnamefont {F.}~\bibnamefont
  {Haake}},\ }\href@noop {} {\emph {\bibinfo {title} {Quantum Signatures of
  Chaos}}}\ (\bibinfo  {publisher} {Springer, 3rd Edition, Berlin},\ \bibinfo
  {year} {2010})\BibitemShut {NoStop}%
\bibitem [{\citenamefont {Akemann}\ \emph {et~al.}(2024)\citenamefont
  {Akemann}, \citenamefont {Chakarov}, \citenamefont {Kr{\"u}ger},
  \citenamefont {Mielke}, \citenamefont {Ottensmann},\ and\ \citenamefont
  {P{\"a}ssler}}]{akemann2024interactions}%
  \BibitemOpen
  \bibfield  {author} {\bibinfo {author} {\bibfnamefont {G.}~\bibnamefont
  {Akemann}}, \bibinfo {author} {\bibfnamefont {N.}~\bibnamefont {Chakarov}},
  \bibinfo {author} {\bibfnamefont {O.}~\bibnamefont {Kr{\"u}ger}}, \bibinfo
  {author} {\bibfnamefont {A.}~\bibnamefont {Mielke}}, \bibinfo {author}
  {\bibfnamefont {M.}~\bibnamefont {Ottensmann}},\ and\ \bibinfo {author}
  {\bibfnamefont {P.}~\bibnamefont {P{\"a}ssler}},\ }\href@noop {} {\bibfield
  {journal} {\bibinfo  {journal} {J. Stat. Mech.}\ }\textbf {\bibinfo {volume}
  {2024}},\ \bibinfo {pages} {053501} (\bibinfo {year} {2024})}\BibitemShut
  {NoStop}%
\bibitem [{\citenamefont {S{\'a}}\ \emph {et~al.}(2020)\citenamefont {S{\'a}},
  \citenamefont {Ribeiro},\ and\ \citenamefont {Prosen}}]{sa2020complex}%
  \BibitemOpen
  \bibfield  {author} {\bibinfo {author} {\bibfnamefont {L.}~\bibnamefont
  {S{\'a}}}, \bibinfo {author} {\bibfnamefont {P.}~\bibnamefont {Ribeiro}},\
  and\ \bibinfo {author} {\bibfnamefont {T.}~\bibnamefont {Prosen}},\
  }\href@noop {} {\bibfield  {journal} {\bibinfo  {journal} {Phys. Rev. X}\
  }\textbf {\bibinfo {volume} {10}},\ \bibinfo {pages} {021019} (\bibinfo
  {year} {2020})}\BibitemShut {NoStop}%
\bibitem [{\citenamefont {Akemann}\ \emph {et~al.}(2026)\citenamefont
  {Akemann}, \citenamefont {Angermann}, \citenamefont {Ayg{\"u}n},
  \citenamefont {Mielke}, \citenamefont {P{\"a}{\ss}ler}, \citenamefont
  {Raitzig},\ and\ \citenamefont {Winkler}}]{akemann2026three}%
  \BibitemOpen
  \bibfield  {author} {\bibinfo {author} {\bibfnamefont {G.}~\bibnamefont
  {Akemann}}, \bibinfo {author} {\bibfnamefont {G.}~\bibnamefont {Angermann}},
  \bibinfo {author} {\bibfnamefont {N.}~\bibnamefont {Ayg{\"u}n}}, \bibinfo
  {author} {\bibfnamefont {A.}~\bibnamefont {Mielke}}, \bibinfo {author}
  {\bibfnamefont {P.}~\bibnamefont {P{\"a}{\ss}ler}}, \bibinfo {author}
  {\bibfnamefont {C.}~\bibnamefont {Raitzig}},\ and\ \bibinfo {author}
  {\bibfnamefont {T.}~\bibnamefont {Winkler}},\ }\href@noop {} {\bibfield
  {journal} {\bibinfo  {journal} {J. Phys. A: Math. Theor.}\ }\textbf {\bibinfo
  {volume} {59}},\ \bibinfo {pages} {255201} (\bibinfo {year}
  {2026})}\BibitemShut {NoStop}%
\bibitem [{\citenamefont {Tekur}\ \emph {et~al.}(2024)\citenamefont {Tekur},
  \citenamefont {Santhanam}, \citenamefont {Agarwalla},\ and\ \citenamefont
  {Kulkarni}}]{tekur2024higher}%
  \BibitemOpen
  \bibfield  {author} {\bibinfo {author} {\bibfnamefont {S.~H.}\ \bibnamefont
  {Tekur}}, \bibinfo {author} {\bibfnamefont {M.}~\bibnamefont {Santhanam}},
  \bibinfo {author} {\bibfnamefont {B.~K.}\ \bibnamefont {Agarwalla}},\ and\
  \bibinfo {author} {\bibfnamefont {M.}~\bibnamefont {Kulkarni}},\ }\href@noop
  {} {\bibfield  {journal} {\bibinfo  {journal} {Phys. Rev. B}\ }\textbf
  {\bibinfo {volume} {110}},\ \bibinfo {pages} {L241410} (\bibinfo {year}
  {2024})}\BibitemShut {NoStop}%
\bibitem [{\citenamefont {Bhosale}\ \emph {et~al.}(2018)\citenamefont
  {Bhosale}, \citenamefont {Tekur},\ and\ \citenamefont
  {Santhanam}}]{bhosale2018scaling}%
  \BibitemOpen
  \bibfield  {author} {\bibinfo {author} {\bibfnamefont {U.~T.}\ \bibnamefont
  {Bhosale}}, \bibinfo {author} {\bibfnamefont {S.~H.}\ \bibnamefont {Tekur}},\
  and\ \bibinfo {author} {\bibfnamefont {M.}~\bibnamefont {Santhanam}},\
  }\href@noop {} {\bibfield  {journal} {\bibinfo  {journal} {Phys. Rev. E}\
  }\textbf {\bibinfo {volume} {98}},\ \bibinfo {pages} {052133} (\bibinfo
  {year} {2018})}\BibitemShut {NoStop}%
\bibitem [{sup()}]{supplementry2026}%
  \BibitemOpen
  \href@noop {} {\bibinfo  {journal} {See the Supplementary Material at [] for
  more plots of higher-order spacing (HOS) distributions for Ginibre,
  $AI^{\dagger}$, and $AII^{\dagger}$ universality classes of local bulk
  statistics, product of independent matrices from GinUE, and non-Hermitian SYK
  model. We also study the HOS statistics of Bernoulli complex symmetric
  matrices and there we have tabulated the results and plot them.}\
  }\BibitemShut {NoStop}%
\bibitem [{\citenamefont {Forrester}(2013)}]{Forrester2013Lyapunov}%
  \BibitemOpen
\bibfield  {journal} {  }\bibfield  {author} {\bibinfo {author} {\bibfnamefont
  {P.~J.}\ \bibnamefont {Forrester}},\ }\href@noop {} {\bibfield  {journal}
  {\bibinfo  {journal} {J. Stat. Phys.}\ }\textbf {\bibinfo {volume} {151}},\
  \bibinfo {pages} {796} (\bibinfo {year} {2013})}\BibitemShut {NoStop}%
\bibitem [{\citenamefont {Comtet}\ \emph {et~al.}(2013)\citenamefont {Comtet},
  \citenamefont {Texier},\ and\ \citenamefont {Tourigny}}]{Comtet2013Lyapunov}%
  \BibitemOpen
  \bibfield  {author} {\bibinfo {author} {\bibfnamefont {A.}~\bibnamefont
  {Comtet}}, \bibinfo {author} {\bibfnamefont {C.}~\bibnamefont {Texier}},\
  and\ \bibinfo {author} {\bibfnamefont {Y.}~\bibnamefont {Tourigny}},\
  }\href@noop {} {\bibfield  {journal} {\bibinfo  {journal} {J. Phys. A: Math.
  Theor.}\ }\textbf {\bibinfo {volume} {46}},\ \bibinfo {pages} {254003}
  (\bibinfo {year} {2013})}\BibitemShut {NoStop}%
\bibitem [{\citenamefont {Pennington}\ \emph {et~al.}(2017)\citenamefont
  {Pennington}, \citenamefont {Schoenholz},\ and\ \citenamefont
  {Ganguli}}]{Pennington2017Resurrecting}%
  \BibitemOpen
  \bibfield  {author} {\bibinfo {author} {\bibfnamefont {J.}~\bibnamefont
  {Pennington}}, \bibinfo {author} {\bibfnamefont {S.}~\bibnamefont
  {Schoenholz}},\ and\ \bibinfo {author} {\bibfnamefont {S.}~\bibnamefont
  {Ganguli}},\ }\href@noop {} {\bibfield  {journal} {\bibinfo  {journal} {in
  Advances in Neural Information Processing Systems}\ } (\bibinfo {year} {MIT
  Press, Cambridge, MA, 2017})}\BibitemShut {NoStop}%
\bibitem [{\citenamefont {Vanneste}(2010)}]{Vanneste2010Estimating}%
  \BibitemOpen
  \bibfield  {author} {\bibinfo {author} {\bibfnamefont {J.}~\bibnamefont
  {Vanneste}},\ }\href@noop {} {\bibfield  {journal} {\bibinfo  {journal}
  {Phys. Rev. E}\ }\textbf {\bibinfo {volume} {81}},\ \bibinfo {pages} {036701}
  (\bibinfo {year} {2010})}\BibitemShut {NoStop}%
\bibitem [{\citenamefont {Burda}\ \emph {et~al.}(2010)\citenamefont {Burda},
  \citenamefont {Janik},\ and\ \citenamefont {Waclaw}}]{Burda2010}%
  \BibitemOpen
  \bibfield  {author} {\bibinfo {author} {\bibfnamefont {Z.}~\bibnamefont
  {Burda}}, \bibinfo {author} {\bibfnamefont {R.}~\bibnamefont {Janik}},\ and\
  \bibinfo {author} {\bibfnamefont {B.}~\bibnamefont {Waclaw}},\ }\href@noop {}
  {\bibfield  {journal} {\bibinfo  {journal} {Phys. Rev. E}\ }\textbf {\bibinfo
  {volume} {81}},\ \bibinfo {pages} {041132} (\bibinfo {year}
  {2010})}\BibitemShut {NoStop}%
\bibitem [{\citenamefont {Akemann}\ and\ \citenamefont
  {Burda}(2012)}]{akemann2012Universal}%
  \BibitemOpen
  \bibfield  {author} {\bibinfo {author} {\bibfnamefont {G.}~\bibnamefont
  {Akemann}}\ and\ \bibinfo {author} {\bibfnamefont {Z.}~\bibnamefont
  {Burda}},\ }\href@noop {} {\bibfield  {journal} {\bibinfo  {journal} {J.
  Phys. A: Math. Theor.}\ }\textbf {\bibinfo {volume} {45}},\ \bibinfo {pages}
  {465201} (\bibinfo {year} {2012})}\BibitemShut {NoStop}%
\bibitem [{\citenamefont {Garc{\'\i}a-Garc{\'\i}a}\ \emph
  {et~al.}(2022{\natexlab{a}})\citenamefont {Garc{\'\i}a-Garc{\'\i}a},
  \citenamefont {Jia}, \citenamefont {Rosa},\ and\ \citenamefont
  {Verbaarschot}}]{Garcia2022Replica}%
  \BibitemOpen
  \bibfield  {author} {\bibinfo {author} {\bibfnamefont {A.~M.}\ \bibnamefont
  {Garc{\'\i}a-Garc{\'\i}a}}, \bibinfo {author} {\bibfnamefont
  {Y.}~\bibnamefont {Jia}}, \bibinfo {author} {\bibfnamefont {D.}~\bibnamefont
  {Rosa}},\ and\ \bibinfo {author} {\bibfnamefont {J.~J.}\ \bibnamefont
  {Verbaarschot}},\ }\href@noop {} {\bibfield  {journal} {\bibinfo  {journal}
  {Phys. Rev. D}\ }\textbf {\bibinfo {volume} {105}},\ \bibinfo {pages}
  {126027} (\bibinfo {year} {2022}{\natexlab{a}})}\BibitemShut {NoStop}%
\bibitem [{\citenamefont {Garc{\'\i}a-Garc{\'\i}a}\ \emph
  {et~al.}(2022{\natexlab{b}})\citenamefont {Garc{\'\i}a-Garc{\'\i}a},
  \citenamefont {S{\'a}},\ and\ \citenamefont
  {Verbaarschot}}]{garcia2022symmetry}%
  \BibitemOpen
  \bibfield  {author} {\bibinfo {author} {\bibfnamefont {A.~M.}\ \bibnamefont
  {Garc{\'\i}a-Garc{\'\i}a}}, \bibinfo {author} {\bibfnamefont
  {L.}~\bibnamefont {S{\'a}}},\ and\ \bibinfo {author} {\bibfnamefont {J.~J.}\
  \bibnamefont {Verbaarschot}},\ }\href@noop {} {\bibfield  {journal} {\bibinfo
   {journal} {Phys. Rev. X}\ }\textbf {\bibinfo {volume} {12}},\ \bibinfo
  {pages} {021040} (\bibinfo {year} {2022}{\natexlab{b}})}\BibitemShut
  {NoStop}%
\bibitem [{\citenamefont {Garc{\'\i}a-Garc{\'\i}a}\ and\ \citenamefont
  {Verbaarschot}(2016)}]{Garcia2016Spectral}%
  \BibitemOpen
  \bibfield  {author} {\bibinfo {author} {\bibfnamefont {A.~M.}\ \bibnamefont
  {Garc{\'\i}a-Garc{\'\i}a}}\ and\ \bibinfo {author} {\bibfnamefont {J.~J.}\
  \bibnamefont {Verbaarschot}},\ }\href@noop {} {\bibfield  {journal} {\bibinfo
   {journal} {Phys. Rev. D}\ }\textbf {\bibinfo {volume} {94}},\ \bibinfo
  {pages} {126010} (\bibinfo {year} {2016})}\BibitemShut {NoStop}%
\bibitem [{\citenamefont {Garc{\'\i}a-Garc{\'\i}a}\ and\ \citenamefont
  {Verbaarschot}(2017)}]{Garcia2017Analytical}%
  \BibitemOpen
  \bibfield  {author} {\bibinfo {author} {\bibfnamefont {A.~M.}\ \bibnamefont
  {Garc{\'\i}a-Garc{\'\i}a}}\ and\ \bibinfo {author} {\bibfnamefont {J.~J.}\
  \bibnamefont {Verbaarschot}},\ }\href@noop {} {\bibfield  {journal} {\bibinfo
   {journal} {Phys. Rev. D}\ }\textbf {\bibinfo {volume} {96}},\ \bibinfo
  {pages} {066012} (\bibinfo {year} {2017})}\BibitemShut {NoStop}%
\bibitem [{\citenamefont {Kitaev}(2015)}]{Kitaev2015}%
  \BibitemOpen
  \bibfield  {author} {\bibinfo {author} {\bibfnamefont {A.}~\bibnamefont
  {Kitaev}},\ }\href@noop {} {\bibinfo {title} {A simple model of quantum
  holography}},\ \bibinfo {howpublished}
  {\url{http://online.kitp.ucsb.edu/online/entangled15/kitaev/};
  \url{http://online.kitp.ucsb.edu/online/entangled15/kitaev2/}} (\bibinfo
  {year} {2015})\BibitemShut {NoStop}%
\bibitem [{\citenamefont {Maldacena}\ and\ \citenamefont
  {Stanford}(2016)}]{Maldacena2016Remarks}%
  \BibitemOpen
  \bibfield  {author} {\bibinfo {author} {\bibfnamefont {J.}~\bibnamefont
  {Maldacena}}\ and\ \bibinfo {author} {\bibfnamefont {D.}~\bibnamefont
  {Stanford}},\ }\href@noop {} {\bibfield  {journal} {\bibinfo  {journal}
  {Phys. Rev. D}\ }\textbf {\bibinfo {volume} {94}},\ \bibinfo {pages} {106002}
  (\bibinfo {year} {2016})}\BibitemShut {NoStop}%
\bibitem [{\citenamefont {Garc{\'\i}a-Garc{\'\i}a}\ \emph
  {et~al.}(2023)\citenamefont {Garc{\'\i}a-Garc{\'\i}a}, \citenamefont
  {S{\'a}},\ and\ \citenamefont {Verbaarschot}}]{garcia2023universality}%
  \BibitemOpen
  \bibfield  {author} {\bibinfo {author} {\bibfnamefont {A.~M.}\ \bibnamefont
  {Garc{\'\i}a-Garc{\'\i}a}}, \bibinfo {author} {\bibfnamefont
  {L.}~\bibnamefont {S{\'a}}},\ and\ \bibinfo {author} {\bibfnamefont {J.~J.}\
  \bibnamefont {Verbaarschot}},\ }\href@noop {} {\bibfield  {journal} {\bibinfo
   {journal} {Phys. Rev. D}\ }\textbf {\bibinfo {volume} {107}},\ \bibinfo
  {pages} {066007} (\bibinfo {year} {2023})}\BibitemShut {NoStop}%
\end{thebibliography}%

\clearpage


\makeatletter
\let\addcontentsline\addcontentslineOriginal
\makeatother


\setcounter{section}{0}
\setcounter{subsection}{0}
\setcounter{equation}{0}
\setcounter{figure}{0}
\setcounter{table}{0}


\renewcommand{\thesection}{S\Roman{section}}
\renewcommand{\theHsection}{supplement.section.\arabic{section}}

\renewcommand{\thesubsection}{S\Roman{section}.\Alph{subsection}}
\renewcommand{\theHsubsection}{supplement.subsection.\arabic{section}.\arabic{subsection}}

\renewcommand{\theequation}{S\arabic{equation}}
\renewcommand{\theHequation}{supplement.equation.\arabic{equation}}

\renewcommand{\thefigure}{S\arabic{figure}}
\renewcommand{\theHfigure}{supplement.figure.\arabic{figure}}

\renewcommand{\thetable}{S\arabic{table}}
\renewcommand{\theHtable}{supplement.table.\arabic{table}}


\newcommand{\suppsection}[2][]{%
    \refstepcounter{section}%
    \ifstrempty{#1}{}{\label{#1}}%
    \section*{\texorpdfstring
        {S\Roman{section}\quad #2}
        {S\Roman{section} #2}}%
    \addcontentsline{toc}{section}
        {S\Roman{section}\quad #2}%
}

\newcommand{\suppsubsection}[2][]{%
    \refstepcounter{subsection}%
    \ifstrempty{#1}{}{\label{#1}}%
    \subsection*{\Alph{subsection}.\quad #2}%
    \addcontentsline{toc}{subsection}
        {\Alph{subsection}.\quad #2}%
}


\vspace*{1cm}
\onecolumngrid

\begin{center}
\textbf{\large Supplementary Material for\\ 
``\textit{Higher-order spacing distributions in non-Hermitian random matrix ensembles and the Sachdev-Ye-Kitaev model}''}
\end{center}

\tableofcontents
\newpage

\suppsection{Introduction}
\label{intro}
In this supplementary material, we present scatter plots of the eigenvalues of the various non-Hermitian random matrix ensembles, a few more plots of HOSD, and a complete study of HOSD of Bernoulli complex symmetric random matrices. These plots provide additional support to our results presented in the main text of the paper.  

\begin{figure*}[h!]
    \centering
    \begin{subfigure}{0.3\textwidth}
        \includegraphics[width=\linewidth]{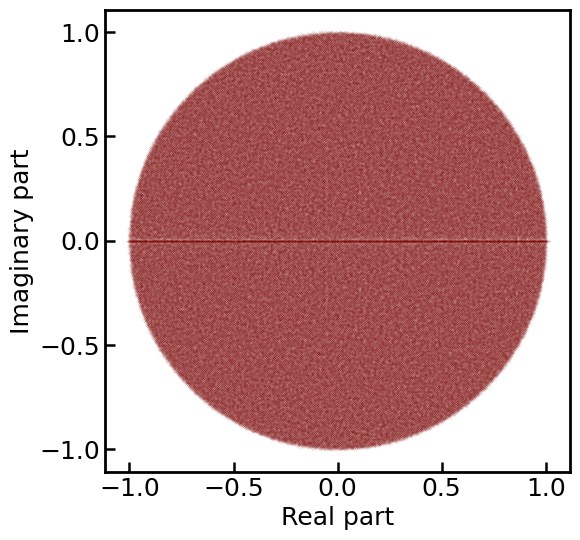}
        \caption{GinOE}
        \label{fig:GinOE_ev}
    \end{subfigure}
    \hfill
    \begin{subfigure}{0.3\textwidth}
        \includegraphics[width=\linewidth]{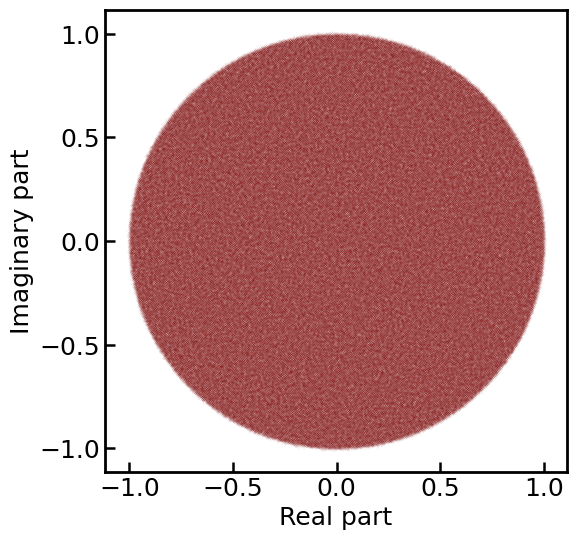}
        \caption{GinUE}
        \label{fig:GinUE_ev}
    \end{subfigure}
    \hfill
    \begin{subfigure}{0.3\textwidth}
        \includegraphics[width=\linewidth]{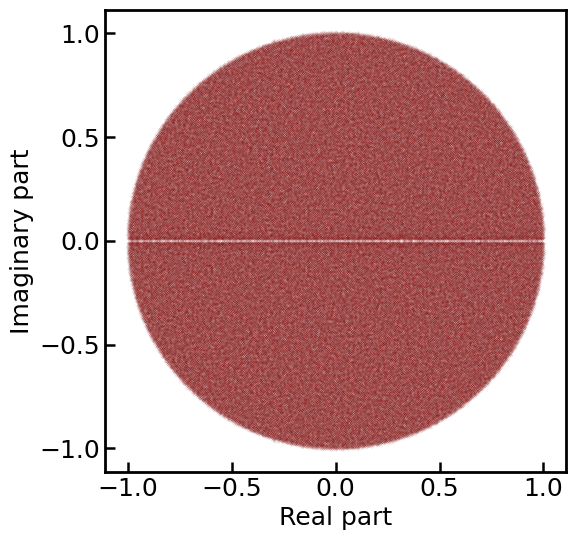}
        \caption{GinSE}
        \label{fig:GinSE_ev}
    \end{subfigure}
\caption{\justifying Scatter plot of complex eigenvalues of Ginibre ensembles. Here, the dimension $N=10000$, and the number of realizations $n=100$ each for GinOE ($AI$) and GinUE ($A$), respectively. For GinSE ($AII$), $2N=10000$ and $n=100$.}
\label{supfig:Ginibre_ensemble_ev}
\end{figure*}

\begin{figure}[h!]
    \centering

    \subfloat[$AI^\dagger$\label{AI_dagger_ev}]{
        \includegraphics[width=0.32\columnwidth]{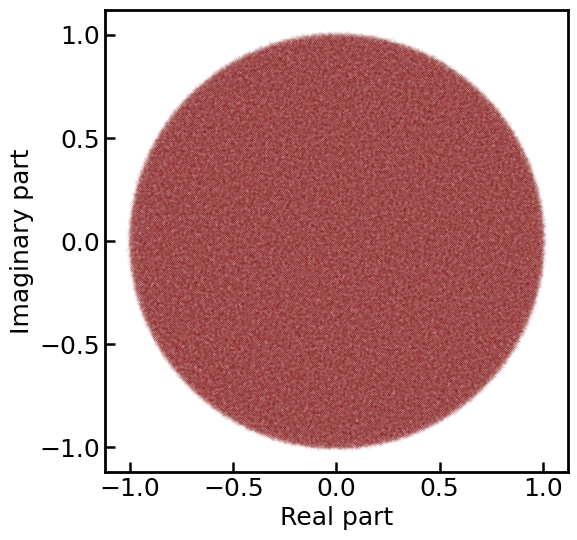}
    }
    \hspace{0.02\columnwidth}
    \subfloat[$AII^\dagger$\label{AII_dagger_ev}]{
        \includegraphics[width=0.32\columnwidth]{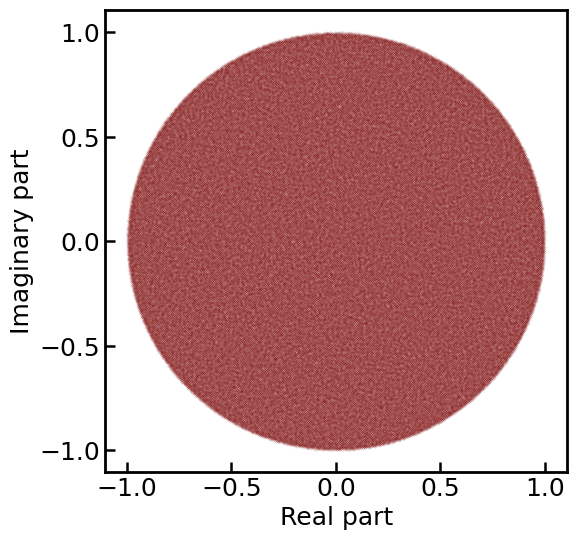}
    }

    \caption{\justifying Scatter plot of complex eigenvalues of the universality classes
    $AI^\dagger$ and $AII^\dagger$. Here, $N=10000$ for
    $AI^\dagger$ and $2N=20000$ for $AII^\dagger$ and $n=100$ for each class.}
    \label{supfig:AI_dagger_AII_dagger_ev}
\end{figure}

\begin{figure}
\begin{center}
\includegraphics*[scale=0.33]{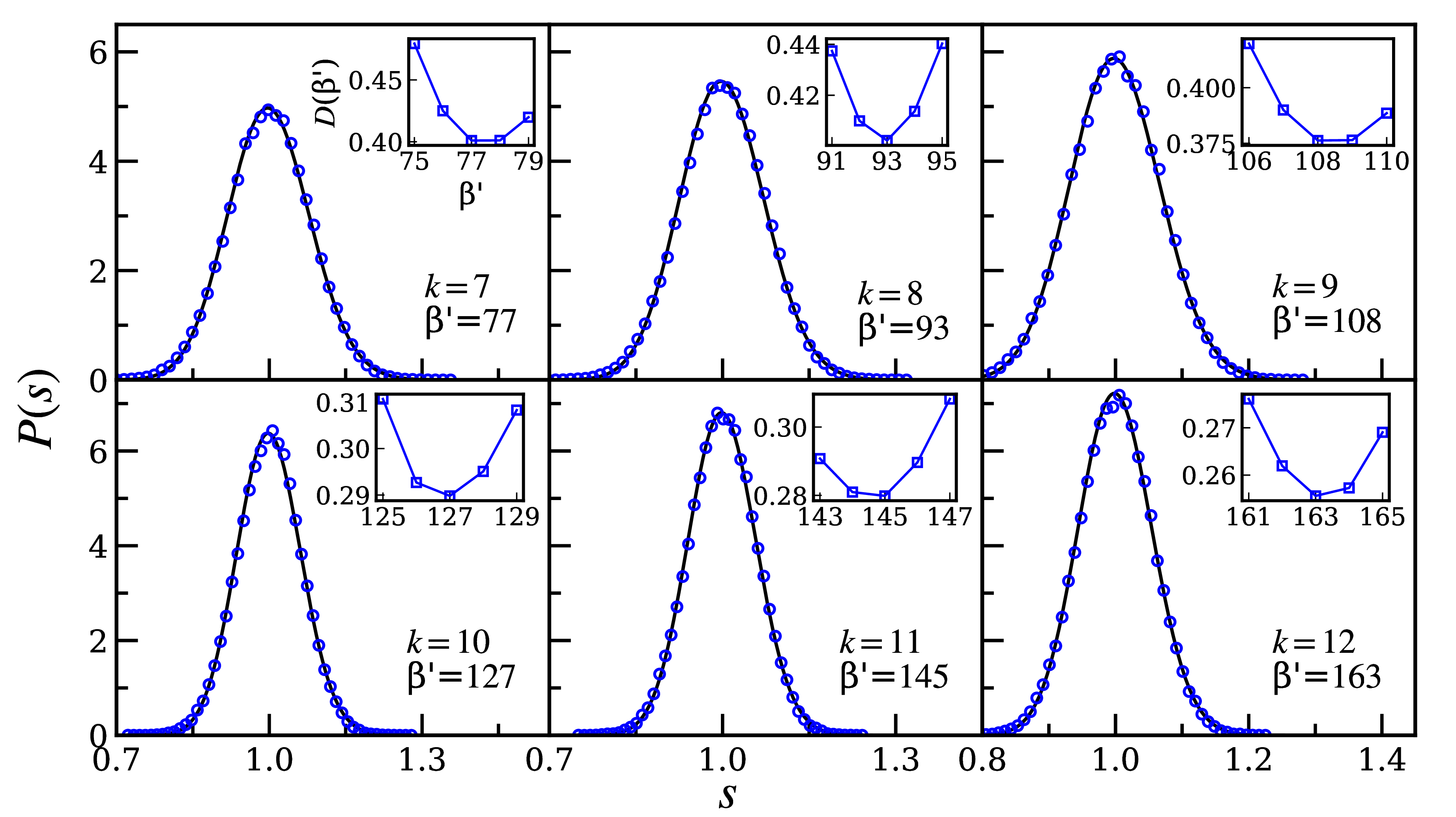}
\caption{\justifying Distribution of the $k$-th order spacings (circles) for the GinOE spectra. Here, $N=10000$ and $n=500$. The solid curve corresponds to $P(s,\beta')$ as given in Eq.~(\ref{SuppEq:PSBetA}), in which $\beta$ is substituted by $\beta'$ and $\beta'$ is given in Table I of the main text. The insets show $D(\beta')$ as a function of $\beta'$.}
\label{fig:GinOE2_SPACINGS}
\end{center}
\end{figure}

\begin{figure}
\begin{center}
\includegraphics*[scale=0.33]{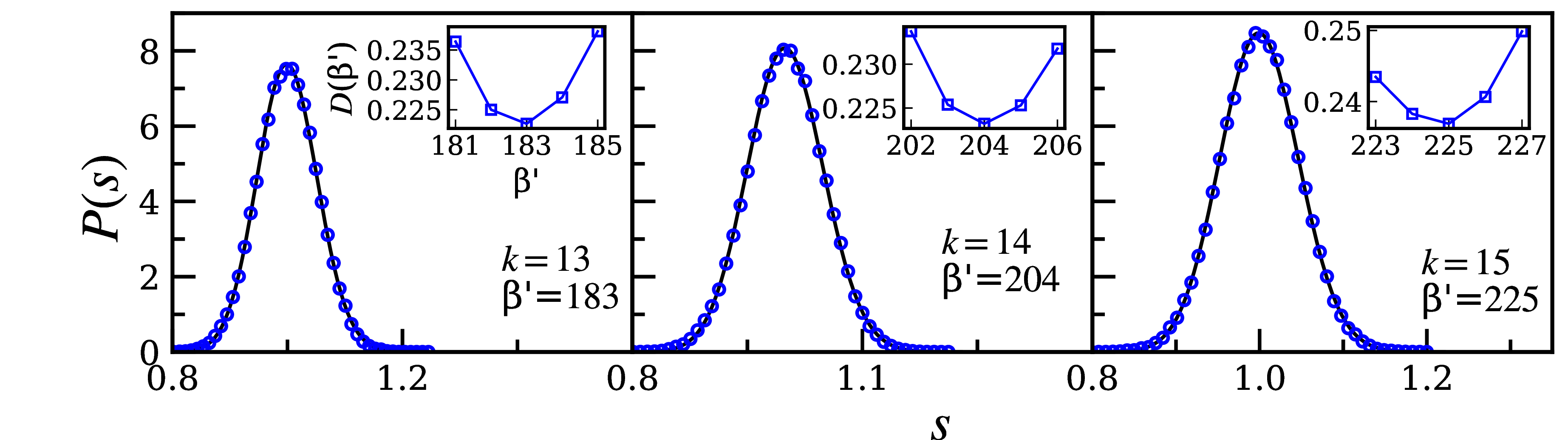} 
\caption{\justifying Same as Fig.~\ref{fig:GinOE2_SPACINGS} but for different values of $k$.}
\label{fig:GinOE3_SPACINGS}
\end{center}
\end{figure}

\begin{figure}
\begin{center}
\includegraphics*[scale=0.33]{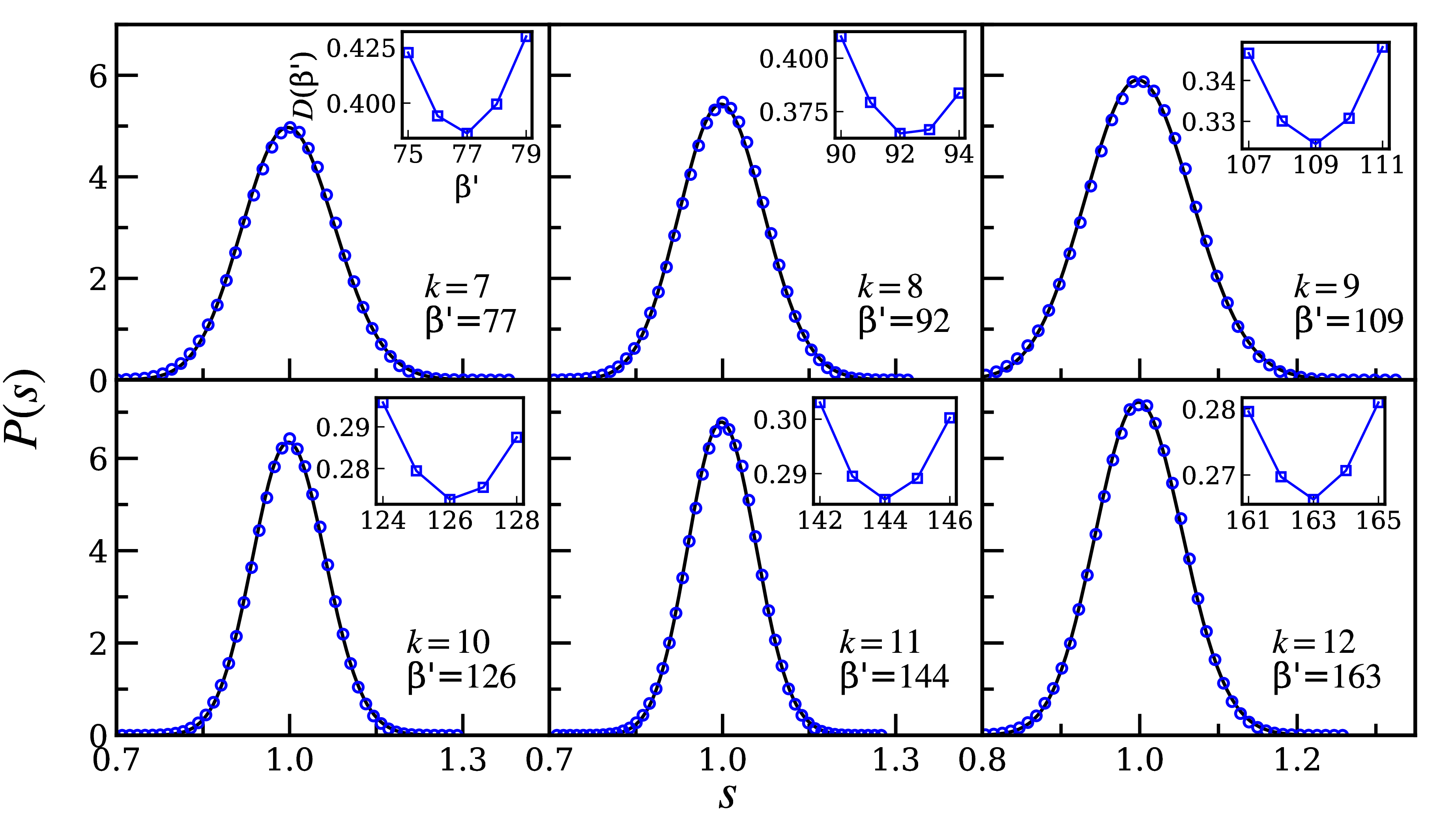}
\caption{\justifying Same as Fig.~\ref{fig:GinOE2_SPACINGS}, but for GinUE.}
\label{fig:GinUE2_SPACINGS}
\end{center}
\end{figure}

\begin{figure}
\begin{center}
\includegraphics*[scale=0.33]{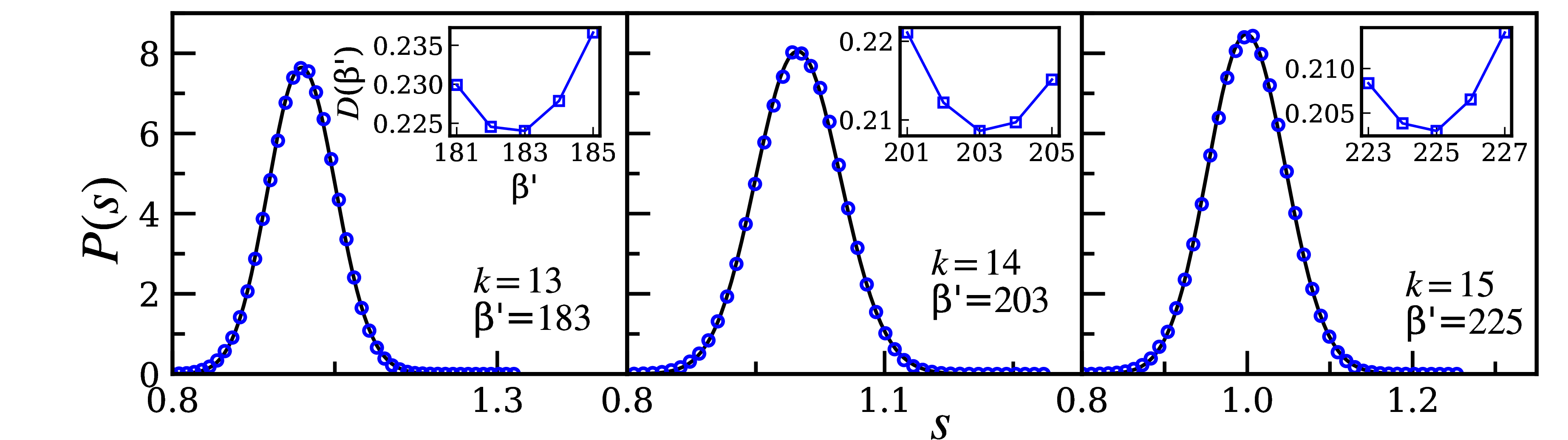}
\caption{\justifying Same as Fig.~\ref{fig:GinUE2_SPACINGS} but for different values of $k$.}
\label{fig:GinUE3_SPACINGS}
\end{center}
\end{figure}

\begin{figure}
\begin{center}
\includegraphics*[scale=0.33]{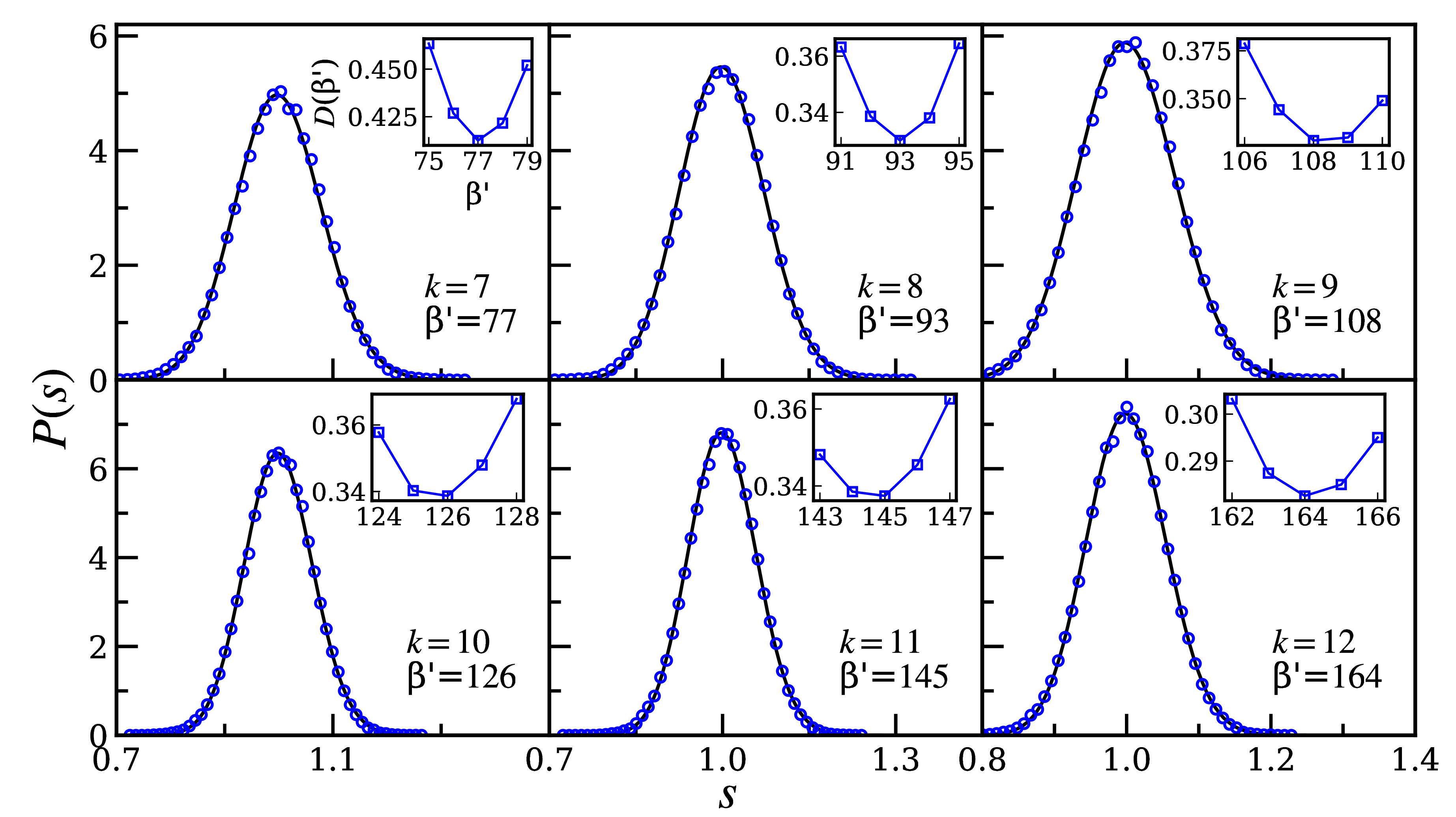}
\caption{\justifying Same as Fig.~\ref{fig:GinOE2_SPACINGS}, but for GinSE. Here, $2N=10000$.}
\label{fig:GinSE2_SPACINGS}
\end{center}
\end{figure}

\begin{figure}
\begin{center}
\includegraphics*[scale=0.33]{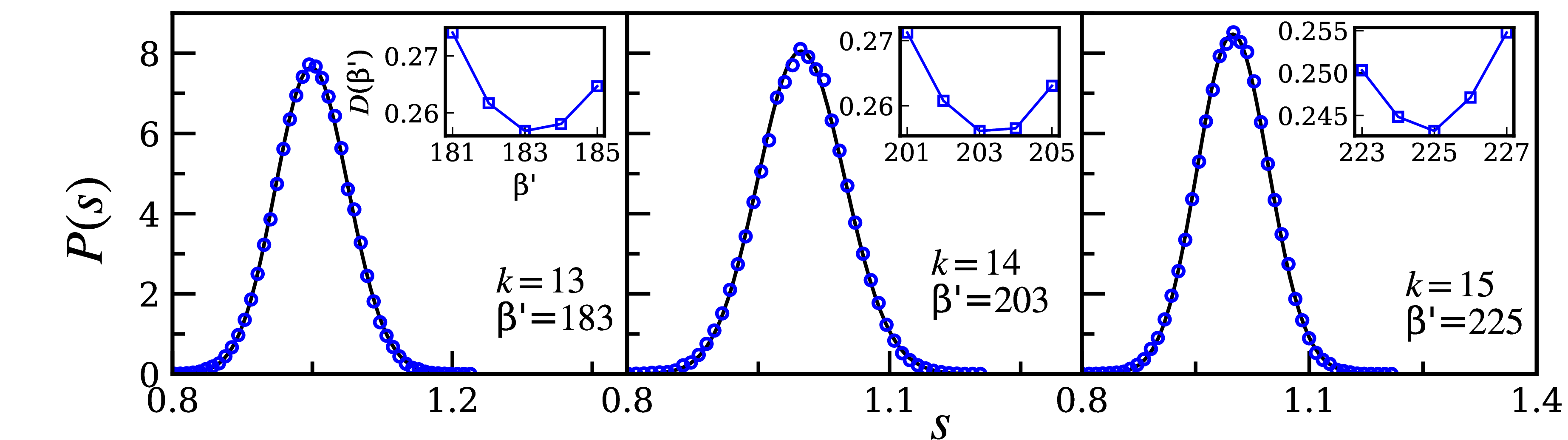}
\caption{\justifying Same as Fig.~\ref{fig:GinSE2_SPACINGS} but for different values of $k$.}
\label{fig:GinSE3_SPACINGS}
\end{center}
\end{figure}

\begin{figure}
\begin{center}
\includegraphics*[scale=0.33]{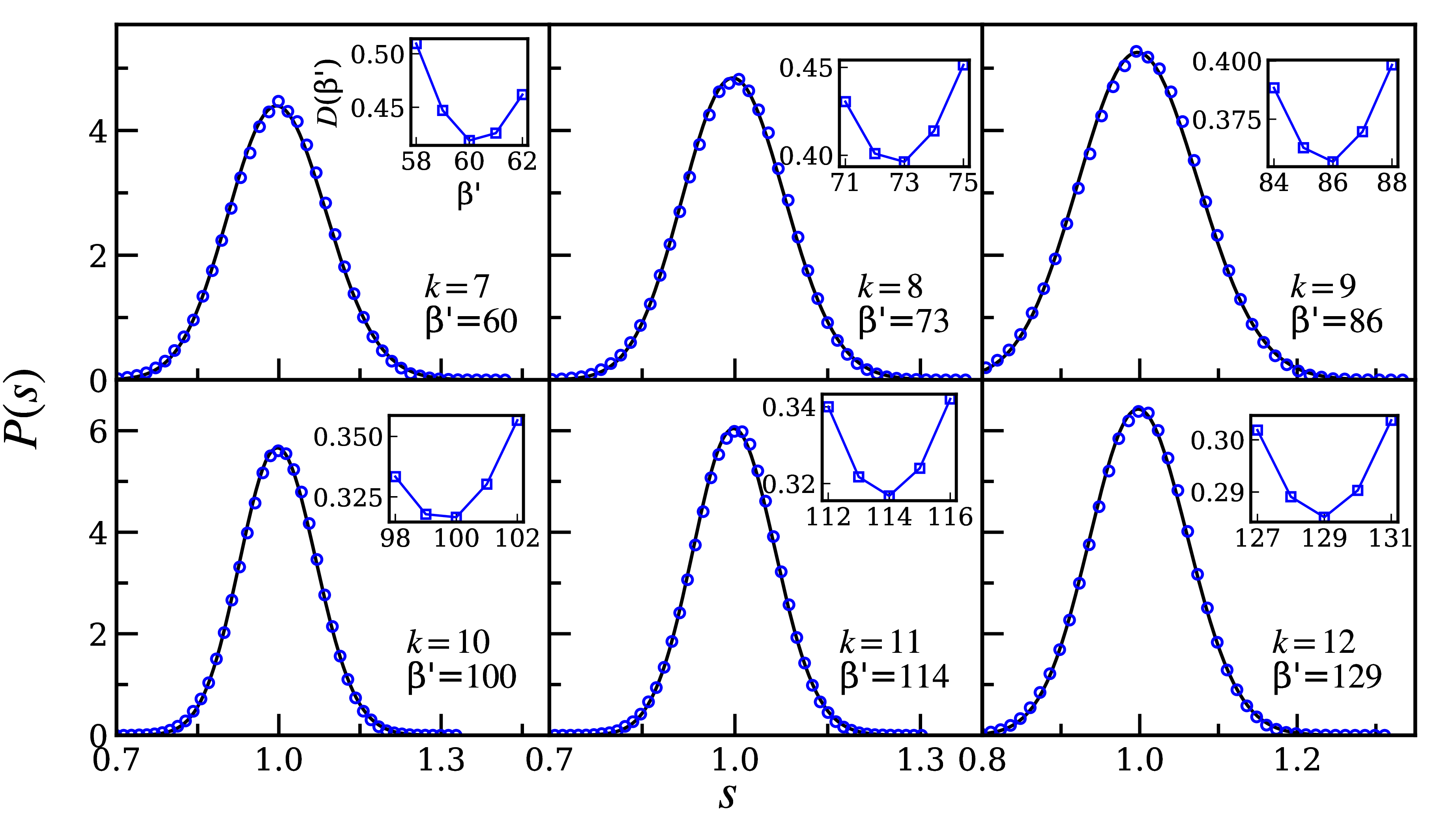}
\caption{\justifying Same as Fig.~\ref{fig:GinOE2_SPACINGS}, but for $AI^{\dagger}$.}
\label{fig: AI_dagger2_SPACINGS}
\end{center}
\end{figure}

\begin{figure}
\begin{center}
\includegraphics*[scale=0.33]{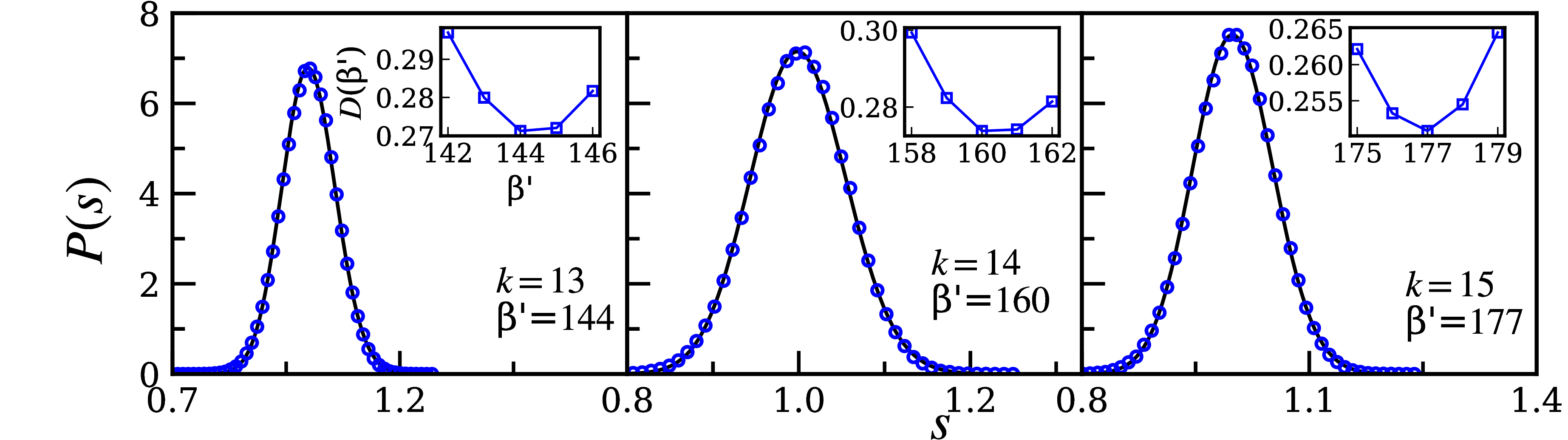}
\caption{\justifying Same as Fig.~\ref{fig: AI_dagger2_SPACINGS} but for different values of $k$.}
\label{fig: AI_dagger3_SPACINGS}
\end{center}
\end{figure}

\begin{figure}
\begin{center}
\includegraphics*[scale=0.33]{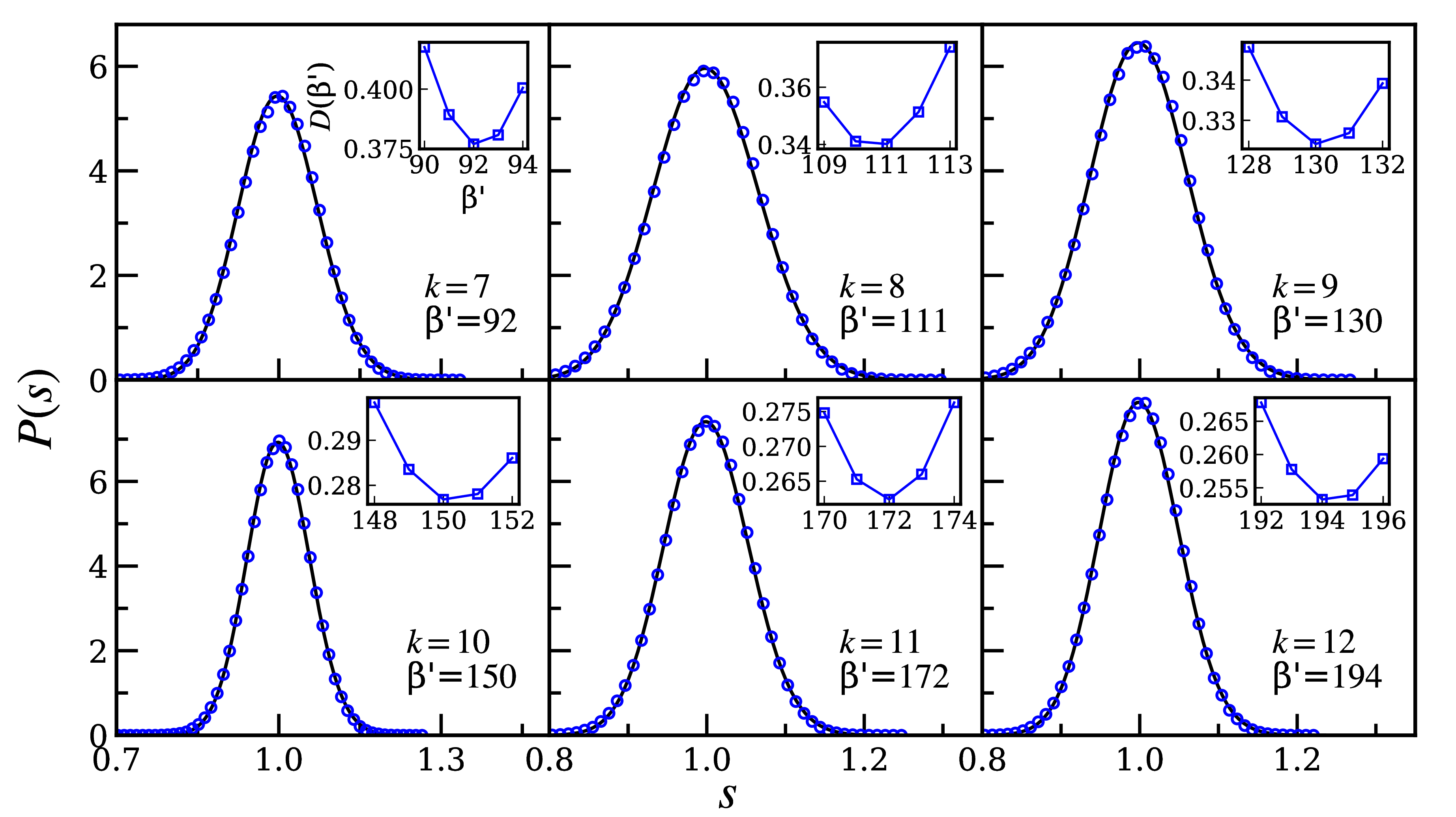}
\caption{\justifying Same as Fig.~\ref{fig:GinOE2_SPACINGS}, but for $AII^{\dagger}$. Here, $2N=20000$.}
\label{fig: AII_dagger2 spacings}
\end{center}
\end{figure}

\begin{figure}
\begin{center}
\includegraphics*[scale=0.33]{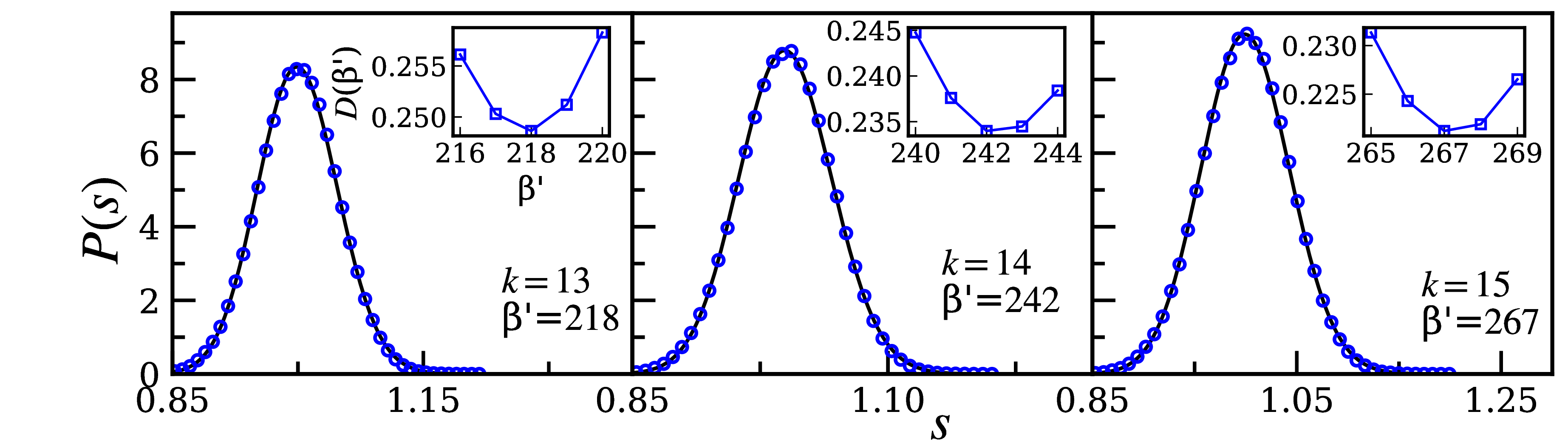}
\caption{\justifying Same as Fig.~\ref{fig: AII_dagger2 spacings} but for different values of $k$.}
\label{fig: AII_dagger3 spacings}
\end{center}
\end{figure}

\begin{figure}
\begin{center}
\includegraphics*[scale=0.33]{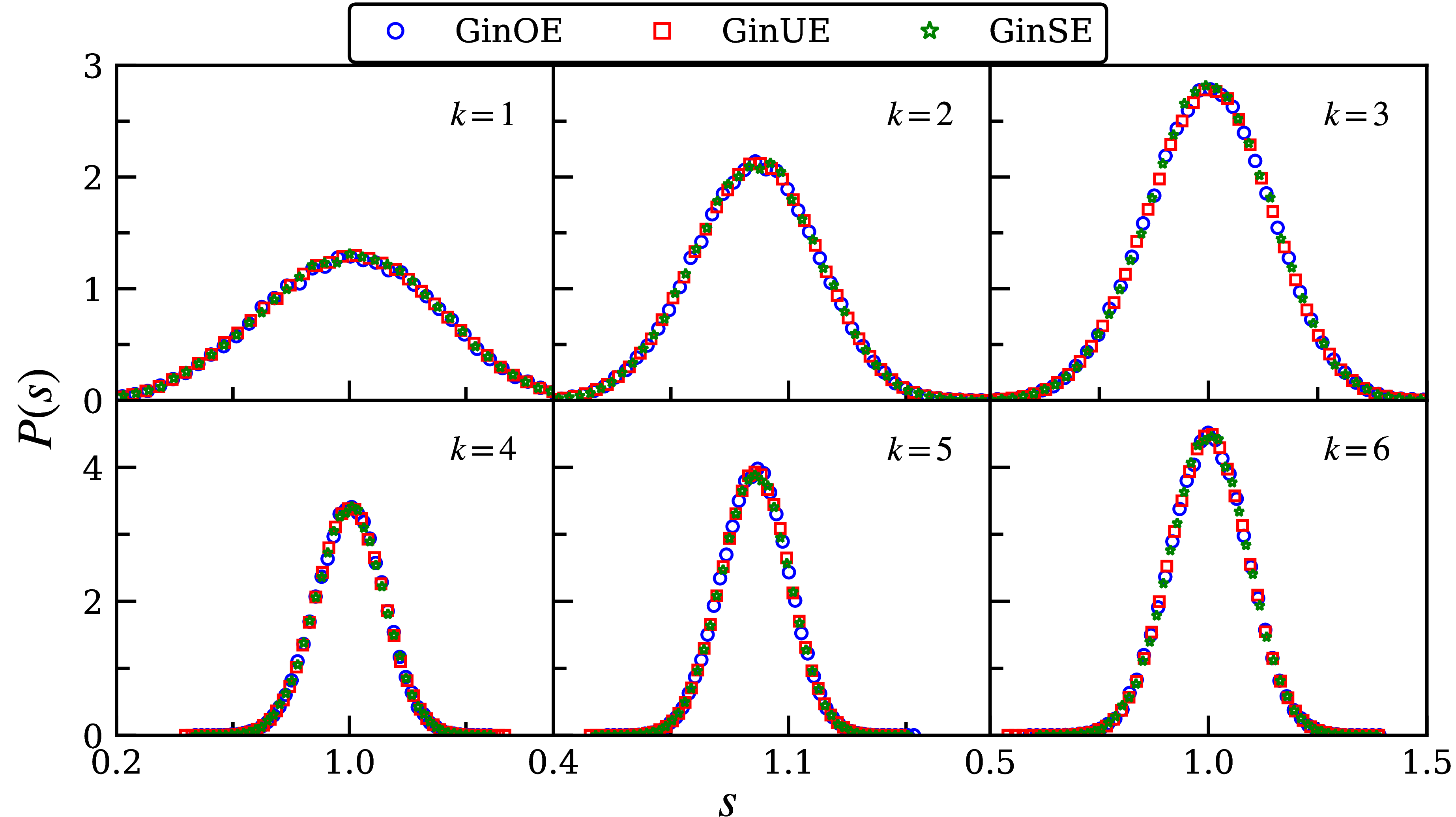}
\caption{\justifying Distribution of the $k$-th order spacings of GinOE, GinUE, and GinSE ensemble.}
\label{fig:k_1_to_6_spacings_Ginibre}
\end{center}
\end{figure}

\begin{figure}
\begin{center}
\includegraphics*[scale=0.36]{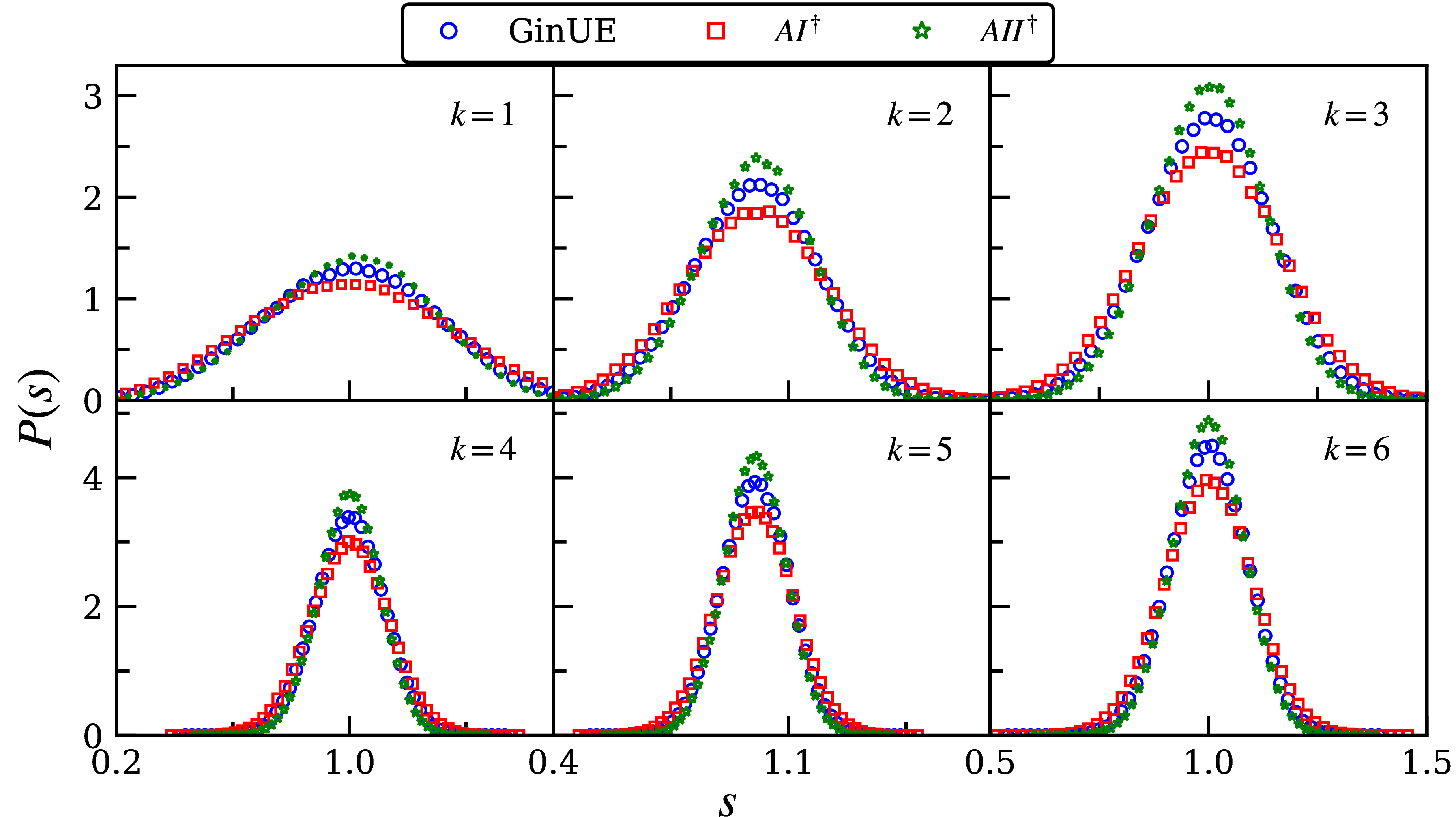}
\caption{\justifying Distribution of the $k$-th order spacings of GinUE, $AI^{\dagger}$, and $AII^{\dagger}$.}
\label{fig:k_1_to_6_spacings_Ginibre_AI_dag_AII_dag}
\end{center}
\end{figure}

\begin{figure}
\begin{center}
\includegraphics*[scale=0.36]{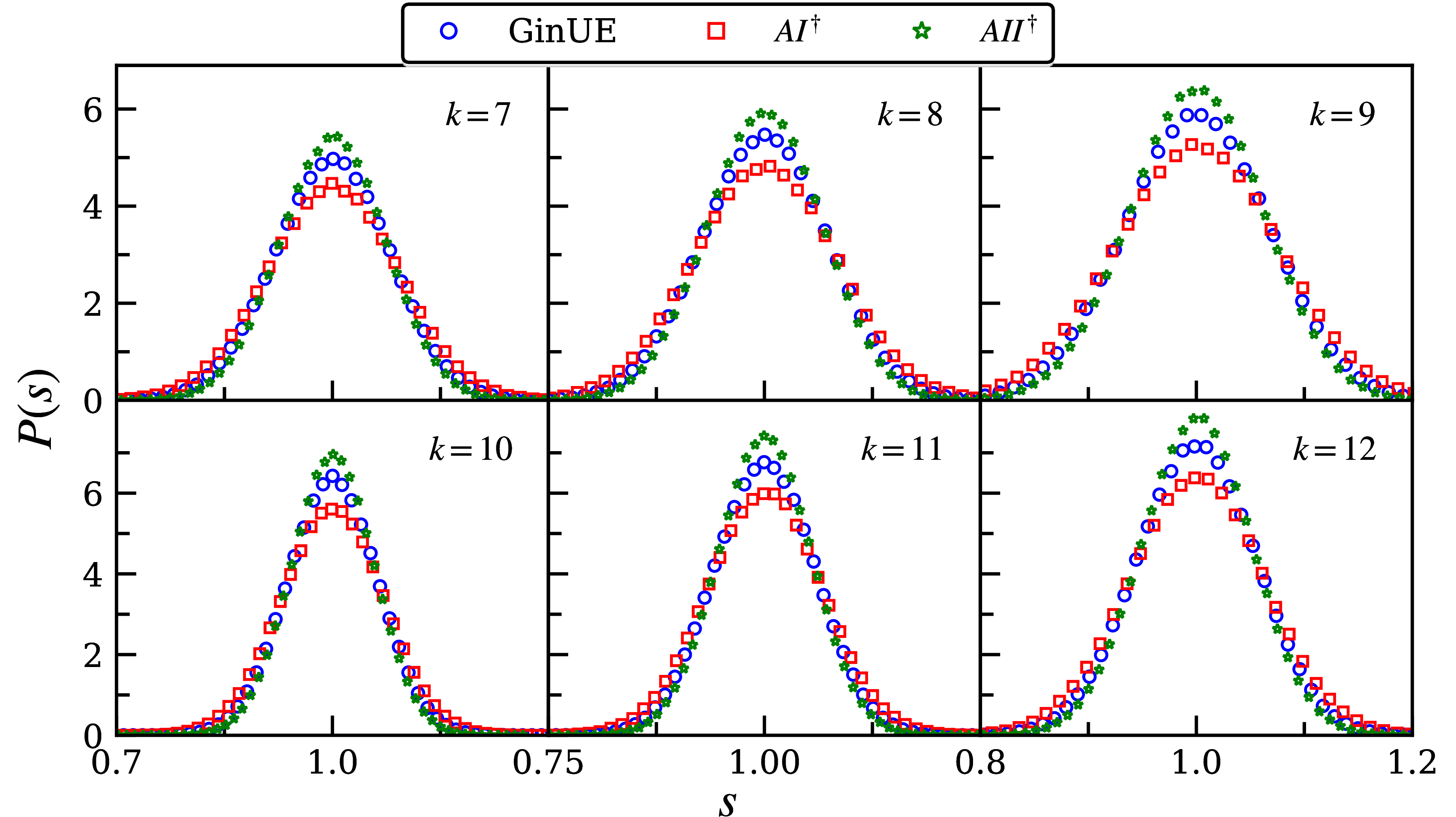}
\caption{\justifying Same as Fig.~\ref{fig:k_1_to_6_spacings_Ginibre_AI_dag_AII_dag}, but for different values of $k$.}
\label{fig:k_7_to_12_spacings_Ginibre_AI_dag_AII_dag}
\end{center}
\end{figure}

\begin{figure}
\begin{center}
\includegraphics*[scale=0.35]{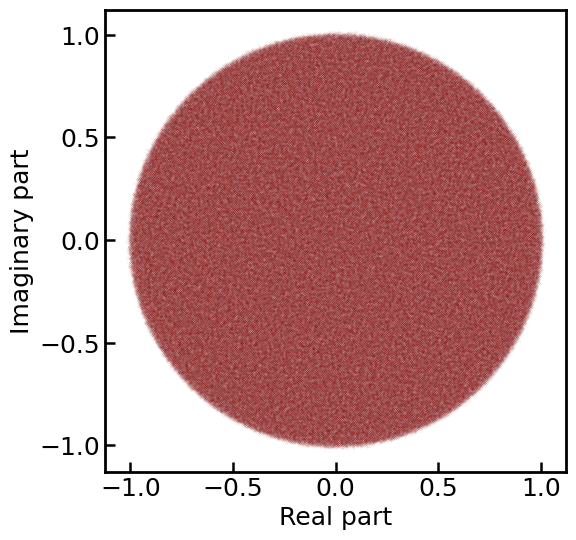}
\caption{\justifying Scatter plot of complex eigenvalues of $100$ Bernoulli complex symmetric matrices of dimension $N=10000$.}
\label{fig:Bernoulli_complex_symmetric}
\end{center}
\end{figure}

\begin{figure}
\begin{center}
\includegraphics*[scale=0.36]{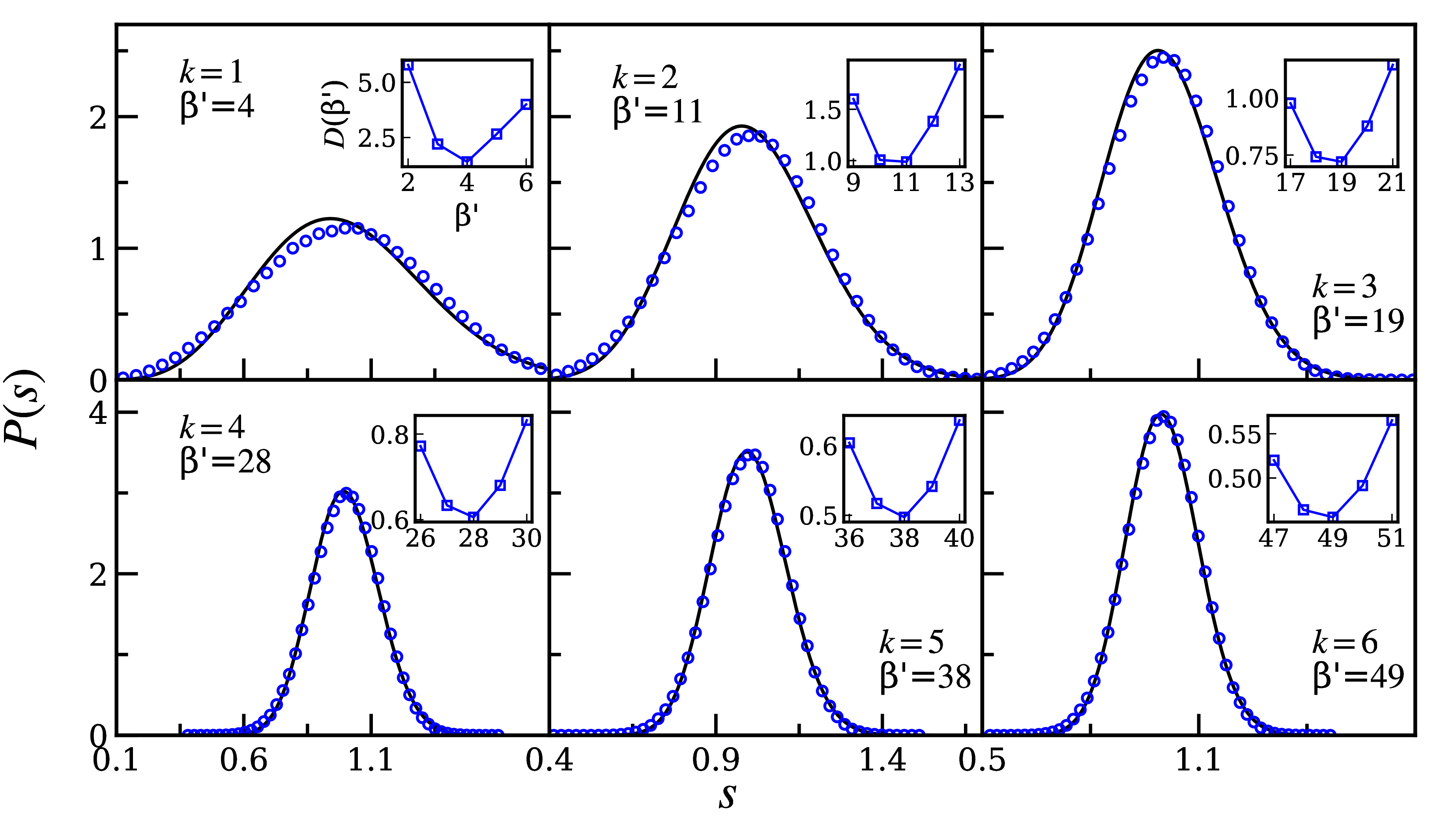}
\caption{\justifying Distribution of the $k$-th order spacings (circles) for the Bernoulli complex symmetric spectra. Here, $N=10000$ and $n=1000$. The solid curve corresponds to $P(s,\beta')$ as given in Eq.~(\ref{SuppEq:PSBetA}), in which $\beta$ is substituted by $\beta'$ and $\beta'$ is given in Table~\ref{Table: HOS_Bernoulli_complex_symmetric}. The insets show $D(\beta')$ as a function of $\beta'$.}
\label{fig: Bernoullispacings1}
\end{center}
\end{figure}

\begin{figure}
\begin{center}
\includegraphics*[scale=0.35]{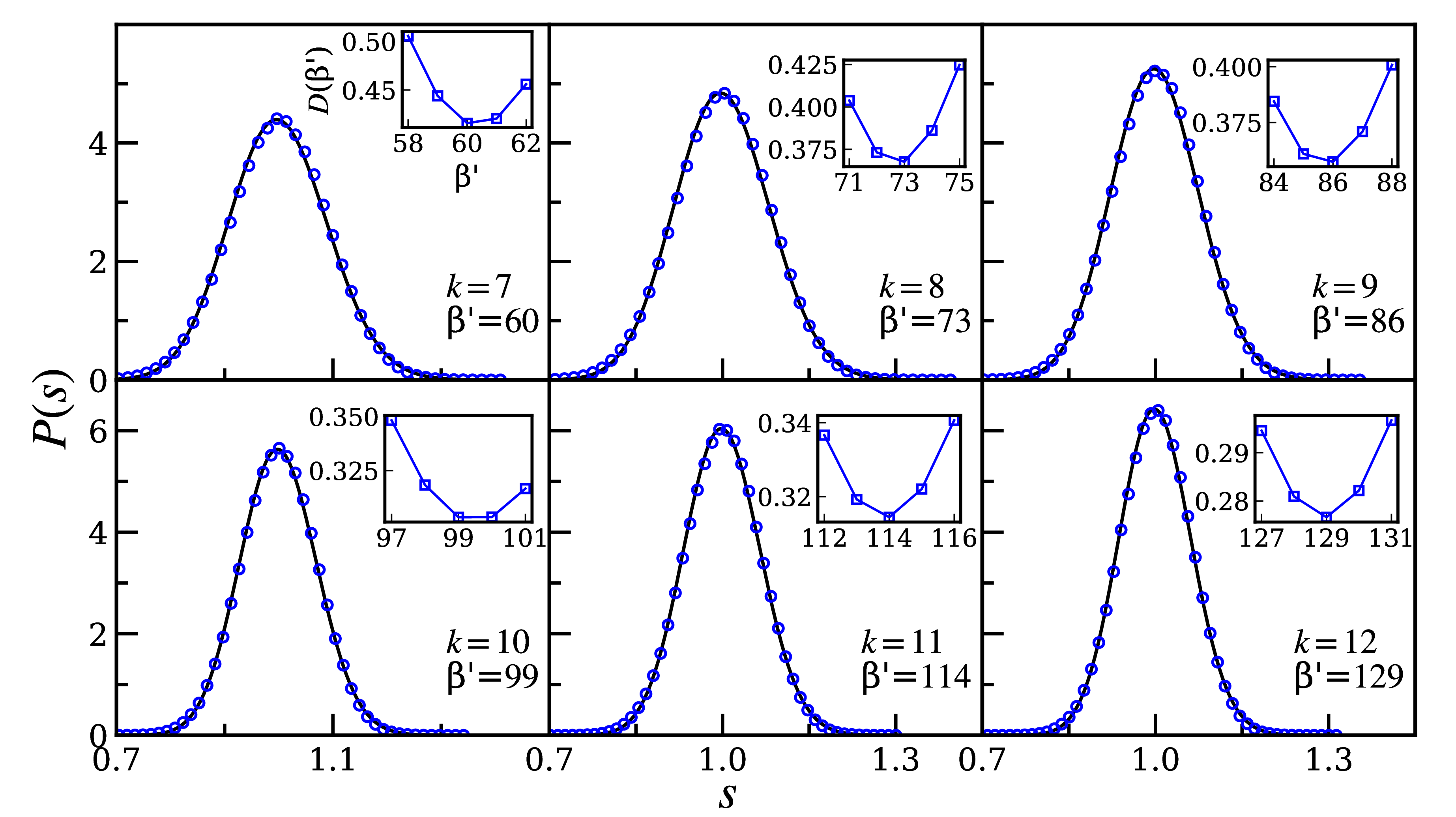}
\caption{\justifying Same as Fig.~\ref{fig: Bernoullispacings1} but for different values of $k$.}
\label{fig: Bernoullispacings2}
\end{center}
\end{figure}

\begin{figure}
\begin{center}
\includegraphics*[scale=0.35]{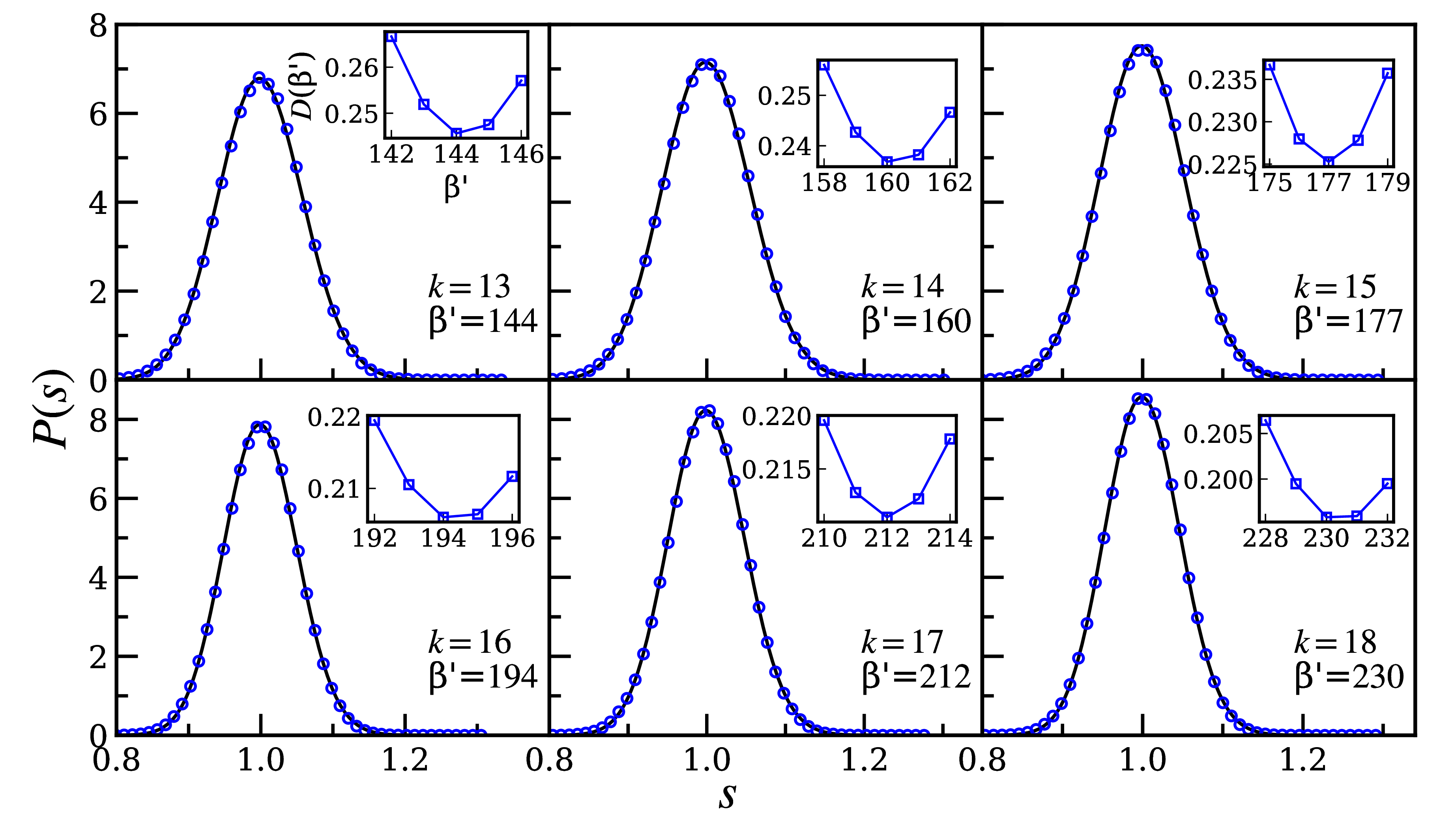}
\caption{\justifying Same as Fig.~\ref{fig: Bernoullispacings1} but for different values of $k$.}
\label{fig: Bernoullispacings3}
\end{center}
\end{figure}

\suppsection{Eigenvalue distributions}
In this section, the Figs.~\ref{supfig:Ginibre_ensemble_ev}-\ref{supfig:AI_dagger_AII_dagger_ev} show the scatter plots corresponding to the three symmetry classes of Ginibre, $AI^\dagger$, and $AII^\dagger$. The plots are uniform for all cases, except that there is an accumulation on the real line for GinOE and a depletion of eigenvalues in the vicinity of the real line for GinSE.
\suppsection{Illustration of our results through additional plots}
\label{supsec: more_plots}
In the case of Hermitian random matrices, NNS is defined as $s_{l}=E_{l+1}-E_{l}$, where $E_{l},\,\,\, l=1,2,3,\ldots$ are the eigenvalues of the corresponding matrix. Here, $E_{l}< E_{l+1}$ for all $l$. Its distribution for random matrices is given by \cite{mehta2004random}:
\begin{equation}
P(s,\beta)=A_{\beta}s^{\beta}\exp(-C_{\beta}s^2),\,\,\,\,\beta=1, 2, 4,
\label{SuppEq:PSBetA}
\end{equation}
where $A_{\beta}$ and $C_{\beta}$ are normalization constants that depend on $\beta$.
\suppsubsection{GinOE case}
\label{supsubsec: GinOE}
In this subsection, we give additional plots of the HOSD of the eigenvalues of GinOE. These are shown in Figs.~\ref{fig:GinOE2_SPACINGS}-\ref{fig:GinOE3_SPACINGS}. Here, we can see excellent agreement between the HOS data of the GinOE ensemble and the analytical distribution function (as per Eq.~\ref{SuppEq:PSBetA}).
\suppsubsection{GinUE case}
\label{supsubsec: GinUE}
In this subsection, we give a few more plots of the HOSD of the eigenvalues of GinUE. These are shown in Figs.~\ref{fig:GinUE2_SPACINGS}-\ref{fig:GinUE3_SPACINGS}. Here, we can see excellent agreement between the HOS data of the GinUE ensemble and the analytical distribution function.
\suppsubsection{GinSE case}
In this subsection, we give additional plots of the HOSD for the eigenvalues of the GinSE. These are shown in Figs.~\ref{fig:GinSE2_SPACINGS}-\ref{fig:GinSE3_SPACINGS}. Here, we can see that the HOS data of GinSE ensemble and the analytical distribution function agree very well with each other.
\label{supsubsec: GinSE}
\suppsubsection{$AI^{\dagger}$ case}
In this subsection, we give additional plots of HOSD the of the eigenvalues of $AI^{\dagger}$. These are shown in Figs.~\ref{fig: AI_dagger2_SPACINGS}-\ref{fig: AI_dagger3_SPACINGS}. Here, we can see excellent agreement between the HOS data of $AI^{\dagger}$ and the analytical distribution function.
\label{supsubsec: AI_dagger spacings}
\suppsubsection{$AII^{\dagger}$ case}
\label{supsubsec: AII_dagger spacings}
In this subsection, we give additional plots for the HOSD of the eigenvalues of $AII^{\dagger}$. These are shown in Figs.~\ref{fig: AII_dagger2 spacings}-\ref{fig: AII_dagger3 spacings}. Here, we can see excellent agreement between the HOS data of $AII^{\dagger}$ and the analytical distribution function.
\suppsubsection{Plotting GinOE, GinUE, and GinSE spacing distributions together}
\label{supsubsec: Plotting together}
In this subsection, the spacing distributions of GinOE, GinUE, and GinSE are plotted together for $k=1$ to $6$ in Fig.~\ref{fig:k_1_to_6_spacings_Ginibre}. It is very clear that the spacing distributions beyond nearest neighbors also agree with each other for the Ginibre ensemble. From Figs.~\ref{fig:k_1_to_6_spacings_Ginibre_AI_dag_AII_dag} and \ref{fig:k_7_to_12_spacings_Ginibre_AI_dag_AII_dag}, it is further observed that the spacing distributions beyond nearest neighbors for the three universality classes are different. 
\suppsection{Bernoulli complex symmetric random matrices}
\label{supsec:Bernoulli complex symmetric}
In this section, we consider complex symmetric random matrices with Bernoulli-distributed entries. The eigenvalue distribution of this matrix is plotted in Fig.~\ref{fig:Bernoulli_complex_symmetric}. They are uniformly distributed on a unit disk. We study HOSD of these eigenvalues and find them obeying the statistics of complex symmetric random matrices with Gaussian-distributed entries ($AI^{\dagger}$). The obtained values of $\beta'$ for both are the same for a given $k$, with a difference of $\pm 1$, for very few $k$ values. These values of $\beta'$ are noted in Table~\ref{Table: HOS_Bernoulli_complex_symmetric}. These observed $\beta'$ values increase with increasing $k$. The results are illustrated in Figs.~\ref{fig: Bernoullispacings1}-\ref{fig: Bernoullispacings3}. Here, we also observe from Fig.~\ref{fig: Bernoullispacings1} that for $k \leq 3$, there is deviation among the distributions, and these deviations decrease with increasing $k$, like other non-Hermitian random matrix ensembles, considered in this paper. 
\begin{table}[t]
\centering
\renewcommand{\arraystretch}{1.5}
\setlength{\tabcolsep}{10pt}
\begin{tabular}{|c|c||c|c||c|c||c|c|}
\hline
\rule{0pt}{12pt}
$k$ & $\beta'$ & $k$ & $\beta'$ & $k$ & $\beta'$& $k$ & $\beta'$\\
\hline
1&4 &6 & 49&11 & 114&16&194\\
2& 11&7 & 60&12 & 129&17&212\\
3& 19&8 & 73&13 & 144&18&230\\
4& 28&9 & 86&14 & 160&19&249\\
5& 38& 10& 99&15 & 177&20&268\\
\hline
\end{tabular}
\caption{\justifying Tabulation of higher-order indices $\beta'$ for spacing distributions of order $k$ for the spectra of Bernoulli complex symmetric matrices. Here, $N=10000$ and $n=1000$.}
\label{Table: HOS_Bernoulli_complex_symmetric}
\end{table}
\suppsection{Product of independent complex Ginibre matrices}
\label{suppsec: Product_GinUE}
In this section, we plot the eigenvalue distributions of the product of independent matrices from GinUE in Fig.~\ref{fig:Product_of_GinUEs_ev}. Here, $M$ represents the number of constituent independent complex Ginibre matrices in a product matrix. We also illustrate a few additional  results, as tabulated in the main text, in Figs.~\ref{fig: M_2_GinUE_Spacings_prosen2}-\ref{fig: M_4_GinUE_Spacings_prosen2}.
\begin{figure*}
    \centering
    \begin{subfigure}{0.29\textwidth}
        \includegraphics[width=\linewidth]{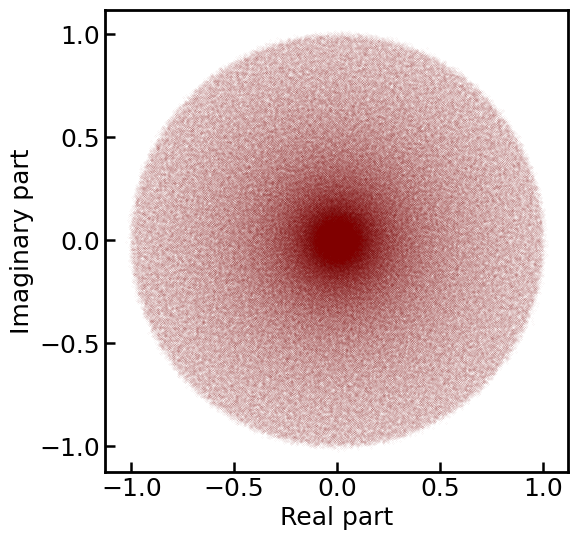}
        \caption{$M=2$}
        \label{fig:M=2_GinUE_ev}
    \end{subfigure}
    \hfill
    \begin{subfigure}{0.29\textwidth}
        \includegraphics[width=\linewidth]{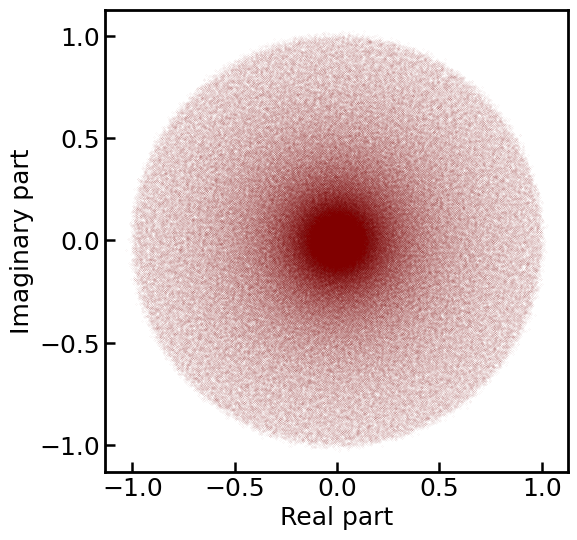}
        \caption{$M=3$}
        \label{fig:M=3_GinUE_ev}
    \end{subfigure}
    \hfill
    \begin{subfigure}{0.29\textwidth}
        \includegraphics[width=\linewidth]{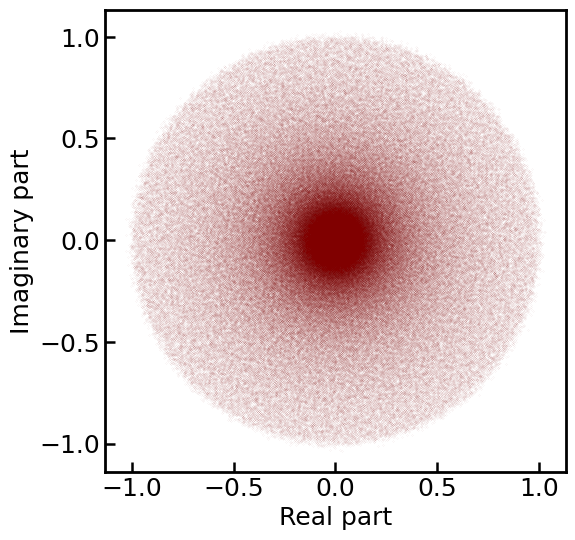}
        \caption{$M=4$}
        \label{fig:M=4_GinUE_ev}
    \end{subfigure}
\caption{\justifying Scatter plot of the complex eigenvalues of the product of $M$ independent complex Ginibre matrices. Here, the dimension of each matrix before product is $N= 5000$ and $n=100$, each for $M=2, 3,$ and $4$.}
\label{fig:Product_of_GinUEs_ev}
\end{figure*}
\begin{figure}
\begin{center}
\includegraphics*[scale=0.33]{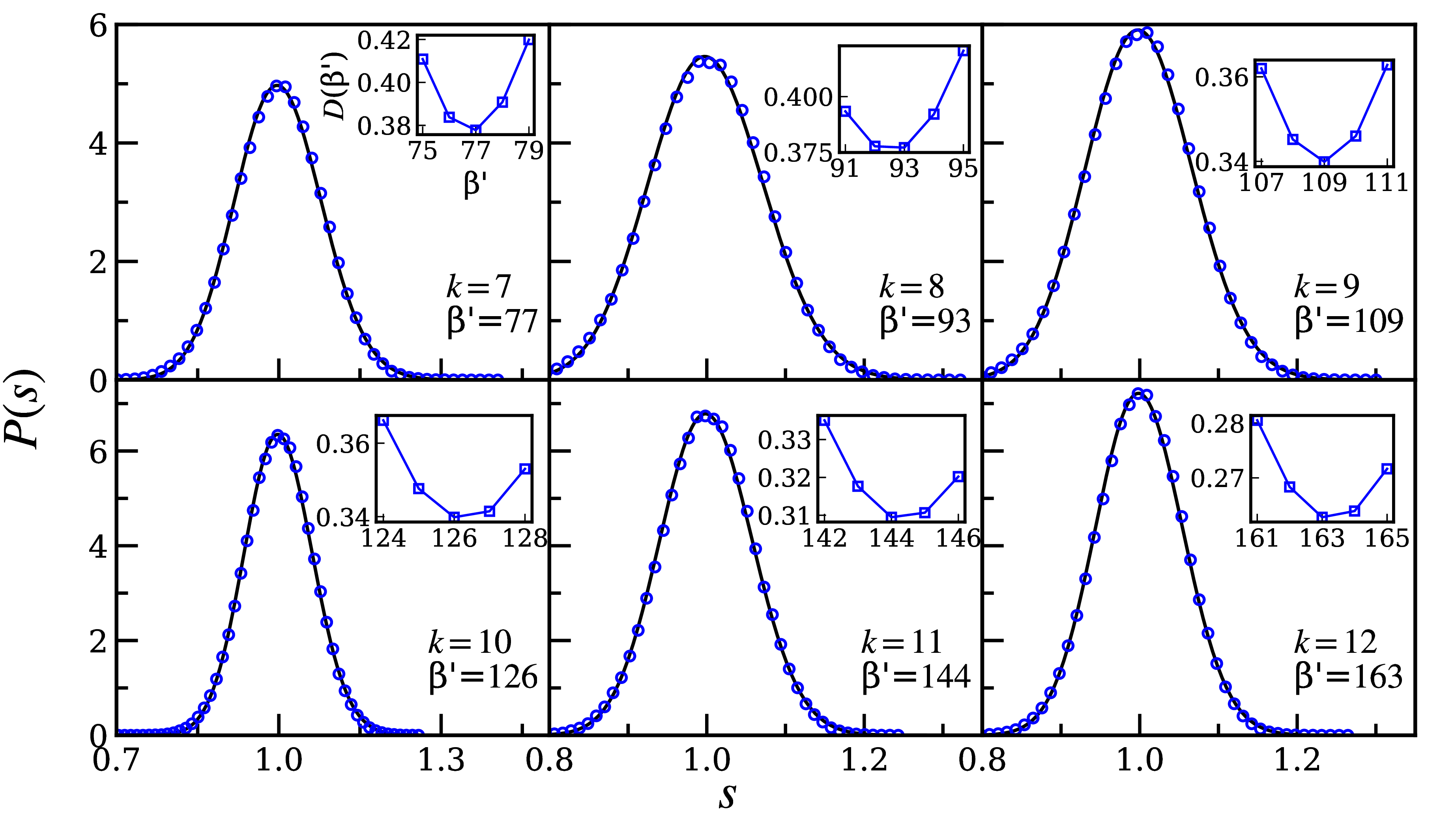}
\caption{\justifying Distribution of the $k$-th order spacings (circles) for the $M=2$ spectra. Here, $N=5000$ and $n=1000$. The solid curve corresponds to $P(s,\beta')$ as given in Eq.~(\ref{SuppEq:PSBetA}), in which $\beta$ is substituted by $\beta'$ and $\beta'$ is given in Table II of the main text. The insets show $D(\beta')$ as a function of $\beta'$.}
\label{fig: M_2_GinUE_Spacings_prosen2}
\end{center}
\end{figure}
\begin{figure}
\begin{center}
\includegraphics*[scale=0.33]{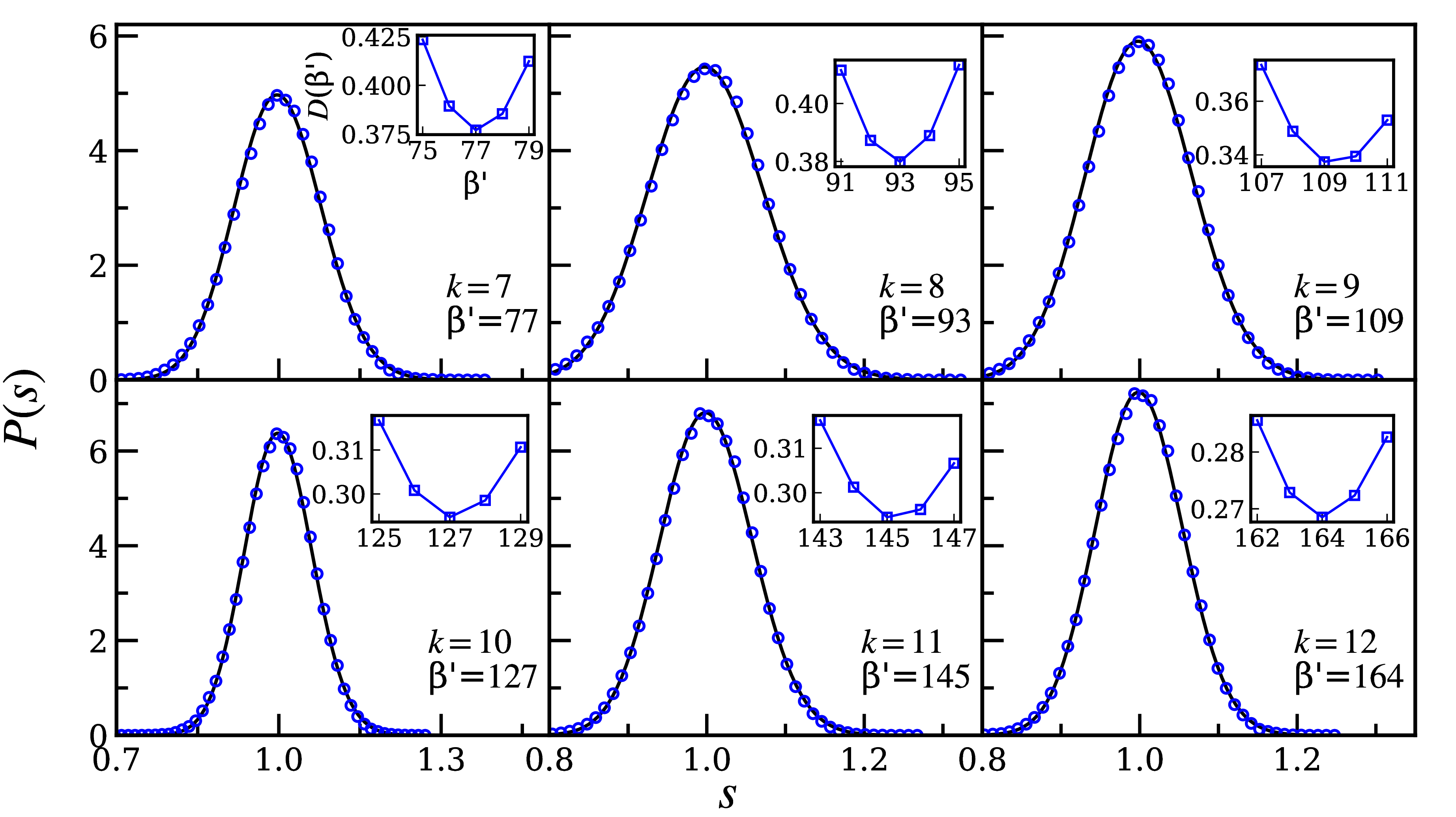}
\caption{\justifying Same as Fig.~\ref{fig: M_2_GinUE_Spacings_prosen2} but for $M=3$.}
\label{fig: M_3_GinUE_Spacings_prosen2}
\end{center}
\end{figure}
\begin{figure}
\begin{center}
\includegraphics*[scale=0.34]{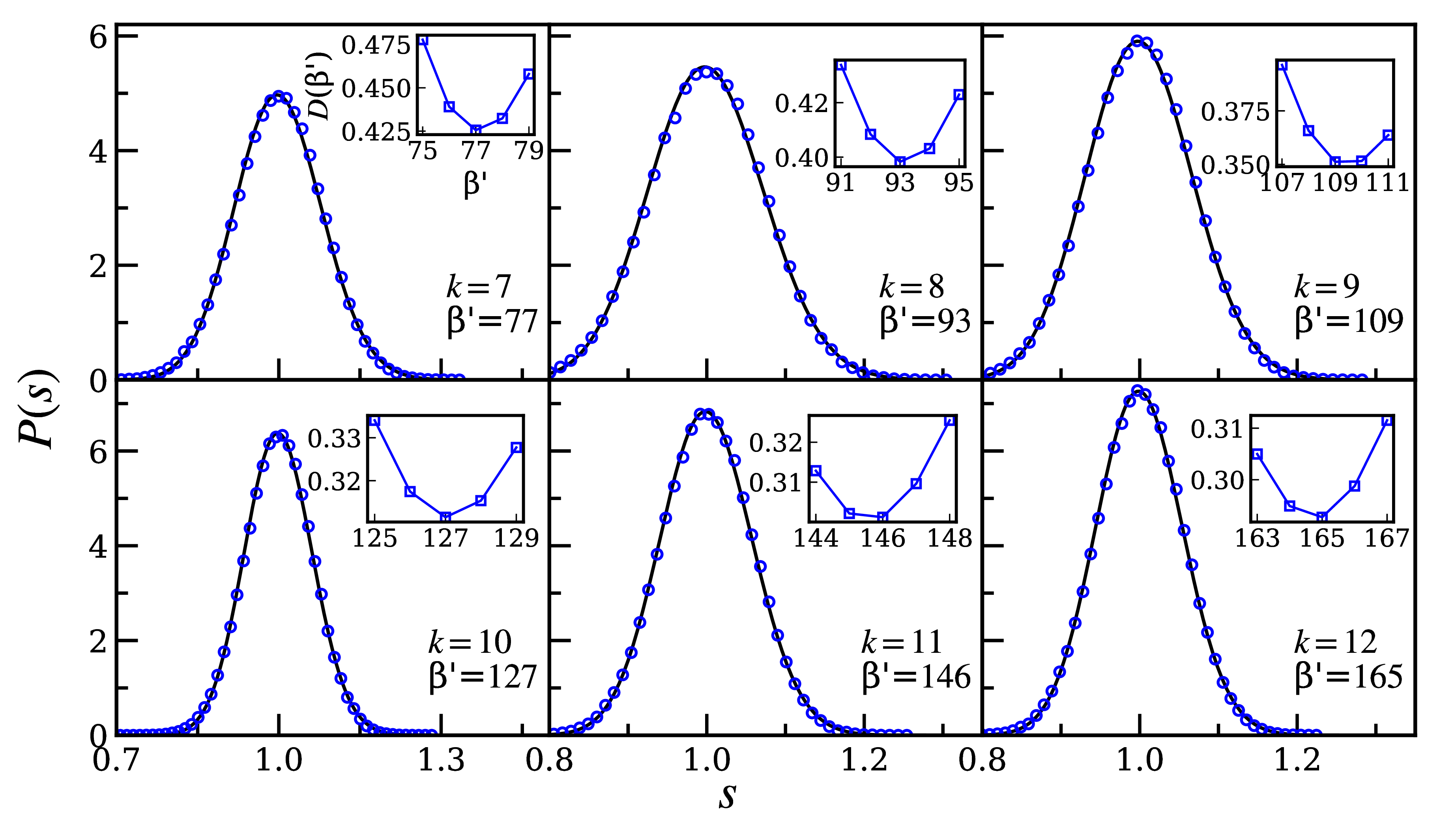}
\caption{\justifying Same as Fig.~\ref{fig: M_2_GinUE_Spacings_prosen2} but for $M=4$.}
\label{fig: M_4_GinUE_Spacings_prosen2}
\end{center}
\end{figure}
\suppsection{Non-Hermitian SYK model}
\label{supsec: nhsyk}
In this section, we have shown the scatter plots of eigenvalues in Fig.~\ref{fig: nhsyk_ev} corresponding to $N=24$ and $q=6$, $N=24$ and $q=8$, and $N=28$ and $q=8$, which belong to the classes GinUE, $AI^{\dagger}$, and $AII^{\dagger}$, respectively. Here, we can see that the eigenvalues are scattered uniformly in the complex plane. In addition to that, we present a few more plots corresponding to the results tabulated in the main text. These are illustrated in Figs.~\ref{fig: k_7_to_12_Nhsyk_GinUE}-\ref{fig: k_7_to_12_Nhsyk_AII_dagger}. 
\begin{figure*}
    \centering
    \begin{subfigure}{0.3\textwidth}
        \includegraphics[width=\linewidth]{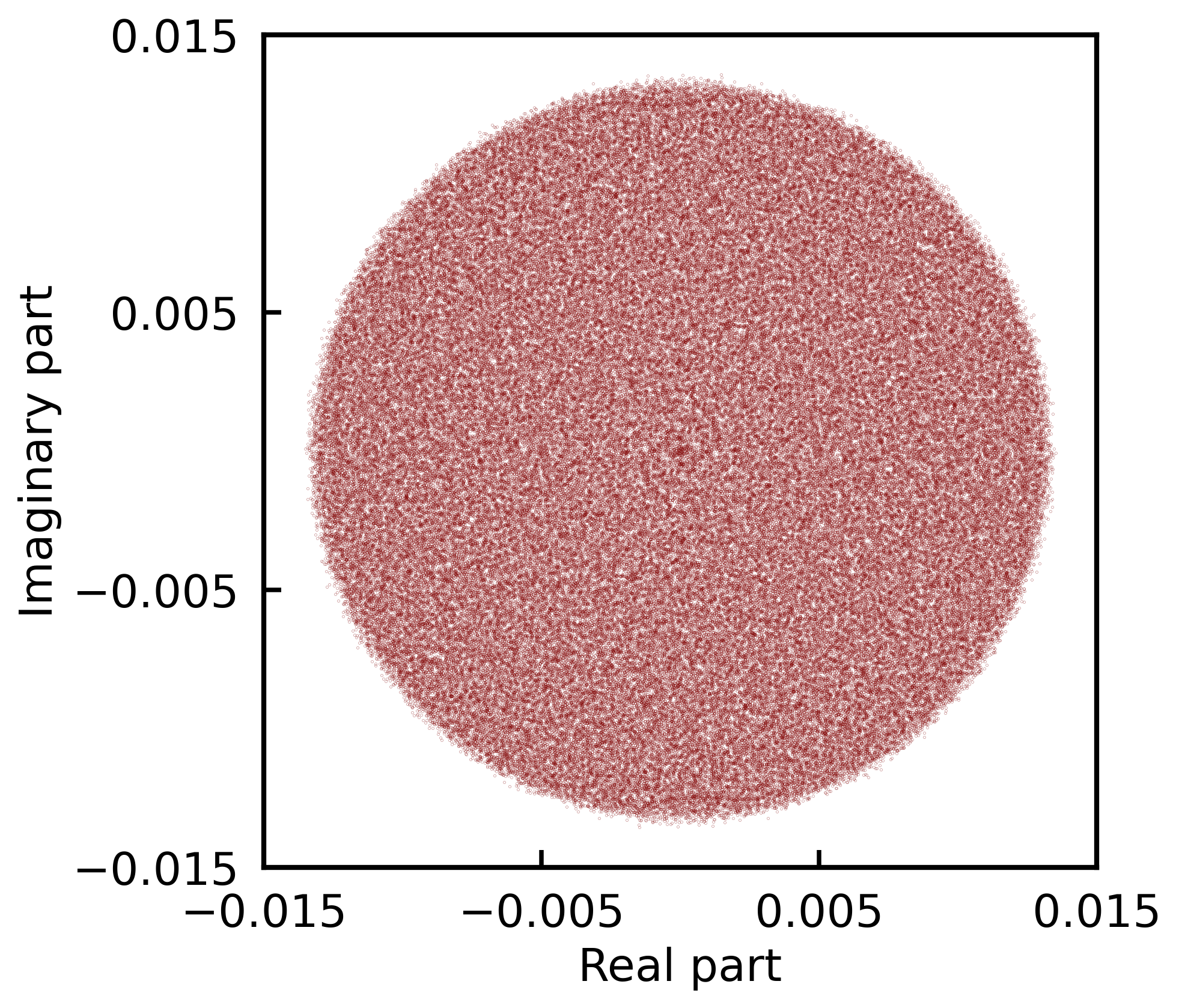}
        \caption{GinUE}
        \label{fig:GinUE_ev}
    \end{subfigure}
    \hfill
    \begin{subfigure}{0.3\textwidth}
        \includegraphics[width=\linewidth]{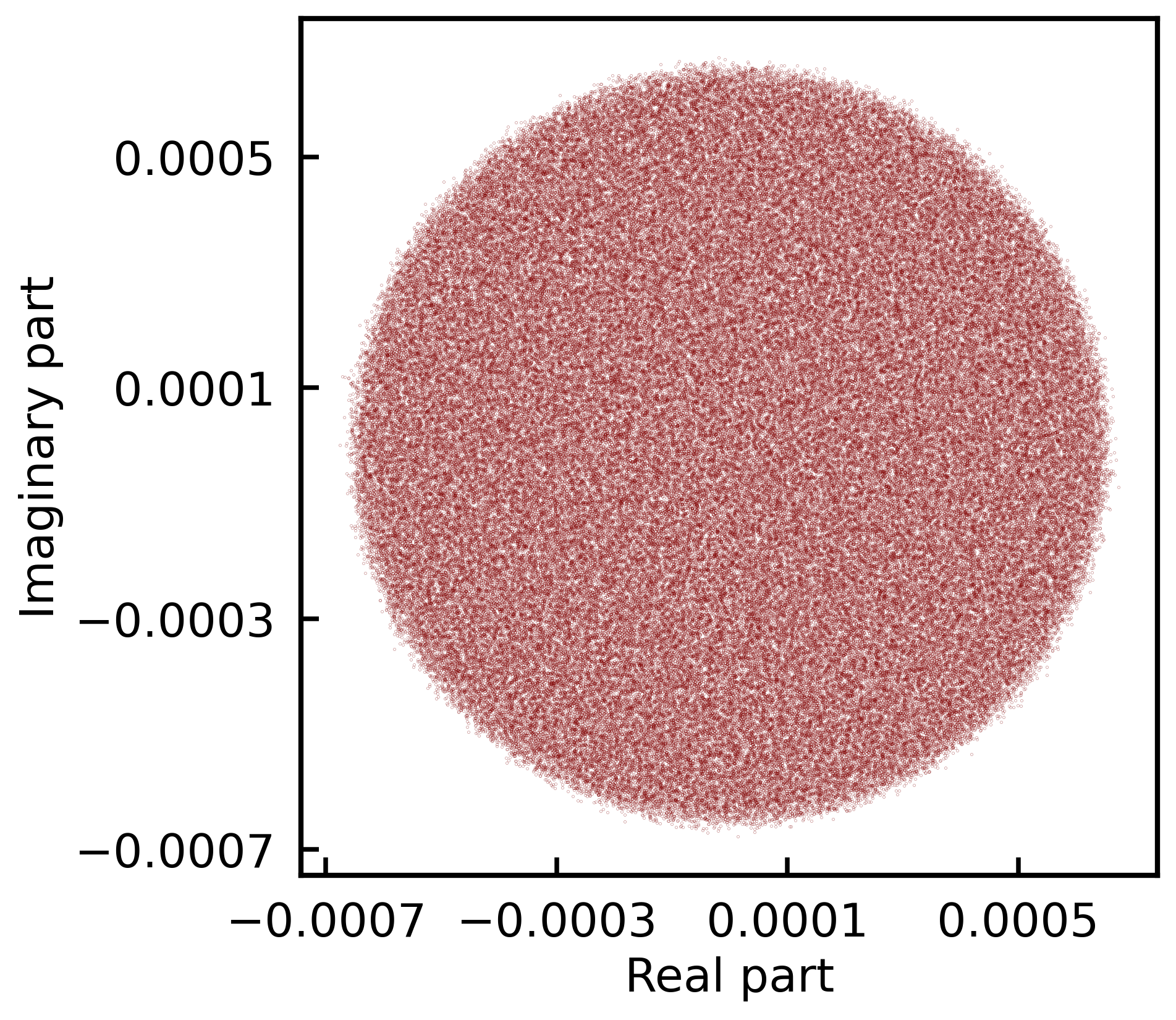}
        \caption{$AI^{\dagger}$}
        \label{fig:AI_dagger_ev}
    \end{subfigure}
    \hfill
    \begin{subfigure}{0.3\textwidth}
        \includegraphics[width=\linewidth]{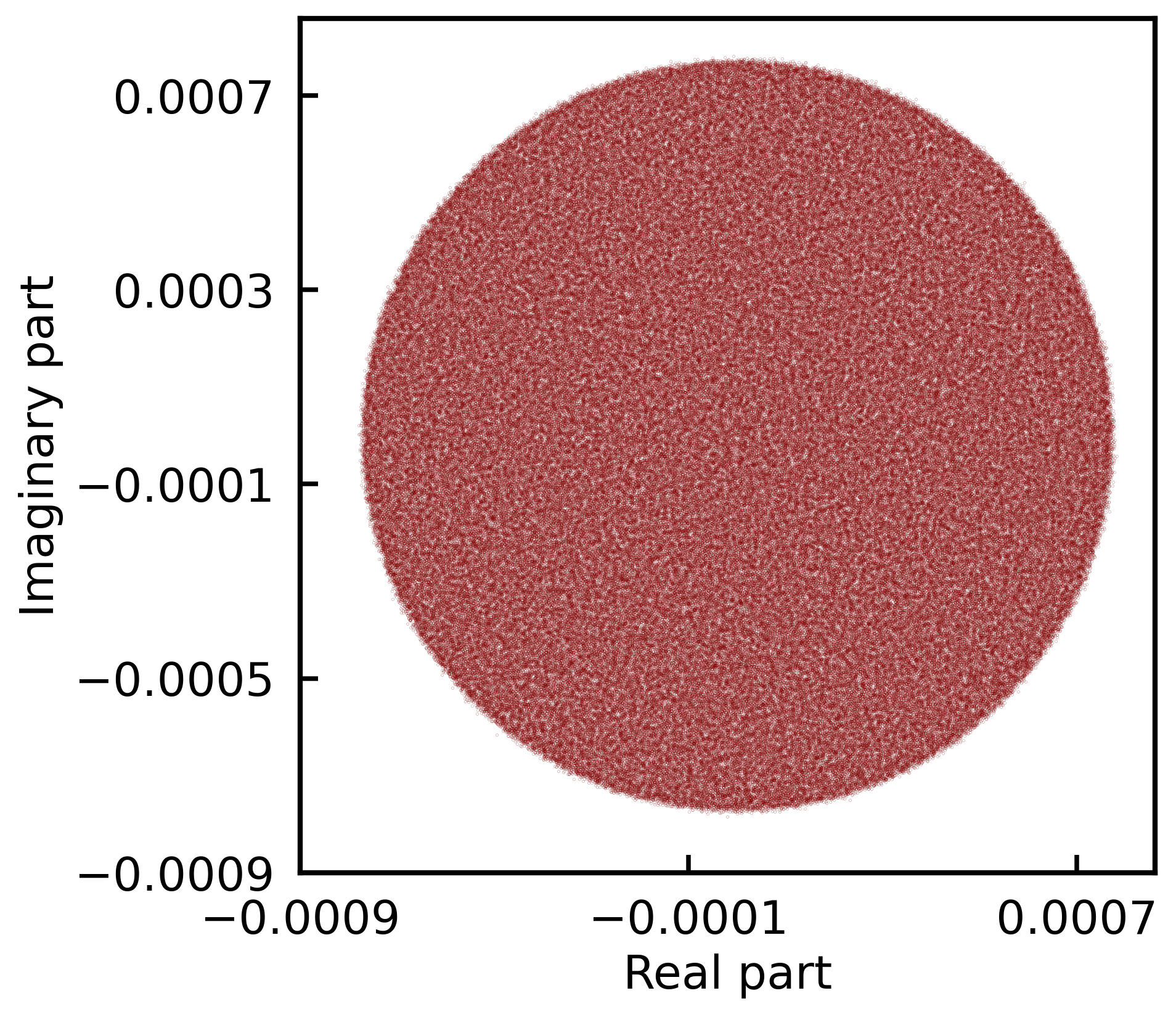}
        \caption{$AII^{\dagger}$}
        \label{fig:AII_dagger_ev}
    \end{subfigure}
\caption{\justifying Scatter plots of eigenvalues corresponding to $N=24$ and $q=6$, $N=24$ and $q=8$, and $N=28$ and $q=8$, which belong to the classes GinUE, $AI^{\dagger}$, and $AII^{\dagger}$, respectively. Here, $n=100$ for each case.}
\label{fig: nhsyk_ev}
\end{figure*}
\begin{figure}
\begin{center}
\includegraphics*[scale=0.33]{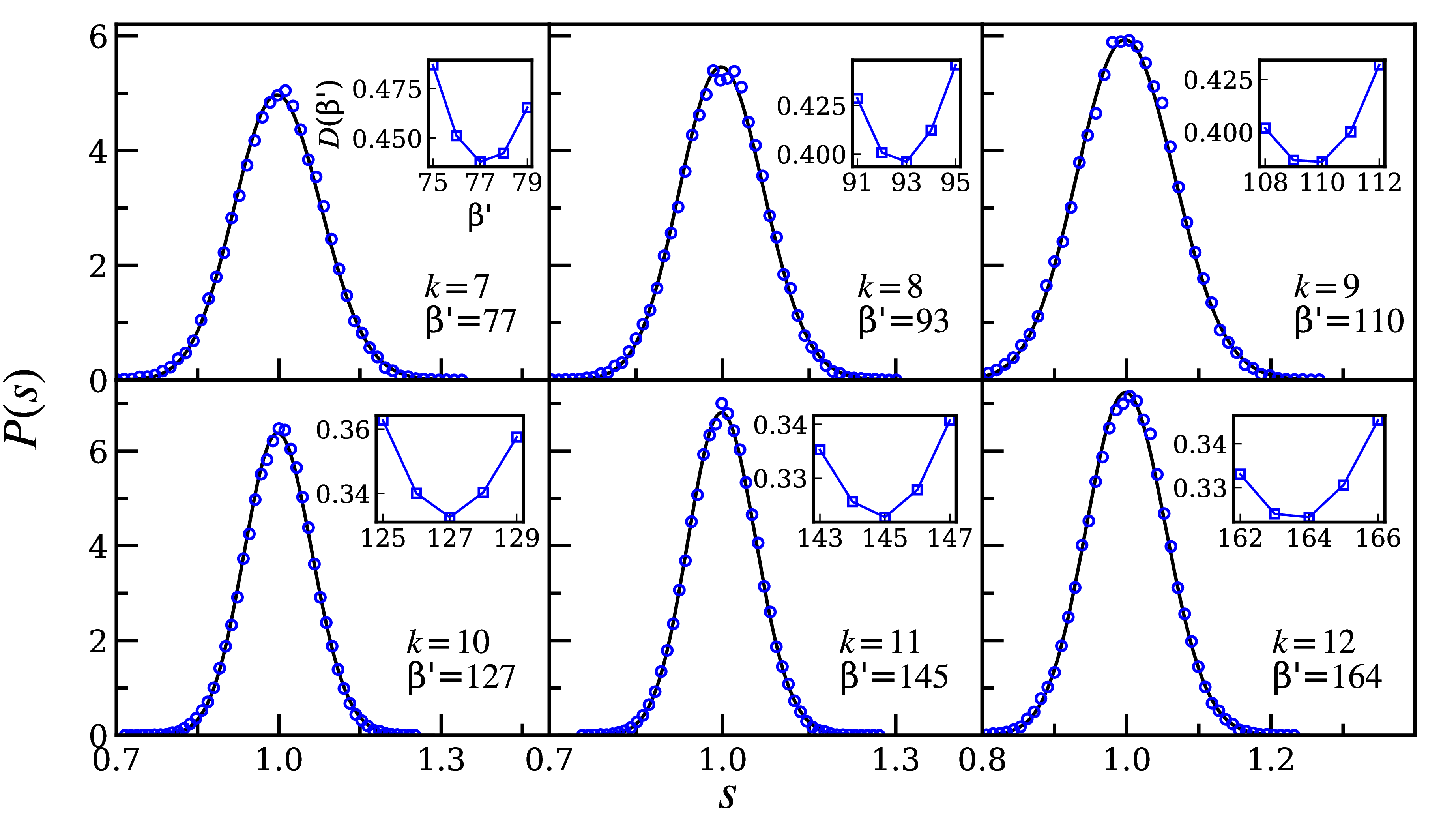}
\caption{\justifying Distribution of the $k$-th order spacings (circles) for the spectra of the nHSYK model for $N=24$ and $q=6$. Here, $n=700$. The solid curve corresponds to $P(s,\beta')$ as given in Eq.~(\ref{SuppEq:PSBetA}), in which $\beta$ is substituted by $\beta'$ and $\beta'$ is given in Table III of the main text. The insets show $D(\beta')$ as a function of $\beta'$.}
\label{fig: k_7_to_12_Nhsyk_GinUE}
\end{center}
\end{figure}
\begin{figure}
\begin{center}
\includegraphics*[scale=0.33]{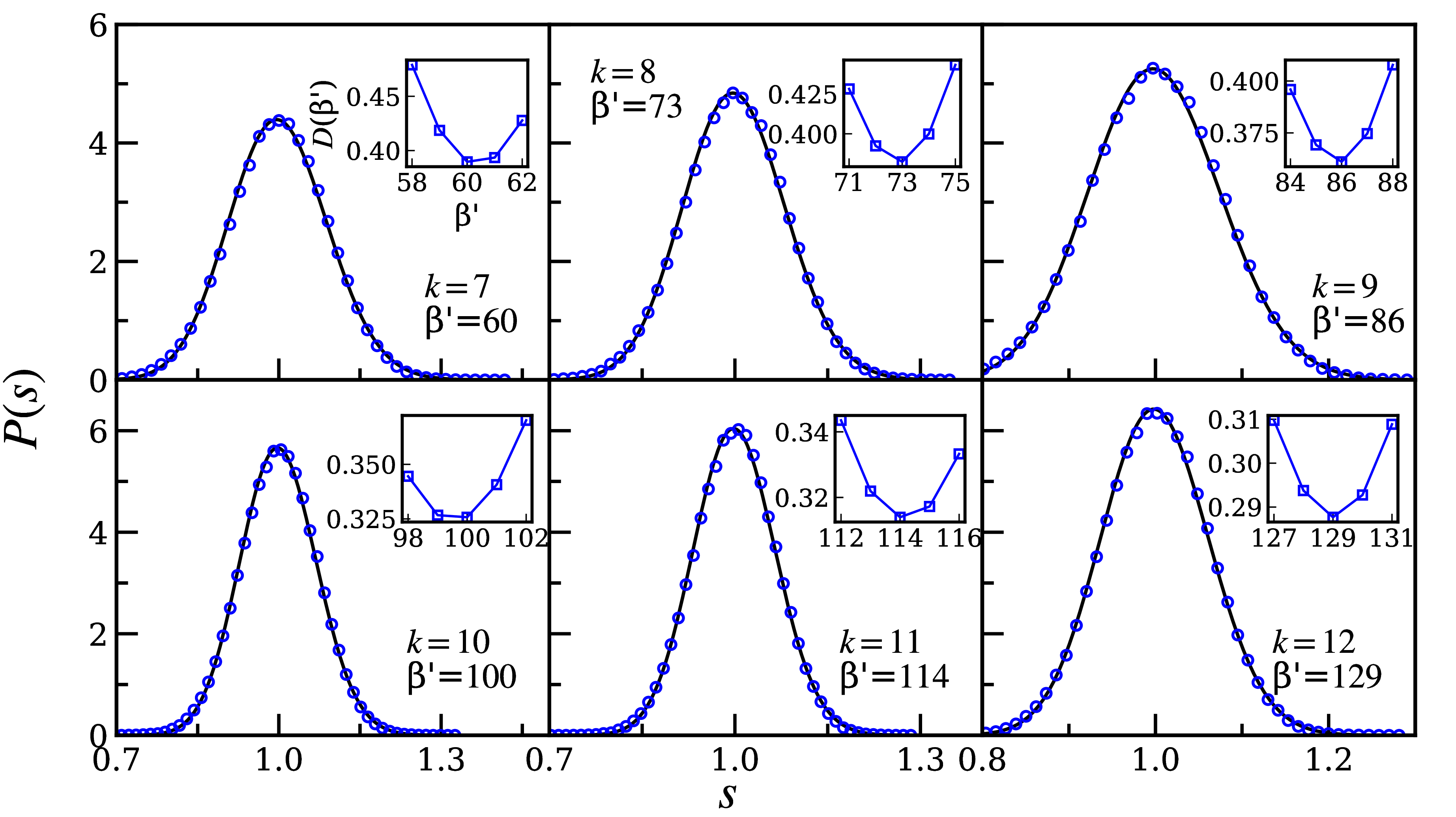}
\caption{\justifying Same as Fig.~\ref{fig: k_7_to_12_Nhsyk_GinUE}, but for $N=24$ and $q=8$. Here, $n=2000$.}
\label{fig: k_7_to_12_Nhsyk_AI_dagger}
\end{center}
\end{figure}
\begin{figure}
\begin{center}
\includegraphics*[scale=0.33]{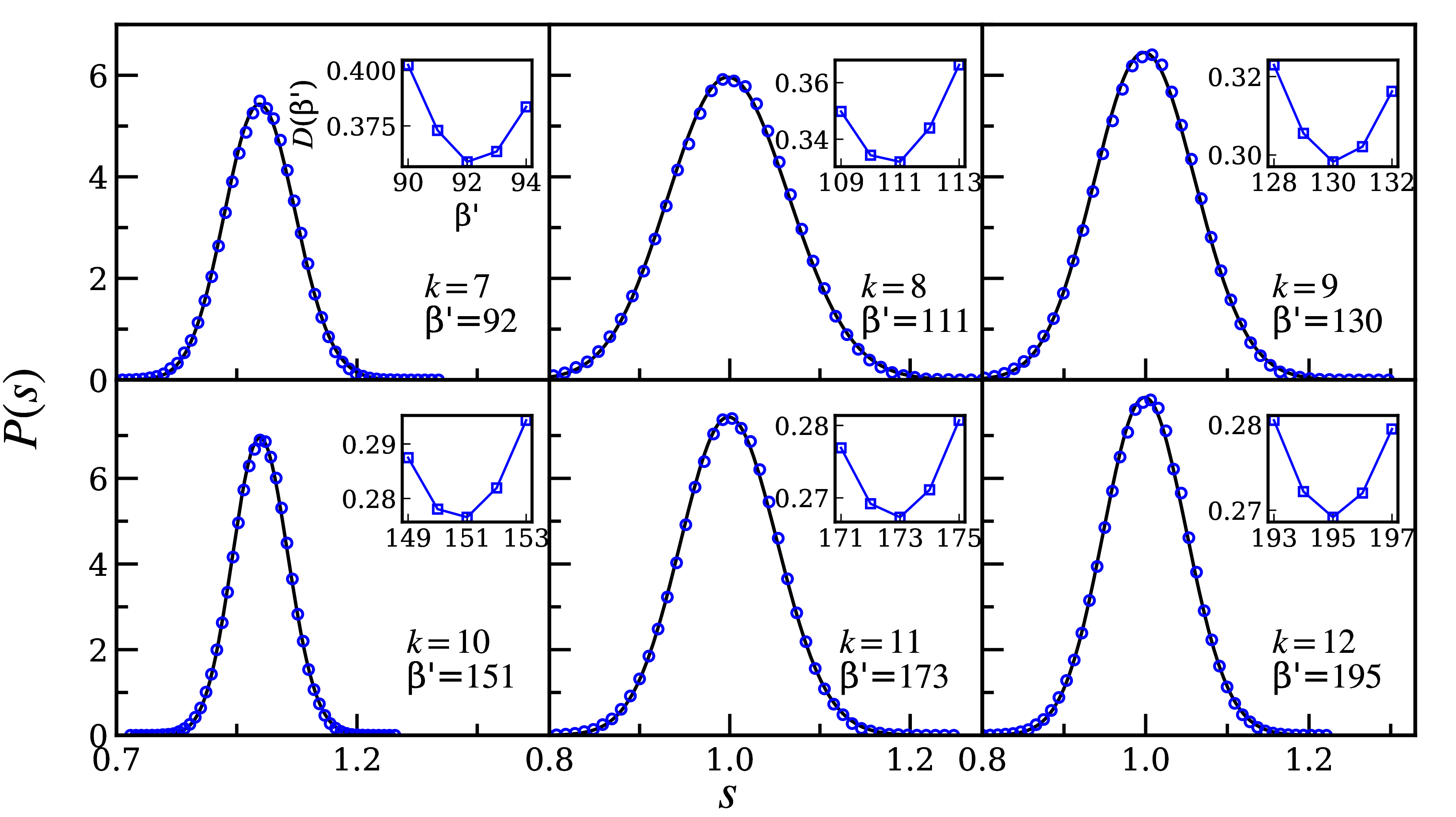}
\caption{\justifying Same as Fig.~\ref{fig: k_7_to_12_Nhsyk_GinUE}, but for $N=28$ and $q=8$. Here, $n=935$.}
\label{fig: k_7_to_12_Nhsyk_AII_dagger}
\end{center}
\end{figure}

\end{document}